\pdfoutput=1

\documentclass[11pt,twoside,a4paper,cmspaper,final,collab]{cms-tdr}
\def\svnVersion{435298P}\def\cmsCernNoTag{CERN-EP-2017-123}\def\cmsCernDate{\today}\def\cmsMessage{Published in the Journal of High Energy Physics as \href{http://dx.doi.org/10.1007/JHEP11(2017)047}{\doi{10.1007/JHEP11(2017)047.}}}

\begin{document}\cmsNoteHeader{HIG-16-041}

\hyphenation{had-ron-i-za-tion}
\hyphenation{cal-or-i-me-ter}
\hyphenation{de-vices}
\RCS$Revision: 427731 $
\RCS$HeadURL: svn+ssh://svn.cern.ch/reps/tdr2/papers/HIG-16-041/trunk/HIG-16-041.tex $
\RCS$Id: HIG-16-041.tex 427731 2017-10-03 18:32:44Z alverson $

\newcommand{\ggH}{\ensuremath{\Pg\Pg\to\PH}\xspace}
\newcommand{\qqH}{\ensuremath{\cPq\cPq\to\cPq\cPq\PH}\xspace}
\newcommand{\WH}{\ensuremath{\PW\PH}\xspace}
\newcommand{\ZH}{\ensuremath{\cPZ\PH}\xspace}
\newcommand{\ttH}{\ensuremath{\ttbar\PH}\xspace}
\newcommand{\qqZZ}{\ensuremath{\qqbar\to\cPZ\cPZ}\xspace}
\newcommand{\ggZZ}{\ensuremath{\Pg\Pg\to\cPZ\cPZ}\xspace}
\newcommand{\HZZfl}{\ensuremath{\PH\to\cPZ\cPZ\to4\ell}\xspace}
\newcommand{\Hllll}{\ensuremath{\PH\to4\ell}\xspace}
\newcommand{\mH}{\ensuremath{m_{\PH}}\xspace}
\newcommand{\mllll}{\ensuremath{m_{4\ell}}\xspace}
\newcommand{\KD}{\ensuremath{\mathcal{D}^\text{kin}_\text{bkg}} \xspace}
\newcommand{\DMeVbfjj}{\ensuremath{\mathcal{D}_\text{2 jet}}\xspace\xspace}
\newcommand{\DMeVbfj}{\ensuremath{\mathcal{D}_\text{1 jet}}\xspace\xspace}
\newcommand{\mlplm}{\ensuremath{m_{\ell^{+}\ell^{-}}}\xspace}
\newcommand{\DMeWh}{\ensuremath{\mathcal{D}_{\WH}}\xspace\xspace}
\newcommand{\DMeZh}{\ensuremath{\mathcal{D}_{\ZH}}\xspace\xspace}
\newcommand{\DMeVh}{\ensuremath{\mathcal{D}_{\mathrm{V}\PH}}\xspace\xspace}
\newcommand{\Zee}{\mathrm{\cPZ\to\Pep\Pem}\xspace}
\newcommand{\MassD}{\mathrm{\mathcal{D}_\text{mass}} \xspace}
\newcommand{\MassDprime}{\mathrm{\mathcal{D}'_\text{mass}} \xspace}
\newcommand{\muV}{\ensuremath{\mu_{\mathrm{VBF},\mathrm{V\PH}}} \xspace}
\newcommand{\muF}{\ensuremath{\mu_{\Pg\Pg\PH,\,\ttbar\PH}} \xspace}
\newcommand{\usedLumi}{35.9\fbinv\xspace}
\newcommand{\valMuAtRunIMass}{\ensuremath{1.05^{+0.19}_{-0.17}}\xspace}
\newcommand{\valMuVAtRunIMass}{\ensuremath{0.00}^{+0.81}_{-0.00}\xspace}
\newcommand{\valMuFAtRunIMass}{\ensuremath{1.19}^{+0.21}_{-0.20}\xspace}
\newcommand{\valMassThreeDRefit}{\ensuremath{125.26\pm 0.20\stat\pm0.08\syst}\xspace}

\cmsNoteHeader{HIG-16-041}
\title{Measurements of properties of the Higgs boson decaying into the four-lepton final state in pp collisions at $\sqrt{s}=13\TeV$}

\date{\today}

\abstract{
Properties of the Higgs boson are measured in the $\PH\to\Z\Z\to4\ell$ ($\ell=\Pe,\PGm$) decay channel.
A data sample of proton-proton collisions at $\sqrt{s}=13\TeV$,
collected with the CMS detector at the LHC and
corresponding to an integrated luminosity of 35.9\fbinv is used.
The signal strength modifier $\mu$, defined as the ratio of the observed Higgs boson rate in the $\PH\to\Z\Z\to4\ell$ decay channel to the standard model expectation,
is measured to be $\mu=1.05^{+0.19}_{-0.17}$ at $m_{\PH}=125.09\GeV$, the combined ATLAS and CMS measurement of the Higgs boson mass.
The signal strength modifiers for the individual Higgs boson production modes are also measured.
The cross section in the fiducial phase space defined by the requirements on lepton kinematics and event topology
is measured to be $2.92~^{+0.48}_{-0.44}\stat~^{+0.28}_{-0.24}\syst\unit{fb}$, which is compatible with the standard model prediction of $2.76\pm0.14\unit{fb}$.
Differential cross sections are reported as a function of the transverse momentum of the Higgs boson, the number of associated jets, and the transverse momentum of the leading associated jet.
The Higgs boson mass is measured to be $m_{\PH}=125.26 \pm 0.21\GeV$ and the width is constrained using the on-shell invariant mass distribution
to be $\Gamma_{\PH}<1.10\GeV$, at 95\% confidence level.
}

\hypersetup{%
pdfauthor={CMS Collaboration},%
pdftitle={
Measurements of properties of the Higgs boson decaying into the four-lepton final state in pp collisions at sqrt(s) = 13 TeV
},%
pdfsubject={CMS},%
pdfkeywords={CMS, physics, Higgs mass, multileptons, fiducial cross section}}

\maketitle

\section{Introduction}
\label{sec:intro}

In 2012, the ATLAS and CMS Collaborations reported
the observation of a new particle with a mass
of approximately 125\GeV and properties consistent with that of the standard model (SM) Higgs boson~\cite{Aad:2012tfa,Chatrchyan:2012ufa,Chatrchyan:2013lba}.
Further studies by the two experiments~\cite{CMS:2014ega, AtlasProperties, CMS:2015kwa}, using the entire LHC Run 1 data set at center-of-mass energies of 7 and 8\TeV indicate agreement within their uncertainties between the measured properties of the new boson and those predicted for the SM Higgs boson~\cite{Englert:1964et,Higgs:1964ia,Higgs:1964pj,Guralnik:1964eu,Higgs:1966ev,Kibble:1967sv}.
The ATLAS and CMS Collaborations have also published a combined measurement of the Higgs boson mass of
$m_{\PH}=125.09\pm0.21\stat\pm0.11\syst\GeV$~\cite{Aad:2015zhl}.

The $\HZZfl$ decay channel ($\ell=\Pe,\PGm$) has a large signal-to-background ratio, and the
precise reconstruction of the final-state decay products allows the complete determination of the kinematics of the Higgs boson.
This makes it one of the most important channels to measure the properties of the Higgs boson.
Measurements performed by the ATLAS and CMS Collaborations using this decay channel with the LHC Run 1 data include the determination of the mass and spin-parity of the boson~\cite{CMSH4lLegacy,CMSH4lSpinParity,CMSH4lAnomalousCouplings,Aad:2014eva,Aad:2015mxa}, its width~\cite{CMSH4lWidth,CMSH4lLifetime,Aad:2015xua}, the fiducial cross sections~\cite{CMSH4lFiducial8TeV,Aad:2015lha}, and the tensor structure of its interaction with a pair of neutral gauge bosons~\cite{CMSH4lAnomalousCouplings,CMSH4lLifetime,Aad:2015mxa}.

In this paper measurements of properties of the Higgs boson decaying into the four-lepton final state in proton-proton (pp) collisions at $\sqrt{s}=13\TeV$ are presented.
Events are classified into categories optimized with respect to those used in Ref.~\cite{CMSH4lLegacy} to provide increased
 sensitivity to subleading production modes of the Higgs boson such as vector boson fusion (VBF) and associated production with
a vector boson ($\WH$, $\ZH$) or top quark pair ($\ttH$).
The signal strength modifier, defined as the ratio of the measured Higgs boson rate in the $\PH\to\Z\Z\to4\ell$ decay channel to the SM expectation, is measured.
The signal strength modifiers for the individual Higgs boson production modes are constrained.
In addition, cross section measurements and dedicated measurements of the mass and width of the Higgs boson are performed.

This paper is structured as follows:
the apparatus and the data samples are described in
Section~\ref{sec:detector} and Section~\ref{sec:datasets}.
Section~\ref{sec:objects} summarizes the event reconstruction and selection.
Kinematic discriminants and event categorization are discussed in Section~\ref{sec:observables}
and Section~\ref{sec:categorization}.
The background estimation and the signal modelling are reported in Section~\ref{sec:bkgd} and Section~\ref{sec:signal}.
We then discuss the systematic uncertainties in Section~\ref{sec:systematics}.
Finally, Section~\ref{sec:results} presents event yields, kinematic distributions, and measured properties.

\section{The CMS detector}
\label{sec:detector}

A detailed description of the CMS detector, together with a definition of the coordinate system used and the relevant kinematic variables, can be found in Ref.~\cite{Chatrchyan:2008zzk}.

The central feature of the CMS apparatus is a superconducting solenoid of 6\unit{m} internal diameter, providing a magnetic field of 3.8\unit{T}.
Within the solenoid volume are a silicon pixel and strip tracker, a lead tungstate crystal electromagnetic calorimeter (ECAL), and a brass and scintillator hadron calorimeter (HCAL), each composed of a barrel and two endcap sections.
Forward calorimeters extend the pseudorapidity ($\eta$) coverage provided by the barrel and endcap detectors.
Muons are detected in gas-ionization chambers embedded in the steel flux-return yoke outside the solenoid.

The silicon tracker measures charged particles within the pseudorapidity range $\abs{\eta}< 2.5$. It consists of 1440 silicon pixel and 15\,148 silicon strip detector modules. For non-isolated particles with transverse momentum \pt between 1 and 10\GeV and $\abs{\eta} < 1.4$, the track resolutions are typically 1.5\% in \pt and 25--90 (45--150)\mum in the transverse (longitudinal) impact parameter \cite{TRK-11-001}.

The electromagnetic calorimeter consists of 75\,848 lead tungstate crystals, which provide coverage in pseudorapidity $\abs{ \eta }< 1.479 $ in the barrel region (EB) and $1.479 <\abs{ \eta } < 3.0$ in the two endcap regions (EE). A preshower detector consisting of two planes of silicon sensors interleaved with a total of $3 X_0$ of lead is located in front of the EE.
The electron momentum is estimated by combining the energy measurement in the ECAL with the momentum measurement in the tracker.
The momentum resolution for electrons with $\pt \approx 45\GeV$ from $\Z \to \Pe \Pe$ decays ranges from 1.7\% for electrons in the barrel region that do not shower in the tracker volume to 4.5\% for electrons in the endcaps that do shower in the tracker volume~\cite{Khachatryan:2015hwa}.

Muons are measured in the pseudorapidity range $\abs{\eta}< 2.4$, with detection planes made using three technologies: drift tubes, cathode strip chambers, and resistive plate chambers. Matching muons to tracks measured in the silicon tracker results in a relative transverse momentum resolution for muons with $20 <\pt < 100\GeV$ of 1.3--2.0\% in the barrel ($\abs{\eta}< 0.9$) and better than 6\% in the endcaps ($\abs{\eta}>0.9$).
The \pt resolution in the barrel is better than 10\% for muons with \pt up to 1\TeV~\cite{Chatrchyan:2012xi}.

The first level (L1) of the CMS trigger system~\cite{CMS-TRG-12-001}, composed of custom hardware processors, uses information from the calorimeters and muon detectors to select the most interesting events in a fixed time interval of less than 4\mus. The high-level trigger (HLT) processor farm further decreases the event rate from around 100\unit{kHz} to less than 1\unit{kHz}, before data storage.

\section{Data and simulated samples }
\label{sec:datasets}

This analysis makes use of $\Pp\Pp$ collision data recorded by the CMS detector in 2016,
corresponding to an integrated luminosity of $\usedLumi$.
Collision events are selected by high-level trigger algorithms that require the presence of leptons passing
loose identification and isolation requirements.
The main triggers of this analysis select either a pair of electrons or muons, or an electron and a muon. The minimal transverse momentum with respect to the beam axis of the leading electron (muon) is 23\,(17)\GeV, while that of the subleading lepton is 12\,(8)\GeV.
To maximize the signal acceptance, triggers requiring three leptons with lower $\pt$ thresholds and
no isolation requirement are also used, as are isolated single-electron and single-muon triggers with the thresholds of 27\GeV and 22\GeV, respectively.
The overall trigger efficiency for simulated signal events that pass the full selection chain of this analysis (described in Section~\ref{sec:objects}) is larger
than 99\%. The trigger efficiency is measured in data with a method based on the ``tag-and-probe'' technique~\cite{CMS:2011aa} using a sample of $4\ell$ events collected by the single-lepton triggers. Leptons passing the single-lepton triggers are used as tags and the other three leptons are used as probes.
The efficiency in data is found to be in agreement with the expectation from simulation.

The Monte Carlo (MC) simulation samples for the signals and the relevant background processes are used to estimate backgrounds,
optimize the event selection, and evaluate the acceptance and systematic uncertainties.
The SM Higgs boson signals are generated at next-to-leading order (NLO) in perturbative quantum chromodynamics (pQCD) with
the \POWHEG 2.0~\cite{Alioli:2008gx,Nason:2004rx,Frixione:2007vw} generator
for the five main production modes: gluon fusion ($\ggH$), vector boson fusion ($\qqH$), and associated
production ($\WH$, $\ZH$, and $\ttH$). For $\WH$ and $\ZH$ the \textsc{minlo hvj}~\cite{Luisoni2013} extension of \POWHEG 2.0 is used.
The cross sections for the various signal processes are taken from Ref.~\cite{YR4}, and in particular the cross section for
the dominant gluon fusion production mode is taken from Ref.~\cite{Anastasiou2016}.
The default set of parton distribution functions (PDFs) used in all simulations is NNPDF30\_nlo\_as\_0118~\cite{Ball2012153}.
The decay of the Higgs boson to four leptons is modeled with \textsc{jhugen}~7.0.2~\cite{Gao:2010qx, Bolognesi:2012mm}.
In the case of $\ZH$ and $\ttH$, the Higgs boson is also allowed to decay as $\PH\to\cPZ\cPZ\to2\ell2$X where X stands for either a quark or a neutrino, thus accounting for four-lepton events where two leptons originate from the decay of the
associated $\cPZ$ boson or top quarks.
In all of the simulated samples, vector bosons are allowed to decay to $\tau$-leptons such that
this contribution is included in all estimations.

To generate a more accurate signal model, the $\pt$ spectrum of the Higgs boson was tuned in the
\POWHEG simulation of the dominant gluon fusion production mode to better match predictions from full phase space
calculations implemented in the \textsc{hres} 2.3 generator~\cite{deFlorian:2012mx, Grazzini:2013mca}.

The SM $\cPZ\cPZ$ background contribution from quark-antiquark annihilation is generated at NLO pQCD with \POWHEG 2.0, while the $\ggZZ$ process is generated at leading order (LO) with \MCFM~\cite{MCFM}.

All signal and background generators are interfaced with \PYTHIA 8.212~\cite{Sjostrand2015159} tune CUETP8M1~\cite{Khachatryan:2015pea} to simulate multiple parton interactions, the underlying event, and the fragmentation and hadronization effects.
The generated events are processed through a detailed simulation of the CMS detector based on \GEANTfour~\cite{Agostinelli:2002hh,GEANT} and are reconstructed with the same algorithms that are used for data.
The simulated events include overlapping $\Pp\Pp$ interactions (pileup) and have been reweighted so that the distribution of the number of interactions per LHC bunch crossing in simulation matches that observed in data.

\section{Event reconstruction and selection}
\label{sec:objects}

Event reconstruction is based on the particle-flow (PF) algorithm~\cite{Sirunyan:2017ulk}, which exploits information from all the CMS subdetectors to identify and reconstruct individual particles in the event.
The PF candidates are classified as charged hadrons, neutral hadrons, photons, electrons, or muons, and they are then used to build higher-level observables such as jets and lepton isolation quantities.

Electrons with $\PT^{\Pe} > 7\GeV$ are reconstructed within the
geometrical acceptance defined by a pseudorapidity $\abs{\eta^{\Pe}} < 2.5$.
Electrons are identified using a multivariate discriminant that includes observables sensitive to
the presence of bremsstrahlung along the electron trajectory, the geometrical and momentum-energy matching between the
electron trajectory and the associated energy cluster in the ECAL, the shape of the electromagnetic shower in the ECAL,
and variables that discriminate against electrons originating from photon conversions such as the number of expected but
missing pixel hits and the conversion vertex fit probability.

Muons within the geometrical acceptance $\abs{\eta^{\Pgm}} < 2.4$ and  $\PT^{\Pgm} > 5\GeV$ are reconstructed by combining
information from the silicon tracker and the muon system~\cite{Chatrchyan:2012xi}.
The matching between the inner and outer tracks proceeds either outside-in, starting from a track in the muon system,
or inside-out, starting from a track in the silicon tracker.
In the latter case, tracks that match track segments in only one or two planes of the muon system are also considered in the analysis
to collect very low-$\pt$ muons that may not have sufficient energy to penetrate the entire muon system.
The muons are selected among the reconstructed muon track candidates by applying minimal requirements on the track in both
the muon system and inner tracker system, and taking into account compatibility with small energy deposits in the calorimeters.

To suppress muons originating from in-flight decays of hadrons and electrons from photon conversions,
we require each lepton track to have the ratio of the impact parameter in three dimensions, computed with respect to the chosen primary vertex position, and its uncertainty to be less than 4.
The primary vertex is defined as the reconstructed vertex with the largest value of summed physics-object $\pt^2$, where the physics objects are the objects returned by a jet finding algorithm~\cite{Cacciari:2008gp,Cacciari:2011ma} applied to all charged tracks associated with the vertex, plus the corresponding associated missing transverse energy, $\ETmiss$,
defined as the magnitude of the vector sum of the transverse momenta of all reconstructed
PF candidates (charged or neutral) in the event.

To discriminate between prompt leptons from $\cPZ$ boson decay and those arising from electroweak decays of hadrons within jets,
an isolation requirement for leptons of ${\cal I}^{\ell}<0.35$ is imposed, where the relative isolation is defined as
\ifthenelse{\boolean{cms@external}}{
\begin{equation}\begin{split}
\label{eqn:pfiso}
{\cal I}^{\ell} \equiv& \Big( \sum \PT^\text{charged} +
                                 \max\left[ 0, \sum \PT^\text{neutral}
                                 +\\
                                  &\sum \PT^{\Pgg}
                                 - \PT^\mathrm{PU}(\ell) \right] \Big)
                                 / \PT^{\ell}.
\end{split}
\end{equation}
}{
\begin{equation}
\label{eqn:pfiso}
{\cal I}^{\ell} \equiv \Big( \sum \PT^\text{charged} +
                                 \max\left[ 0, \sum \PT^\text{neutral}
                                 +
                                  \sum \PT^{\Pgg}
                                 - \PT^\mathrm{PU}(\ell) \right] \Big)
                                 / \PT^{\ell}.
\end{equation}
}

The isolation sums involved are all restricted to a volume bounded by a
cone of angular radius  $\Delta R=0.3$ around the lepton direction at the
primary vertex, where the angular distance between two particles
$i$ and $j$ is $\Delta R(i,j) = \sqrt{\smash[b]{(\eta^i-\eta^j)^{2} + (\phi^i-\phi^j)^{2}}}$.
The $\sum \PT^\text{charged}$ is the scalar sum of the
transverse momenta of charged hadrons originating from
the chosen primary vertex of the event.
The $\sum \PT^\text{neutral}$ and $\sum \PT^{\Pgg}$ are the
scalar sums of the transverse momenta for neutral hadrons and photons, respectively.
Since the isolation variable is particularly sensitive to energy deposits from pileup interactions, a $\PT^\text{PU}(\ell)$ contribution is subtracted, using two different techniques.
For muons, we define $\PT^\mathrm{PU}(\Pgm) \equiv 0.5 \, \sum_i \PT^{\mathrm{PU}, i}$, where $i$ runs over the momenta of the charged hadron PF candidates not originating from the primary vertex, and the factor of 0.5 corrects for the different fraction of charged and neutral particles in the cone.
For electrons, the \FASTJET technique~\cite{Cacciari:2007fd,Cacciari:2008gn,Cacciari:2011ma} is used, in which $\PT^\mathrm{PU}(\Pe) \equiv \rho \, A_\text{eff}$, where the effective area $ A_\text{eff}$ is the geometric area of the isolation cone scaled by a factor that accounts for the residual dependence of the average pileup deposition on the $\eta$ of the electron,
and $\rho$ is the median of the $\PT$ density distribution of neutral particles within the area of any jet in the event.

An algorithm is used to recover the final-state radiation (FSR) from leptons.
Photons reconstructed by the PF algorithm within $\abs{\eta_\Pgg}< 2.4$ are considered as FSR candidates if they pass
$\PT^{\Pgg} > 2\GeV$ and $\mathcal{I}^{\Pgg}< 1.8$, where the photon relative isolation $\mathcal{I}^{\Pgg}$ is defined as
for the leptons in Eq.~\ref{eqn:pfiso}. Associating every such photon to the closest selected lepton in the event,
we discard photons that do not satisfy $\Delta R(\cPgg,\ell)/(\PT^{\Pgg})^{2}<0.012\GeV^{-2}$ and $\Delta R(\cPgg,\ell)<0.5$.
We finally retain the lowest-$\Delta R(\cPgg,\ell)/(\PT^{\Pgg})^{2}$ photon candidate of every lepton, if any.
Photons thus identified are excluded from any isolation computation.

The momentum scale and resolution for electrons and muons are calibrated in bins of $\pt^\ell$ and $\eta^\ell$
using the decay products of known dilepton resonances.
The electron momentum scale is corrected with a $\Zee$ sample by matching the peak of the reconstructed
dielectron mass spectrum in data to the one in simulation.
A pseudorandom Gaussian smearing is applied
to electron energies in simulation to make the $\Zee$ mass resolution match the one in data~\cite{CMS:EGM-14-001}.
Muon momenta are calibrated using a Kalman filter approach~\cite{Fruhwirth:1987fm}, using $\PJGy$ meson and $\cPZ$ boson decays.

A ``tag-and-probe'' technique based on inclusive samples of $\cPZ$ boson events in data and simulation
is used to measure the efficiency of the reconstruction and selection
for prompt electrons and muons in several bins of $\PT^\ell$ and $\eta^\ell$.
The difference in the efficiencies measured in simulation and data, which on average is 1\% (4\%) per muon (electron),
is used to rescale the selection efficiency in the simulated samples.

Jets are reconstructed from the PF candidates, clustered by the anti-$\kt$ algorithm~\cite{Cacciari:2008gp, Cacciari:2011ma} with a distance parameter of 0.4, and with the constraint that the charged particles are compatible with the primary vertex.
The jet momentum is determined as the vector sum of all particle momenta in the jet, and is found in the simulation to reproduce the true momentum at the 5 to 10\% level over the whole \pt spectrum and detector acceptance.
Jet energy scale corrections are derived from the simulation and confirmed with measurements examining the energy balance in dijet, multijet, $\gamma+\text{jet}$, and leptonic $\cPZ/\gamma+\text{jet}$ events~\cite{Chatrchyan:2011ds,Khachatryan:2016kdb}.
Jet energies in simulation are smeared to match the resolution in data.
To be considered in the analysis, jets must satisfy $\pt^{\text{jet}}>30\GeV$
and $\abs{\eta^{\text{jet}}}<4.7$, and be separated from all selected lepton candidates and any selected
FSR photon by $\Delta R(\ell/\cPgg,\text{jet})>0.4$.

For event categorization, jets are tagged as b-jets using the Combined Secondary Vertex algorithm~\cite{Chatrchyan:2012jua,CMS-PAS-BTV-15-001} which combines information about the impact parameter significance,
the secondary vertex and the jet kinematics.
The variables are combined using a multilayer perceptron approach to compute the b tagging discriminator.
Data-to-simulation scale factors for the b tagging efficiency are applied as a function of jet $\pt$, $\eta$, and flavor.
The $\ETmiss$ is also used for the event categorization.

The event selection is designed to extract signal candidates from events containing at least four well-identified and isolated leptons, each originating from the primary vertex and possibly accompanied by an FSR photon candidate.
In what follows, unless otherwise stated, FSR photons are included in invariant mass computations.

First, $\cPZ$ boson candidates are formed with pairs of leptons ($\Pep\Pem$, $\PGmp\PGmm$) of the same flavor and opposite sign (OS) and required to pass $12 < \mlplm  < 120\GeV$.
They are then combined into $\cPZ\cPZ$ candidates, wherein we denote as $\cPZ_1$ the $\cPZ$ candidate with an invariant mass closest to the nominal $\cPZ$ boson mass ($m_{\cPZ}$)~\cite{Zmass}, and as $\cPZ_2$ the other one.
The flavors of the leptons involved define three mutually exclusive subchannels: $4\Pe$, $4\PGm$, and $2\Pe 2\Pgm$.

To be considered for the analysis, $\cPZ\cPZ$ candidates have to pass a set of kinematic requirements that improve the sensitivity to Higgs boson decays.
The $\cPZ_1$ invariant mass must be larger than 40\GeV.
All leptons must be separated in angular space by at least $\Delta R(\ell_i, \ell_j) > 0.02$.
At least two leptons are required to have $\pt > 10\GeV$ and at least one is required to have $\pt > 20\GeV$.
For $\cPZ_1\cPZ_2$ candidates composed of four same flavor leptons, an alternative pairing $\cPZ_a \cPZ_b$ can be formed out of the same four leptons.
We discard the $\cPZ_1\cPZ_2$ candidate if $m(\cPZ_a)$ is closer to $m_{\Z}$ than $m(\cPZ_1)$ and $m(\cPZ_b)<12\GeV$.
This protects against events that contain an on-shell $\cPZ$ and a low-mass dilepton resonance.
In events with only four leptons this requirement leads to the event being discarded, while in events with more than four leptons
other $\cPZ\cPZ$ candidates are considered.
To further suppress events with leptons originating from hadron decays in jet fragmentation or from the decay of low-mass hadronic resonances, all four OS lepton pairs that can be built with the four leptons (irrespective of flavor) are required to satisfy $m_{\ell^{+}\ell'^{-}} > 4\GeV$, where selected FSR photons are disregarded in the invariant mass computation.
Finally, the four-lepton invariant mass $\mllll$ must be larger than 70\GeV.

In events where more than one $\cPZ\cPZ$ candidate passes the above selection, the candidate with the highest value of $\KD$ (defined in Section~\ref{sec:observables}) is retained, except when two candidates consist of the same four leptons in which case the candidate with  the ${\Z_1}$ mass closest to $m_{\Z}$ is retained.
The additional leptons that do not form the ZZ candidate but pass identification, vertex compatibility, and isolation
requirements are used in the event categorization, see Section~\ref{sec:categorization}.

\section{Kinematic discriminants and event-by-event mass uncertainty}
\label{sec:observables}

The full kinematic information from each event using either the Higgs boson decay products or associated
particles in its production is extracted using matrix element calculations and is used to form several kinematic discriminants.
These computations rely on the \textsc{mela} package~\cite{Gao:2010qx,Bolognesi:2012mm,Anderson:2013afp}
and use \textsc{jhugen} matrix elements for the signal and \MCFM matrix elements for the background.
The decay kinematics of the scalar $\PH$ boson and the production kinematics of gluon fusion
in association with one jet ($\PH$+1 jet) or two jets ($\PH$+2 jets), VBF, $\ZH$, and $\WH$ associated production are explored
in this analysis.
The kinematics of the full event are described by decay observables $\vec\Omega^{\PH\to4\ell}$ or observables describing
associated production $\vec\Omega^{\PH+\mathrm{JJ}}$. The definition of these observables can be found in Refs.~\cite{Gao:2010qx,Bolognesi:2012mm,Anderson:2013afp}.

The discriminant sensitive to the $\Pg\Pg/{\qqbar}\to4\ell$ kinematics is calculated as~\cite{Chatrchyan:2012ufa,CMSH4lAnomalousCouplings}
\begin{equation}
\label{eq:ggmela}
\mathcal{D}^\text{kin}_\text{bkg} =
\left[1+
\frac{ \mathcal{P}^{\Pq \Paq }_\text{bkg} (\vec\Omega^{\PH\to4\ell} | m_{4\ell})  }
{ \mathcal{P}^{\cPg\cPg}_\text{sig}(\vec\Omega^{\PH\to4\ell} | m_{4\ell}) }
\right]^{-1},
\end{equation}
where $\mathcal{P}^{\cPg\cPg}_\text{sig}$ is the probability density for an event to be consistent with the signal
and $\mathcal{P}^{\Pq \Paq }_\text{bkg}$ is the corresponding probability density for the dominant $\qqZZ\to4\ell$ background process, all calculated either with the \textsc{jhugen} or \MCFM matrix elements within the \textsc{mela} framework.

Four discriminants
are used to enhance the purity of event categories
as described in Section~\ref{sec:categorization}.
\DMeVbfjj is the discriminant sensitive to the VBF signal topology with two associated jets,
\DMeVbfj is the discriminant sensitive to the VBF signal topology with one associated jet,
and \DMeWh or \DMeZh are the discriminants sensitive to the ZH or WH signal topologies with two associated jets from the decay of the Z$\to{\qqbar}$ or the W$\to{\qqbar^\prime}$:
\begin{gather}
\label{eqn:prodmela}
\begin{split}
\DMeVbfjj =
\left[1+
\frac{ \mathcal{P}_{\PH\mathrm{JJ}} (\vec\Omega^{\PH+\mathrm{JJ}} | m_{4\ell}) }
{\mathcal{P}_\mathrm{VBF}  (\vec\Omega^{\PH+\mathrm{JJ}} | m_{4\ell})  }
\right]^{-1}
\,~~~~~~~~~
\DMeVbfj =
\left[1+
\frac{ \mathcal{P}_{\PH\mathrm{J}} (\vec\Omega^{\PH+\mathrm{J}} | m_{4\ell}) }
{\int \rd\eta_\mathrm{J}\mathcal{P}_\mathrm{VBF}  (\vec\Omega^{\PH+\mathrm{JJ}} | m_{4\ell}) }
\right]^{-1}
\,\\
\DMeWh =
\left[1+
\frac{ \mathcal{P}_{\PH\mathrm{JJ}} (\vec\Omega^{\PH+\mathrm{JJ}} | m_{4\ell}) }
{\mathcal{P}_{\PW\PH}  (\vec\Omega^{\PH+\mathrm{JJ}} | m_{4\ell})  }
\right]^{-1}
\,~~~~~~~~~
\DMeZh =
\left[1+
\frac{ \mathcal{P}_{\PH\mathrm{JJ}} (\vec\Omega^{\PH+\mathrm{JJ}} | m_{4\ell}) }
{\mathcal{P}_{\Z\PH}  (\vec\Omega^{\PH+\mathrm{JJ}} | m_{4\ell})  }
\right]^{-1}
\,~~~~~ \\
\end{split}
\end{gather}
where $\mathcal{P}_\mathrm{VBF}$, $\mathcal{P}_{\PH\mathrm{JJ}}$, $\mathcal{P}_{\PH\mathrm{J}}$, and $\mathcal{P}_{\mathrm{V}\PH}$ are probability densities obtained from the \textsc{jhugen} matrix elements for
the VBF, $\PH+\text{2 jets}$, $\PH+\text{1 jet}$,
and VH ($\mathrm{V}=\PW,\cPZ$) processes, respectively.
The expression $\int \rd\eta_\mathrm{J}\mathcal{P}_\mathrm{VBF}$ is the integral of the two-jet VBF matrix element probability density
discussed above over the $\eta_J$ values of the unobserved jet with the constraint that the total
transverse momentum of the $\PH+\text{2 jets}$ system is zero.
By construction, all discriminants defined in Eqs.~\ref{eq:ggmela} and~\ref{eqn:prodmela} have values bounded between 0 and 1.

The uncertainty in the momentum measurement can be predicted for each lepton.
For muons, the full covariance matrix is obtained from the muon track fit, and the directional uncertainties are negligibly small.
For the electrons, the momentum uncertainty is estimated from the combination of the ECAL and tracker measurements, neglecting the uncertainty in the track direction.
The uncertainty in the kinematics at the per-lepton level is then
propagated to the four-lepton candidate to predict the mass uncertainty ($\MassD$) on an
event-by-event basis.
For FSR photons, a parametrization obtained from
simulation is used for the uncertainty in the photon $\pt$.
The per-lepton momentum uncertainties are corrected in data and simulation using $\cPZ$ boson events.
Events are divided into different categories based on the predicted dilepton mass resolution.
A Breit--Wigner parameterization convolved with a double-sided Crystal
Ball function~\cite{CrystalBall} is then fit to the dilepton mass
distribution in each category to extract the resolution and compare it to the predicted resolution.
Corrections to the lepton momentum uncertainty are derived through an iterative procedure
in different bins of lepton $\pt$ and $\eta$.
After the corrections are derived, a closure test of the agreement between the predicted and fitted $4\ell$ mass resolution is performed in data and in simulation, in bins of the predicted $4\ell$ mass resolution, confirming that the calibration brings it close to the fitted value.
A systematic uncertainty of 20\% in the $4\ell$ mass resolution is assigned to cover the residual differences between the predicted and fitted
resolutions.

\section{Event categorization}
\label{sec:categorization}

To improve the sensitivity to the various Higgs boson production mechanisms, the selected events are classified into mutually exclusive categories.
The category definitions exploit the jet multiplicity, the number of b-tagged jets, the number of additional leptons (defined as leptons that pass identification, vertex compatibility, and isolation requirements, but do not form the ZZ candidate), and requirements on the kinematic discriminants described in Section~\ref{sec:observables}.

Seven categories are defined, using the criteria applied in the following order (\ie an event is considered for the subsequent
category only if it does not satisfy the requirements of the previous category):

\begin{itemize}
\item { The VBF-2jet-tagged category} requires exactly four leptons. In addition, there must be either two or three jets of which at most one is b tagged, or four or more jets none of which are b-tagged. Finally, $\mathcal{D}_\text{2 jet}>0.5$ is required.
\item { The VH-hadronic-tagged category} requires exactly four leptons. In addition, there must be two or three jets, or four or more jets none of which are b-tagged. $\mathcal{D}_{\mathrm{V}\PH} \equiv \max(\mathcal{D}_{\Z\PH},\mathcal{D}_{\PW\PH})>0.5$ is required.
\item { The VH-leptonic-tagged category} requires no more than three jets and no b-tagged jets in the event,
and exactly one additional lepton or one additional pair of OS, same-flavor leptons.
This category also includes events with no jets and at least one additional lepton.
\item { The $\ttbar\PH$-tagged category} requires at least four jets of which at least one is b tagged, or at least one additional lepton.
\item { The VH-\ETmiss-tagged category} requires exactly four leptons, no more than one jet and $\ETmiss$ greater than 100\GeV.
\item { The VBF-1jet-tagged category} requires exactly four leptons, exactly one jet and $\mathcal{D}_\text{1 jet}>0.5$.
\item { The Untagged category} consists of the remaining selected events.
\end{itemize}

\begin{figure}[!htb]
\centering
\includegraphics[width=0.8\textwidth]{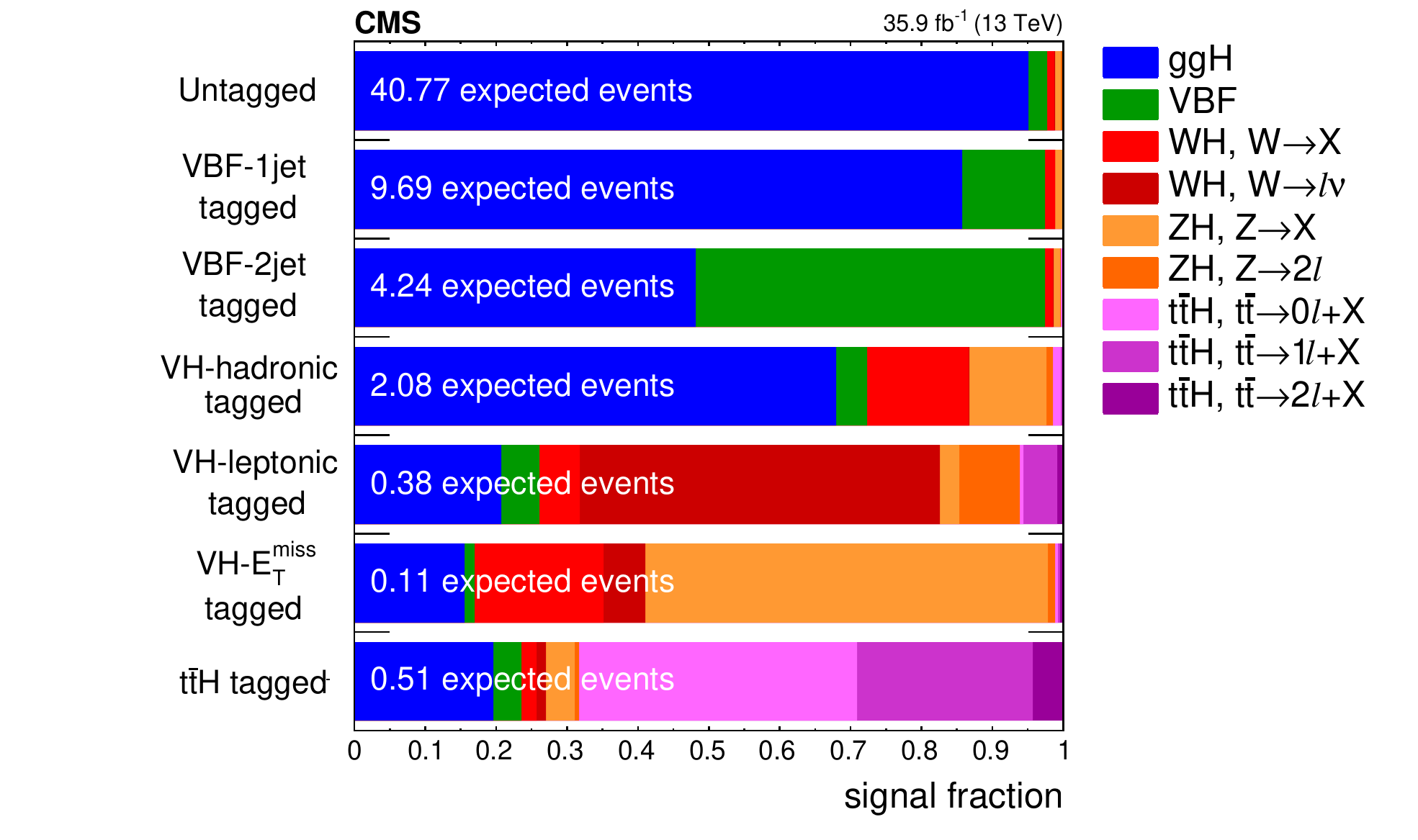}
\caption{Relative signal purity in the seven event categories in terms of the five main production mechanisms of the Higgs boson in the $118<\mllll<130\GeV$ mass window are shown. The $\WH$, $\ZH$, and $\ttH$ processes are split according to the decay of the associated particles, where X denotes anything other than an electron or a muon.
Numbers indicate the total expected signal event yields in each category.
\label{fig:categ-purity}}

\end{figure}

The definitions of the categories were chosen to achieve high signal purity whilst maintaining high efficiency for each of the main Higgs boson production mechanisms.
The order of the categories is chosen to maximize the signal purity target in each category.
Figure~\ref{fig:categ-purity} shows the relative signal purity of the seven event categories for the various Higgs boson production processes.
The VBF-1jet-tagged and VH-hadronic-tagged categories are expected to have substantial contamination from gluon fusion, while the purity of the VBF process in the VBF-2jet-tagged category is expected to be about 49\%.

\section{Background estimation}
\label{sec:bkgd}

\subsection{Irreducible backgrounds}
\label{sec:irrbkgd}

The irreducible backgrounds to the Higgs boson signal in the $4\ell$ channel, which come from the
production of $\cPZ\cPZ$ via $\Pq\Paq$ annihilation or gluon fusion, are estimated using simulation.
The fully differential cross section for the \qqZZ~ process has been computed at next-to-next-to-leading order (NNLO)~\cite{Grazzini2015407},
and the NNLO/NLO K-factor as a function of $m_{\cPZ\cPZ}$ has been applied to the \POWHEG sample.
This K-factor varies from 1.0 to 1.2 and is 1.1 at $m_{\cPZ\cPZ}=125\GeV$.
Additional NLO electroweak corrections, which depend on the initial state quark flavor and kinematics,
are also applied in the region $m_{\cPZ \cPZ}>2m_{\cPZ}$ where the corrections have been computed~\cite{Bierweiler:1312}.
The uncertainty due to missing electroweak corrections in the region $m_{\cPZ \cPZ}<2m_{\cPZ}$ is expected to
be small compared to the uncertainties in the pQCD calculation.

The production of $\cPZ\cPZ$ via gluon fusion contributes at NNLO in pQCD.
It has been shown~\cite{Bonvini:1304.3053}  that the soft-collinear approximation is able to
describe the background cross section and the interference term at NNLO\@.
Further calculations also show that at NLO the K-factor for the signal and background~\cite{Melnikov:2015laa}
and at NNLO the K-factor for the signal and interference terms~\cite{Li:2015jva} are very similar.
Therefore, the same K-factor used for the signal is also used for the background~\cite{Passarino:1312.2397v1}.
The NNLO K-factor for the signal is obtained as a function of $m_{\cPZ\cPZ}$ using the \textsc{hnnlo}~v2 program~\cite{Catani:2007vq,Grazzini:2008tf,Grazzini:2013mca} by calculating the NNLO and LO $\Pg\Pg\to\PH\to2\ell2\ell^\prime$ cross sections at the small $\PH$ boson decay width of 4.1\MeV and taking their ratios.
The NNLO/LO K-factor for \ggZZ~varies from 2.0 to 2.6 and is 2.27 at $m_{\cPZ\cPZ}=125\GeV$; a systematic uncertainty of 10\% in its determination when applied to the background process is used in the analysis.

\subsection{Reducible backgrounds}
\label{sec:redbkgd}

Additional backgrounds to the Higgs boson signal in the $4\ell$ channel arise
from processes in which heavy flavor jets produce secondary leptons, and
also from processes in which decays of heavy flavor hadrons, in-flight decays of
light mesons within jets, or (for electrons) the decay of charged
hadrons overlapping with $\pi^0$ decays, are misidentified as prompt leptons.
We denote these reducible backgrounds as ``Z+X'' since the dominant process producing them is $\cPZ+\text{jets}$,
while subdominant processes in order of importance are $\ttbar+\text{jets}$, $\cPZ\gamma+\text{jets}$,
$\PW\cPZ+\text{jets}$, and $\PW\PW+\text{jets}$. In the case of $\cPZ\gamma+\text{jets}$, the photon may convert to an $\Pep\Pem$
pair with one of the decay products not being reconstructed, giving rise to a signature with three prompt leptons.
The contribution from the reducible background is estimated
using two independent methods having dedicated control regions in data.
The control regions are defined by a dilepton pair satisfying all the requirements of a $\cPZ_1$ candidate and two additional leptons, OS or same-sign (SS), satisfying certain relaxed identification requirements when compared to those used in the analysis.
These four leptons are then required to pass the $\cPZ\cPZ$ candidate selection.
The event yield in the signal region is obtained by weighting the control region events
by the lepton misidentification probability (or misidentification rate) $f$, defined as the fraction of
nonsignal leptons that are identified by the analysis selection criteria.

The lepton misidentification rates $f_{\Pe}$ and $f_{\mu}$ are determined from data, separately for the SS and OS methods, using a control region defined by a $\cPZ_1$ candidate and exactly one additional lepton passing the relaxed selection.
The $\cPZ_1$ candidate consists of a pair of leptons, each of which passes the selection requirements used in the analysis.
For the OS method, the mass of the $\cPZ_1$ candidate is required  to satisfy
 $\abs{m(\cPZ_1) - m_{\Z}} < 7$\GeV  to reduce the contribution of (asymmetric) photon conversions, which
is estimated separately.
In the SS method, the contribution from photon conversions is estimated by determining an
average misidentification rate.
Furthermore the $\ETmiss$ is required to be less than 25\GeV  to
suppress contamination from $\PW\cPZ$ and $\ttbar$ processes.
The fraction of these events in which the additional lepton passes the selection requirements used in the analysis
gives the lepton misidentification rate $f$. The lepton misidentification rates is measured as a function
of $\pt^{\ell}$ and $\abs{\eta^{\ell}}$ and is assumed to be independent of the presence of any additional leptons.

\subsubsection{Method using OS leptons}
\label{sec:xzos}

The control region for the OS method consists of events with a
$\cPZ_1$ candidate and two additional OS leptons of the same-flavor.
The expected yield in the signal region
is obtained from two categories of events.

The first category is composed of events with two leptons
that pass (P) the tight lepton identification requirements and
two leptons that pass the loose identification but fail (F)
the tight identification, and is denoted as the 2P2F region.
Backgrounds, which intrinsically have only two prompt leptons, such as
$\cPZ+\text{jets}$ and $\ttbar$, are estimated with this control region.
To obtain the expected yield in the signal region, each event $i$ in the
2P2F region is weighted by a factor $[f^i_{3}/(1-f^i_{3})]
[f^i_{4}/(1-f^i_{4})]$, where $f^i_{3}$ and $f^i_{4}$ are the
misidentification rates for the third and fourth lepton, respectively.

The second category consists of events
where exactly one of the two additional leptons passes the analysis selection, and is referred
to as the 3P1F region.
Backgrounds with three prompt leptons, such as $\PW\cPZ+\text{jets}$ and
$\cPZ\gamma+\text{jets}$ with the photon converting to $\Pep\Pem$,
are estimated using this region. To obtain the expected yield in the signal region, each event
$j$ in the 3P1F region is weighted by a factor
$f^j_4/(1-f^j_4)$, where $f^j_4$ is the misidentification rate for the lepton that does not pass the analysis selection.
The contribution from $\cPZ\cPZ$ events to the 3P1F region ($N^{\cPZ\cPZ}_\mathrm{3P1F}$),
which arises from events where a prompt lepton fails the identification requirements,  is estimated from simulation
and scaled with a factor $w_{\cPZ\cPZ}$ appropriate to the integrated luminosity of the analyzed data set.

The contamination of 2P2F-type processes in the 3P1F region is estimated
as $\sum_i \{ [f^i_3/(1-f^i_3)] + [f^i_4/(1-f^i_4)]\}$
and contributes an amount equal to $\sum_i \{2 [f^i_3/(1-f^i_3)][f^i_4/(1-f^i_4)]\}$ to the expected yield in the signal region.
This amount is subtracted from the total background estimate to avoid double counting.

The total reducible background estimate in the signal region coming from the two categories 2P2F and 3P1F without double counting, $N^\text{reducible}_\mathrm{SR}$, can be written as:
\begin{equation}
  \label{eq:PredictionSR}
  N^\text{reducible}_\mathrm{SR}=
  \sum_j^{N_\mathrm{3P1F}} \frac{f^j_4}{1-f^j_4}
  \, \, - \, \,
  w_{\cPZ\cPZ} \sum_j^{N^{\cPZ\cPZ}_\mathrm{3P1F}} \frac{f^j_4}{1-f^j_4}
  \, \, - \, \,
  \sum_i^{N_\mathrm{2P2F}} \frac{f^i_3}{1-f^i_3} \frac{f^i_4}{1-f^i_4},
\end{equation}
where $N_\mathrm{3P1F}$ and $N_\mathrm{2P2F}$ are the number of events in the 3P1F and 2P2F regions, respectively.

\subsubsection{Method using SS leptons}
\label{sec:zxss}

The control region for the SS method, referred to as the $\mathrm{2P2L_{SS}}$ region,
consists of events with a $\cPZ_1$ candidate and two additional SS leptons of
same-flavor.  These two additional leptons are required to pass the loose
selection requirements for leptons.

The contribution of photon conversions to the electron misidentification probability $f$ is estimated.  Its linear dependence on the fraction of loose electrons in the sample with tracks having one missing hit in the pixel detector, $r_\text{miss}$, is used to derive a corrected misidentification rate $\tilde f$.
The dependence is determined by measuring $f$ in samples with different values of $r_\text{miss}$ formed by
varying the requirements on $\vert m_{\ell_{1}\ell_{2}} - m_\cPZ \vert$ and
$\abs{ m_{\ell_{1}\ell_{2} \Pe_\text{loose}} - m_\cPZ }$.
Here $\ell_{1}$ and $\ell_{2}$ are the leptons which form the $\cPZ_1$ candidate and $\Pe_\text{loose}$
is the additional electron passing the loose selection.

The expected number of reducible background events in the signal
region can then be written as:
\begin{equation}
  N^\text{reducible}_\mathrm{SR}=
  r_\text{OS/SS}   \,
  \sum_i^{N_\mathrm{2P2L_{SS}}}  \tilde f^i_3  \,  \tilde f^i_4 \,,
\end{equation}
where the ratio $r_\text{OS/SS}$ between the number of events in the
$\mathrm{2P2L_{OS}}$ and $\mathrm{2P2L_{SS}}$ control regions is obtained
from simulation. The $\mathrm{2P2L_{OS}}$ region is defined analogously to the
$\mathrm{2P2L_{SS}}$ region but with an OS requirement for the additional pair of loose leptons.

\subsubsection{Prediction and uncertainties}
\label{sec:zxcomb}

The predicted yield in the signal region of the reducible background
from the two methods are in agreement within their statistical uncertainties, and
since they are mutually independent, the results of the two methods are combined.
The final estimate is obtained by weighting the individual mean values of both methods according to their corresponding variances.
The shape of the $m_{4\ell}$ distribution for the reducible background
is obtained by combining the prediction from the OS and SS methods and fitting the distributions
with empirical functional forms built from Landau~\cite{Landau:1944if} and
exponential distributions.

The dominant systematic uncertainty in the reducible background estimation
arises from the limited number of events
in the control regions as well as in the region where the misidentification rate is applied.
Additional sources of systematic uncertainty, estimated using simulated samples,
come from the fact that the composition of the regions
used to compute the misidentification rates typically differs from that of control regions where they are applied.
The subdominant systematic uncertainty in the $m_{4\ell}$ shape is determined by taking the envelope of
differences among the shapes from the OS and SS methods in the three different final states.
The combined systematic uncertainty is estimated to be about 40\%.

\section{Signal modeling}
\label{sec:signal}

The signal shape of a narrow resonance around $\mH\sim125$\GeV is parametrized using a double-sided Crystal Ball function.
The signal shape is parametrized as a function of $\mH$
by performing a simultaneous fit of several mass points for $\ggH$ production around 125\GeV.
Each parameter of the double-sided Crystal Ball function is given a linear dependence on $\mH$ for a total of 12 free parameters.
Of these parameters, 10 are left free in the simultaneous fits.
The parameters that control the prominence of the tails in the two Crystal Ball functions are forced to have a unique value
at all $\mH$ values, to remove large correlations and because they are constant within the uncertainty. This parameterization, derived
separately for each $4\ell$ final state, is found to provide a good description of the resonant part of the signal for all production modes
and event categories. An additional non-resonant contribution from $\WH$, $\ZH$, and $\ttH$ production arises when one of the leptons from
the Higgs boson decay is lost or is not selected. This contribution is modeled by a Landau distribution which is added to the total
probability density function for those production modes.

For the measurement of the width the signal shape for a broad resonance around $\mH\sim125$\GeV is parameterized in the following way.
First, the gluon fusion or electroweak (VBF and VH)  signal production is treated jointly with the corresponding background and their interference as:
\begin{equation}
\mathcal{P}^i(m_{4\ell}; \mH, \Gamma_H) =
\mu_i \, \mathcal{P}^i_\text{sig}(m_{4\ell}; \mH, \Gamma_\PH)
+ \mathcal{P}^i_\text{bkg} (m_{4\ell})
+ \sqrt{\mu_i}  \mathcal{P}_\text{int} (m_{4\ell}; \mH, \Gamma_\PH),
\label{eq:psig}
\end{equation}
where $\mu_i$ is the signal strength in the  production type $i$, gluon fusion or electroweak, and the small $\ttH$ contribution is treated jointly with gluon fusion.
The general parameterization of the probability density function in Eq.~(\ref{eq:psig}) is based on the
framework of \MCFM + \textsc{jhugen} + \textsc{hnnlo} within \textsc{mela}.
The ideal parameterization is based on the matrix element calculation with the $\PH$ boson propagator removed from the cross section scans as a function of $m_{4\ell}$.
The propagator is included analytically with $\mH$ and $\Gamma_\PH$ as unconstrained parameters of the model.
Detector effects are included via the multiplicative efficiency function $\mathcal{E}(m_{4\ell})$ and convolution for the mass resolution $\mathcal{R}(m_{4\ell}|m^\text{truth}_{4\ell})$, both extracted from the full simulation in the same way as for the narrow resonance discussed above.
The resulting distribution is
\begin{equation}
\mathcal{P}^\text{reco}(m_{4\ell})=
\left(\mathcal{E}(m_{4\ell}^\text{truth})\, \mathcal{P}(m^\text{truth}_{4\ell};\mH,\Gamma_\PH)\right) \otimes \mathcal{R}(m_{4\ell}|m^\text{truth}_{4\ell}).
\label{eq:signalpdf1D}
\end{equation}

\section{Systematic uncertainties}
\label{sec:systematics}

The experimental uncertainties common to all measurements include the uncertainty
in the integrated luminosity measurement (2.5\%)~\cite{CMS-PAS-LUM-17-001} and the uncertainty in the lepton identification and reconstruction efficiency
(ranging from 2.5 to 9\% on the overall event yield for the $4\mu$ and $4\Pe$ channels), which affect both
signal and background. Experimental uncertainties in the reducible background estimation, described in Section~\ref{sec:redbkgd},
vary between 36\% ($4\mu$)  and 43\% ($4\Pe$).

\begin{figure}[!htb]
\centering
\includegraphics[width=0.4\linewidth]{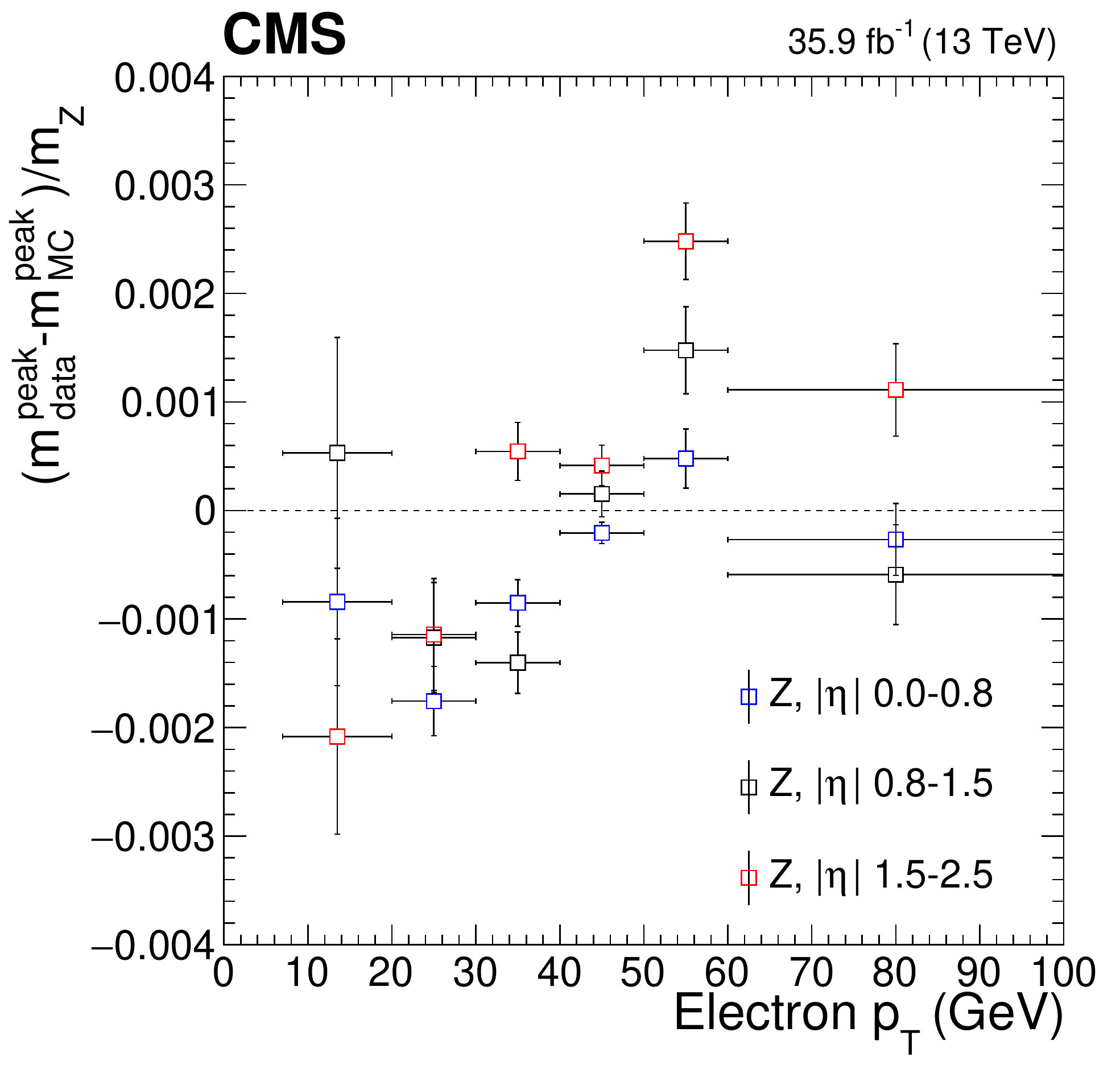}
\includegraphics[width=0.4\linewidth]{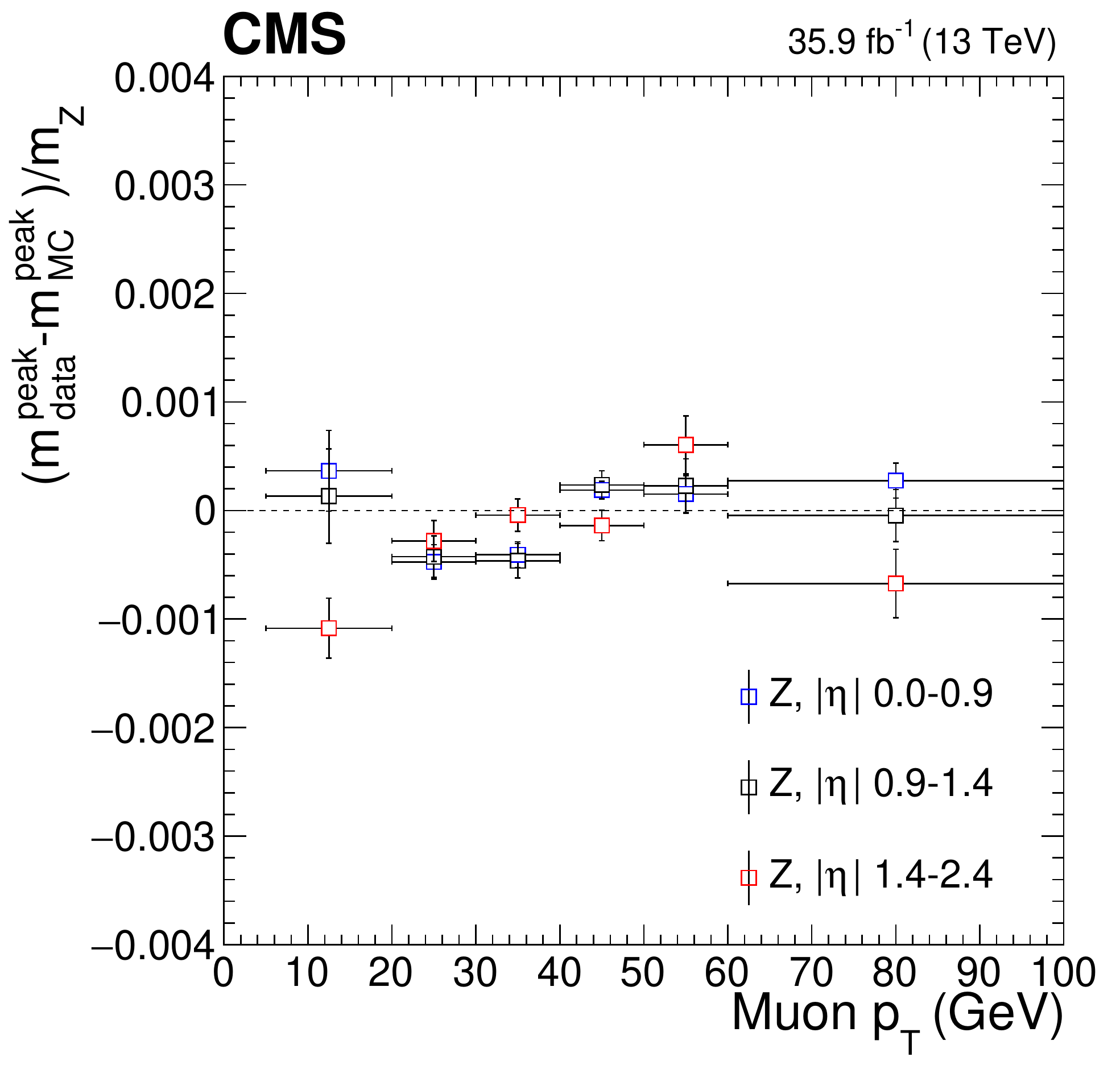}
\caption{ Difference between the $\Z\to\ell\ell$ mass peak positions
in data (${m}^\text{peak}_\text{data}$) and simulation (${m}^\text{peak}_\mathrm{MC}$) normalized by the
nominal $\cPZ$ boson mass (${m}_\Z$), as a function of the $\pt$ and $\abs{\eta}$ of one of the leptons regardless of the second
for electrons (left) and muons (right).
\label{fig:lepScale}}

\end{figure}

The uncertainty in the lepton energy scale, which is the dominant source of
systematic uncertainty in the Higgs boson mass measurement, is determined by considering the
$Z\to\ell\ell$ mass distributions in data and simulation.
Events are separated into categories based on the $\pt$ and $\eta$ of one of the two leptons, selected randomly, and integrating over the other.
A Breit--Wigner parameterization convolved with a double-sided Crystal Ball function is then fit to the dilepton mass distributions.
The offsets in the measured peak position with respect to the nominal $\cPZ$ boson mass in data and simulation are extracted, and the results are shown in Fig.~\ref{fig:lepScale}.
In the case of electrons, since the same data set is used to derive and validate the momentum scale corrections,
the size of the corrections is taken into account for the final value of the uncertainty.
The $4\ell$ mass scale uncertainty is determined to be 0.04\%, 0.3\%, and 0.1\% for the  $4\mu$, $4\Pe$, and $2\Pe2\mu$ channels, respectively.
The uncertainty in the $4\ell$ mass resolution coming from the uncertainty
in the per-lepton energy resolution is 20\%, as described in Section~\ref{sec:observables}.

Theoretical uncertainties that affect both the signal and background estimation
include uncertainties from the renormalization and the factorization scales and the choice of the PDF set.
The uncertainty from the renormalization and factorization scale is determined by varying these scales between
0.5 and 2 times their nominal value while keeping their ratio between 0.5 and 2.
The uncertainty from the PDF set is determined following the PDF4LHC recommendations~\cite{Butterworth:2015oua}.
An additional uncertainty of 10\% in the K factor used for the $\ggZZ$ prediction is applied as
described in Section~\ref{sec:irrbkgd}. A systematic uncertainty of 2\%~\cite{YR4} in the $\Hllll$ branching fraction only affects the signal yield.
The theoretical uncertainties in the background yield are included for all measurements, while the theoretical uncertainties in the overall signal
yield are not included in the measurement uncertainties when cross sections, rather than signal strength modifiers, are extracted.

In the case of the measurements which use event categorization, experimental and theoretical uncertainties that account for possible migration of signal and background events between categories are included.
The main sources of uncertainty in the event categorization include the renormalization and factorization scales, PDF set, and the modeling of the fragmentation, hadronization, and the underlying event.
These uncertainties amount to 4--20\% for the signal and 3--20\% for the background, depending on the category,
and are largest for the prediction of the $\ggH$ yield in the VBF-2jet-tagged category.
Additional uncertainties come from the imprecise knowledge of the jet energy scale (from 2\% for the $\ggH$ yield in the
untagged category to 15\% for the $\ggH$ yield in the VBF-2jet-tagged category) and b tagging efficiency and mistag rate (up to 6\% in
the $\ttH$-tagged category).

\section{Results}
\label{sec:results}

\begin{figure}[!htb]
\vspace*{0.3cm}
\centering
\includegraphics[height=0.4\textwidth]{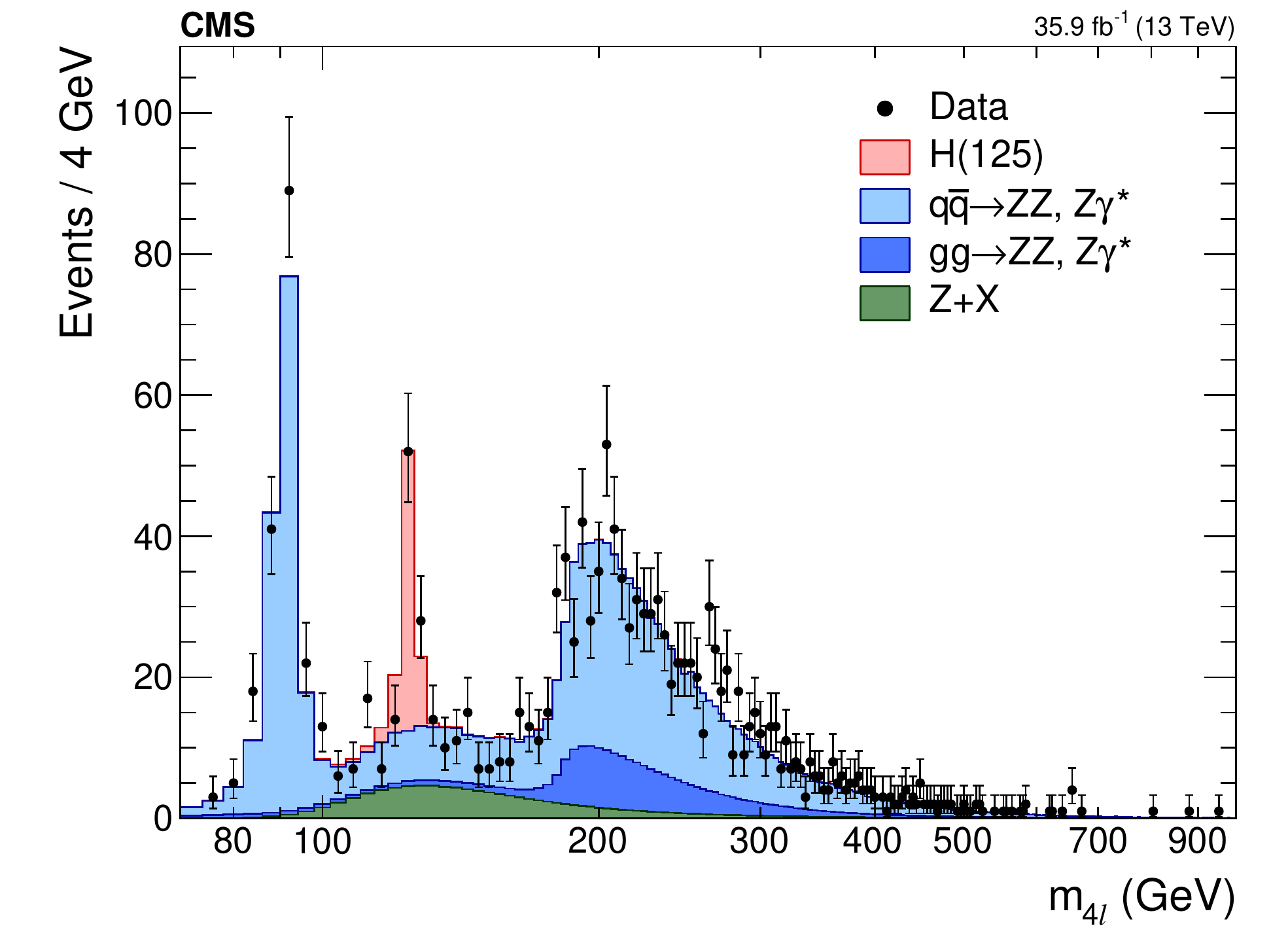}
\includegraphics[height=0.4\textwidth]{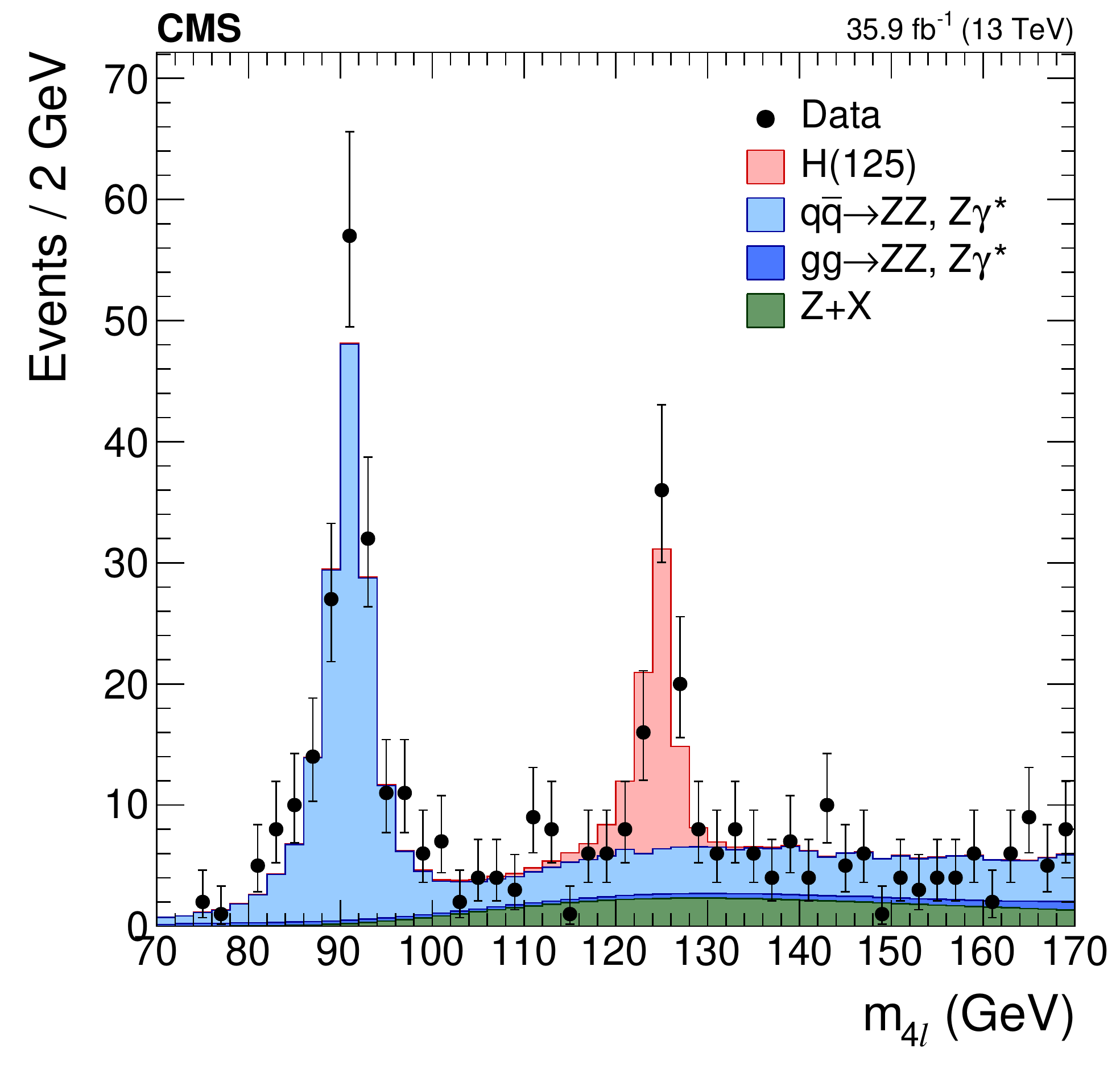}
\caption{Distribution of the reconstructed four-lepton invariant mass $\mllll$ in the full mass range (left) and the low-mass range (right). Points with error bars represent the data and stacked histograms represent expected signal and background distributions. The SM Higgs boson signal with $\mH=125\GeV$, denoted as $\PH(125)$, and the $\cPZ\cPZ$ backgrounds are normalized to the SM expectation, whilst the $\cPZ$+X background is normalized to the estimation from data.
The order in perturbation theory used for the normalization of the irreducible backgrounds is described in Section~\ref{sec:irrbkgd}.
No events are observed with $\mllll>1\TeV$.
\label{fig:FullM4l}}

\end{figure}

\begin{figure}[!htbp]
\vspace*{0.3cm}
\centering
{ \includegraphics[width=0.32\textwidth]{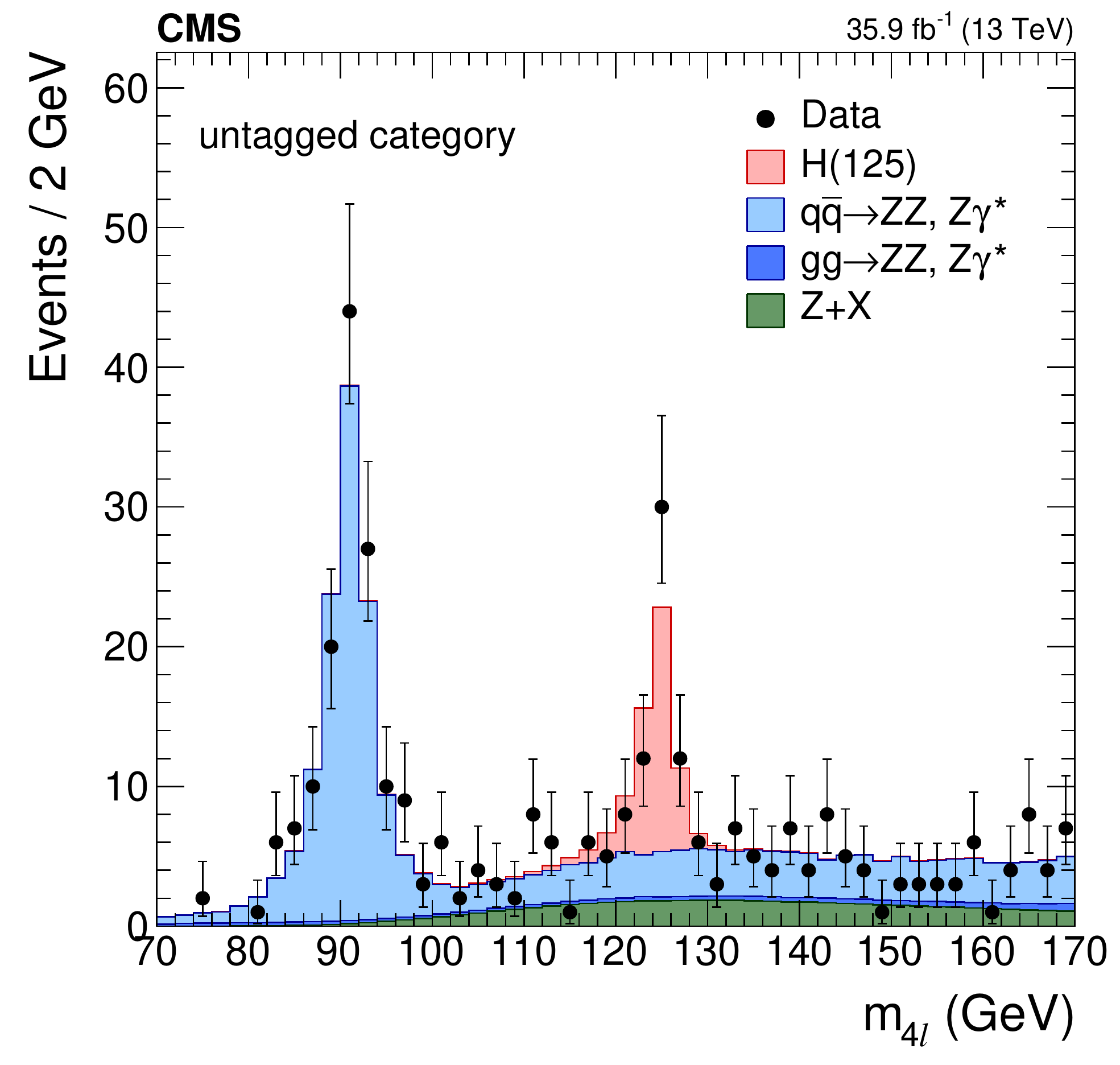}}
{ \includegraphics[width=0.32\textwidth]{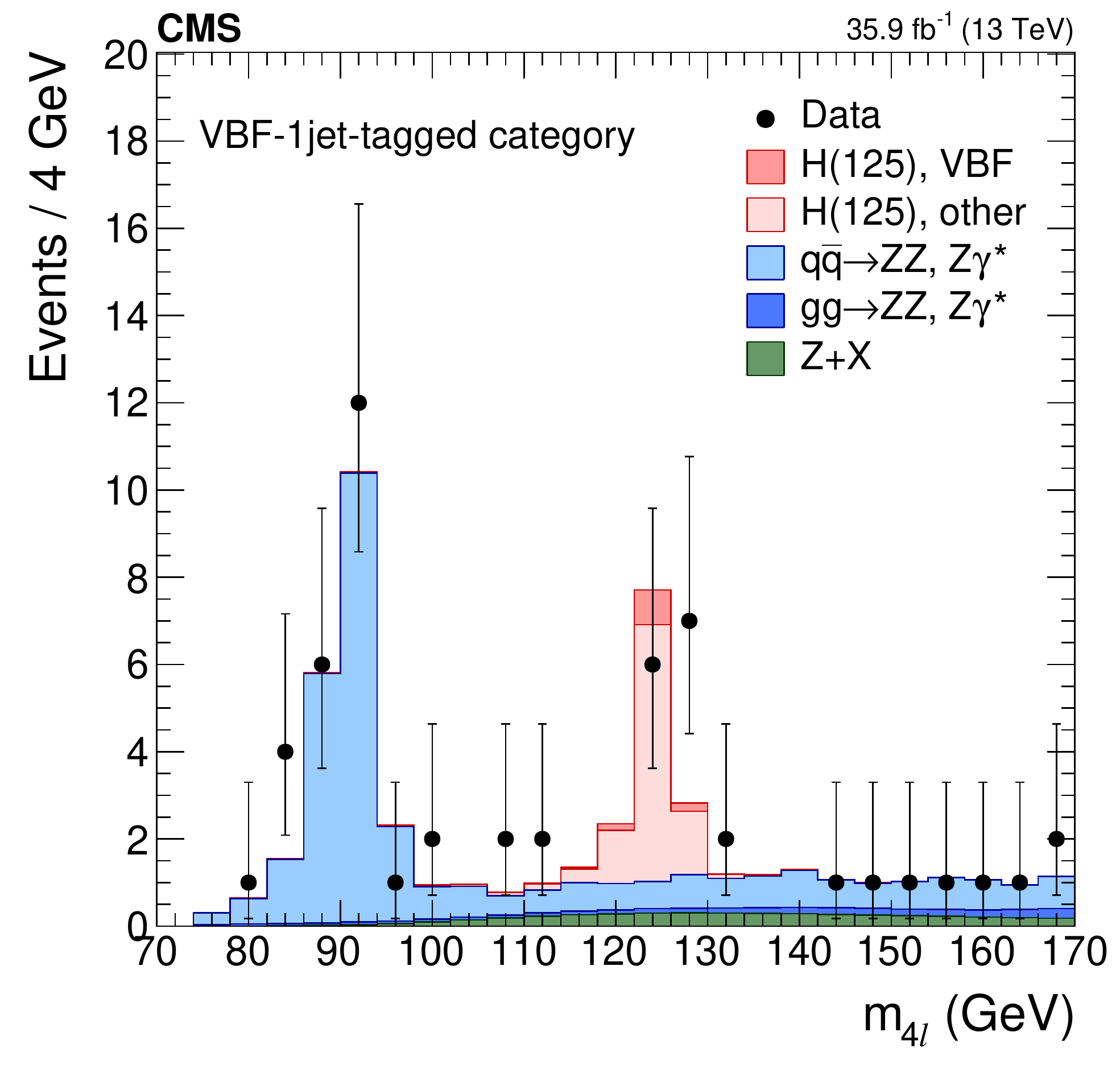}}\\
{ \includegraphics[width=0.32\textwidth]{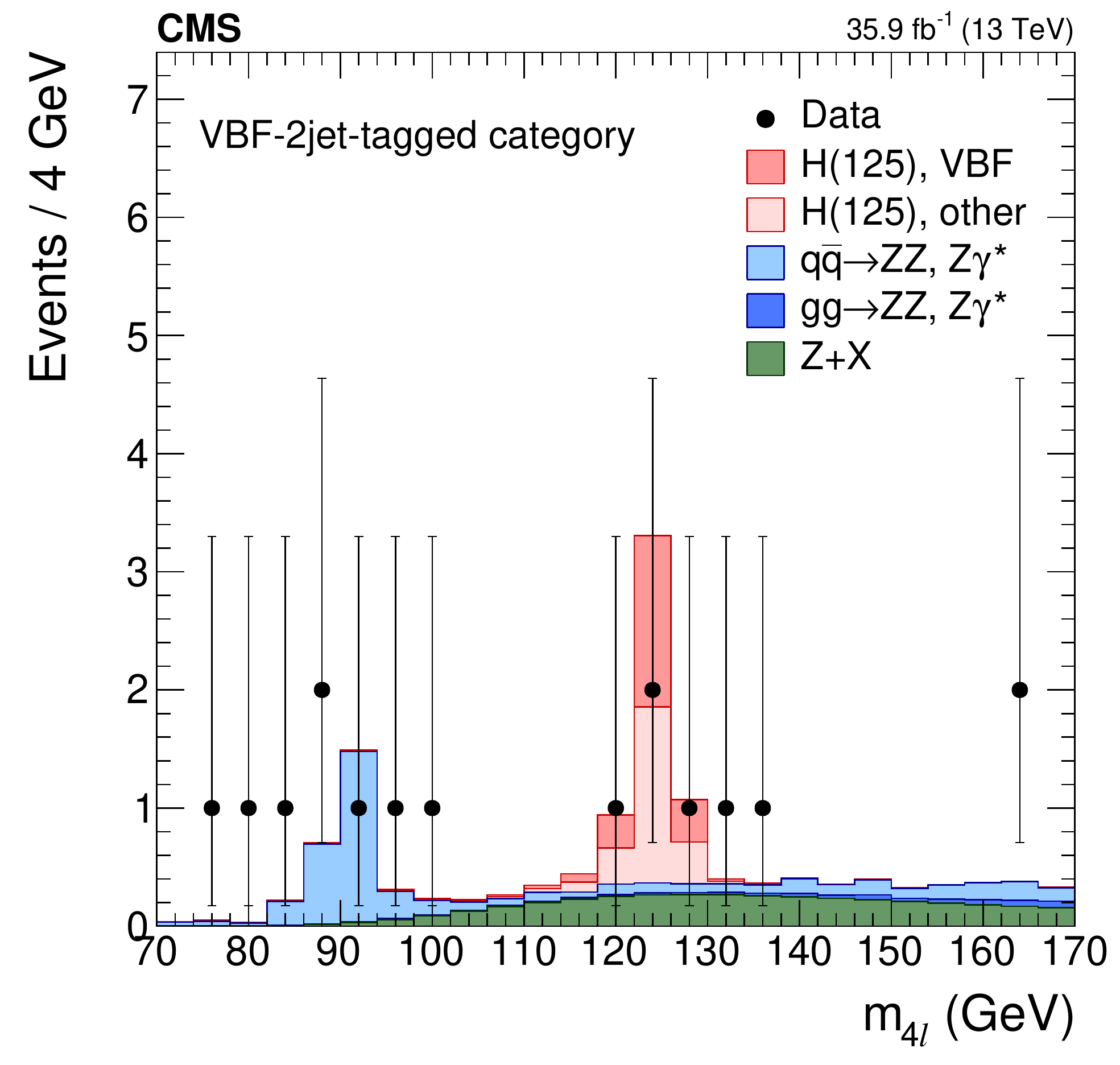}}
{ \includegraphics[width=0.32\textwidth]{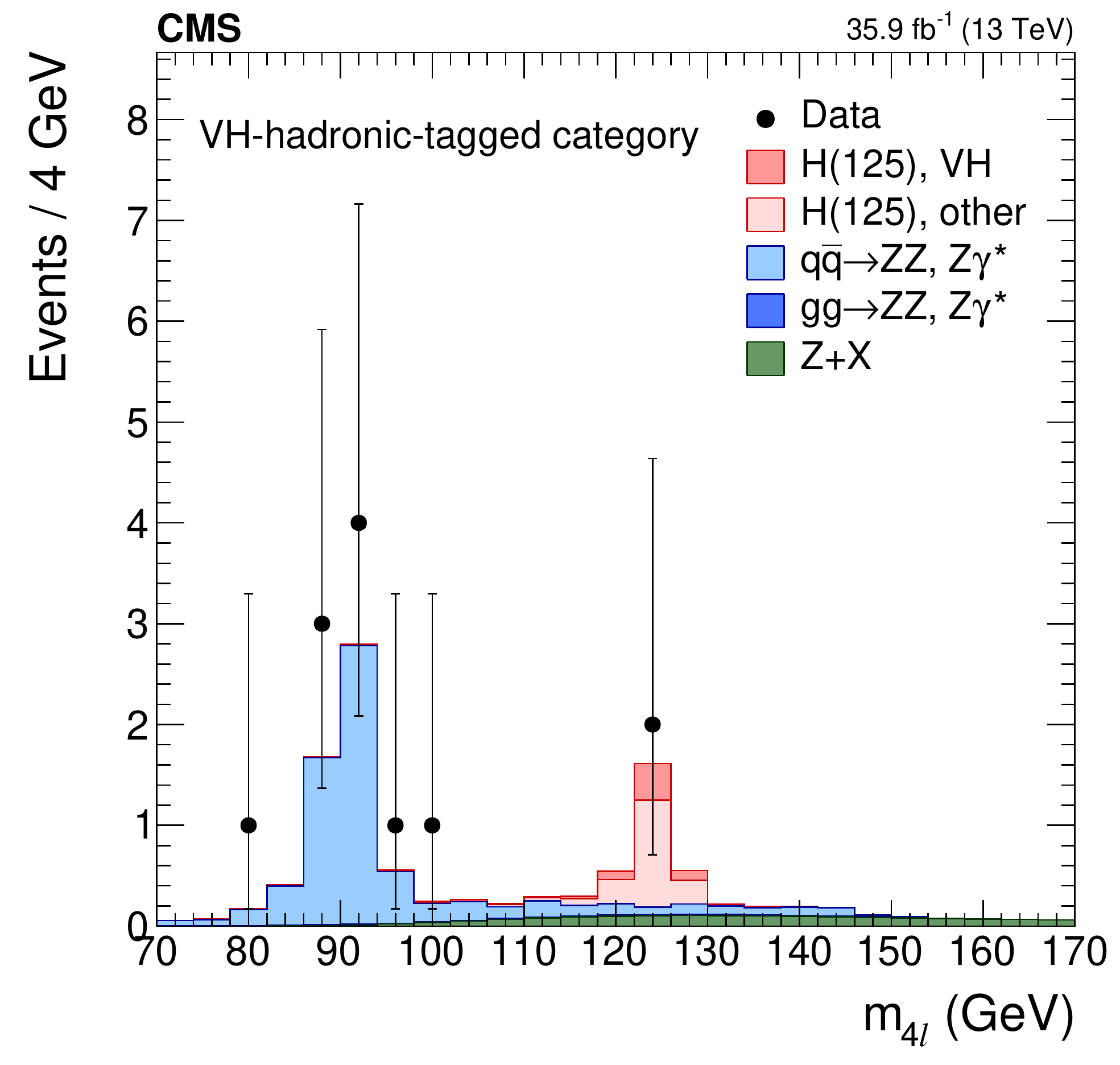}}\\
{ \includegraphics[width=0.32\textwidth]{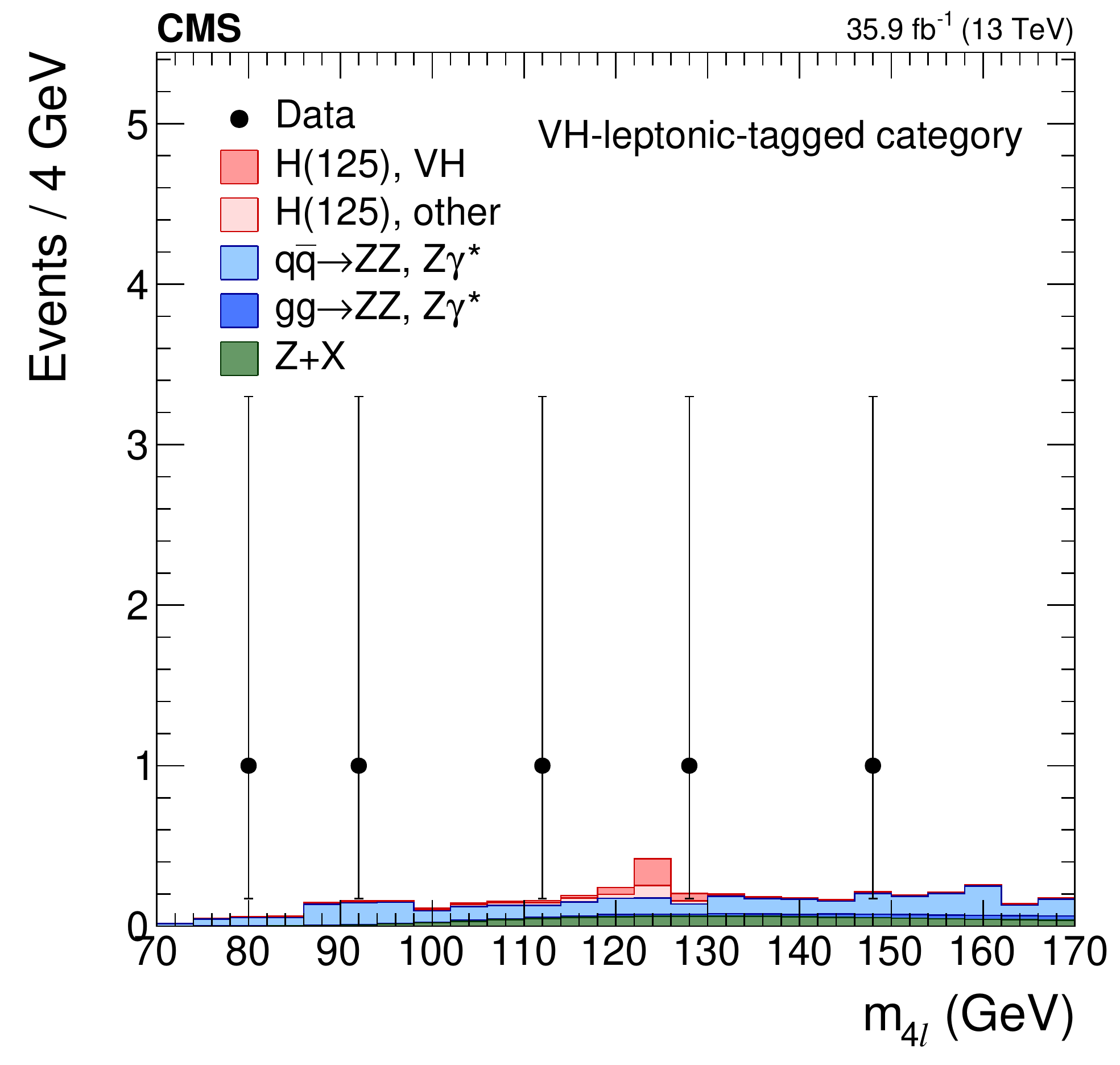}}
{ \includegraphics[width=0.32\textwidth]{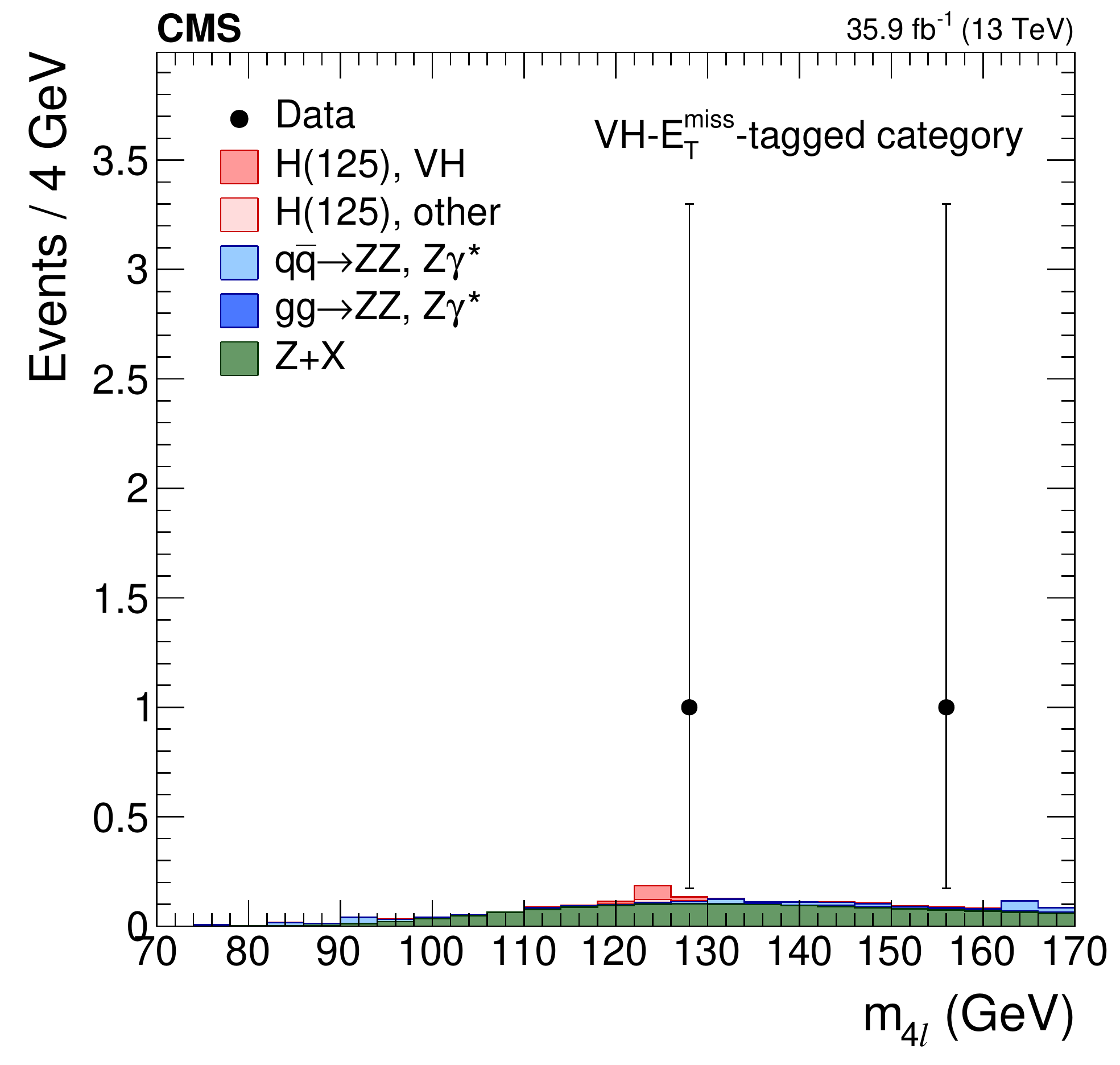}}
{ \includegraphics[width=0.32\textwidth]{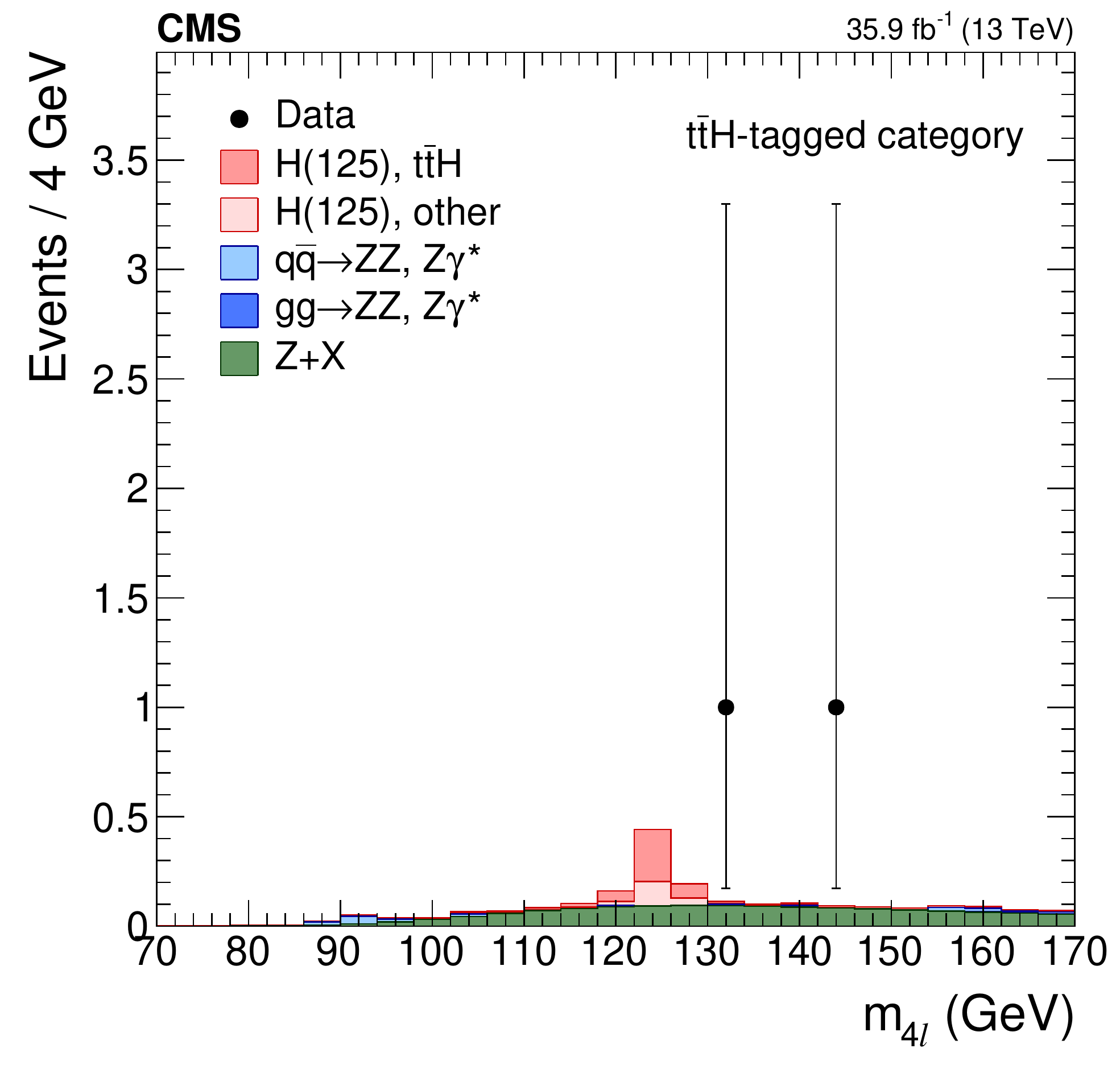}}
\caption{Distribution of the reconstructed four-lepton invariant mass in the seven event categories for the low-mass range.
(Top left) untagged category. (Top right) VBF-1jet-tagged category. (Center left) VBF-2jet-tagged category. (Center right) VH-hadronic-tagged category. (Bottom left) VH-leptonic-tagged category. (Bottom middle) VH-E$_T^\text{miss}$-tagged category. (Bottom right) \ttH-tagged category.
Points with error bars represent the data and stacked histograms represent expected signal and background distributions. The SM Higgs boson signal with $\mH=125\GeV$, denoted as $\PH(125)$, and the $\cPZ\cPZ$ backgrounds are normalized to the SM expectation,  whilst the $\cPZ$+X background is normalized to the estimation from data.
For the categories other than the untagged category, the SM Higgs boson signal is separated into two components: the production mode that is targeted by the specific
category, and other production modes, where the gluon fusion dominates.
The order in perturbation theory used for the normalization of the irreducible backgrounds is described in Section~\ref{sec:irrbkgd}.
\label{fig:Mass4l-3}}

\end{figure}

The reconstructed four-lepton invariant mass distribution is shown in Fig.~\ref{fig:FullM4l} for the sum of the  $4\Pe$, $4\Pgm$, and $2\Pe2\Pgm$ channels, and compared with the expectations from signal and background processes.
The error bars on the data points correspond to the so-called Garwood confidence intervals at 68\% confidence level (CL)~\cite{Garwood}.
The observed distribution agrees with the expectation within the statistical uncertainties over the whole spectrum.
In Fig.~\ref{fig:Mass4l-3}, the reconstructed four-lepton invariant mass distributions are split by event category, for the low-mass range.

The number of candidates observed in data and the expected yields for the backgrounds and the Higgs boson signal after the full event selection are reported in Table~\ref{tab:EventYieldsFull} for $\mllll>70\GeV$.
Table~\ref{tab:EventYieldsPeakCateg} shows the expected and observed yields for each of the seven event categories and their total.

\begin{table}[htb]
\centering
\topcaption{The numbers of expected background and signal events
and the number of observed candidate events after the full selection, for each final state,
for $\mllll>70\GeV$.
The signal and ZZ backgrounds are estimated from simulation,
while the $\cPZ$+X event yield is estimated from data.
Uncertainties include statistical and systematic sources.
\label{tab:EventYieldsFull}}
\begin{tabular}{lcccc}
 \hline
Channel & $4\Pe$ & $4\Pgm$ & $2\Pe2\Pgm$& $4\ell$ \\
\hline
\qqZZ                   & $ 193^{+19}_{-20}$ & $ 360^{+25}_{-27}$ & $ 471^{+33}_{-36}$ & $1024^{+69}_{-76}$ \\
\ggZZ                   & $  41.2^{+ 6.3}_{- 6.1}$ & $  69.0^{+ 9.5}_{- 9.0}$ & $ 102^{+14}_{-13}$ & $ 212^{+29}_{-27}$ \\
$\cPZ$+X               & $  21.1^{+ 8.5}_{-10.4}$ & $  34^{+14}_{-13}$ & $  60^{+27}_{-25}$ & $ 115^{+32}_{-30}$ \\
\hline

Sum of backgrounds      & $ 255^{+24}_{-25}$ & $ 463^{+32}_{-34}$ & $ 633^{+44}_{-46}$ & $1351^{+86}_{-91}$ \\
\hline

Signal   & $  12.0^{+ 1.3}_{- 1.4}$ & $  23.6\pm2.1$ & $  30.0\pm2.6$  & $  65.7\pm5.6$        \\
\hline

Total expected          & $ 267^{+25}_{-26}$ & $ 487^{+33}_{-35}$ & $ 663^{+46}_{-47}$ & $1417^{+89}_{-94}$ \\
\hline
Observed                & 293 & 505 & 681 & 1479 \\
\hline
\end{tabular}

\end{table}

\begin{table}[htb]
\centering
\topcaption{The numbers of expected background and signal events and the number of observed candidate events after the full selection, for each event category, for the mass range $118<\mllll<130\GeV$.
The yields are given for the different production modes.
The signal and ZZ backgrounds yields are estimated from simulation,
while the $\cPZ$+X yield is estimated from data.
\label{tab:EventYieldsPeakCateg}}
\newcolumntype{.}{D{.}{.}{2.2}}
\resizebox{\textwidth}{!}{
\begin{tabular}{ll*{8}.}
\hline
               && \multicolumn{7}{c}{Event category} &  \\ \cline{3-9}\\[-2ex]
               && \multicolumn{1}{c}{ Untagged } & \multicolumn{1}{c}{ VBF-1j } & \multicolumn{1}{c}{ VBF-2j } & \multicolumn{1}{c}{ VH-hadr. } & \multicolumn{1}{c}{ VH-lept. } & \multicolumn{1}{c}{ VH-\ETmiss } & \multicolumn{1}{c}{ $\ttH$ } & \multicolumn{1}{c}{  Inclusive}\\
\hline
&{\qqZZ}  & 19.18 & 2.00 & 0.25 & 0.30 & 0.27 & 0.01 & 0.01 & 22.01 \\
&{\ggZZ}  & 1.67 & 0.31 & 0.05 & 0.02 & 0.04 & 0.01 & \multicolumn{1}{c}{$<$0.0} & 2.09 \\
&{$\cPZ$+X}  & 10.79 & 0.88 & 0.78 & 0.31 & 0.17 & 0.30 & 0.27 & 13.52 \\
\multicolumn{2}{l}{Sum of backgrounds} & 31.64 & 3.18 & 1.08 & 0.63 & 0.49 & 0.32 & 0.28 & 37.62 \\
\multicolumn{2}{r}{uncertainties} & \multicolumn{1}{c}{$^{+4.30}_{-3.42}$} & \multicolumn{1}{c}{$^{+0.37}_{-0.32}$} & \multicolumn{1}{c}{$^{+0.29}_{-0.21}$} & \multicolumn{1}{c}{$^{+0.13}_{-0.09}$} & \multicolumn{1}{c}{$^{+ 0.07}_{-0.07}$} & \multicolumn{1}{c}{$^{+0.14}_{-0.11}$} & \multicolumn{1}{c}{$^{+0.09}_{-0.07}$}& \multicolumn{1}{c}{$^{+5.19}_{-4.18}$} \\
\hline
&$\ggH$ & 38.78 & 8.31 & 2.04 & 1.41 & 0.08 & 0.02 & 0.10 & 50.74 \\
&VBF    & 1.08 & 1.14 & 2.09 & 0.09 & 0.02 & \multicolumn{1}{c}{$<$0.01} & 0.02 & 4.44 \\
&WH     & 0.43 & 0.14 & 0.05 & 0.30 & 0.21 & 0.03 & 0.02 & 1.18 \\
&ZH     & 0.41 & 0.11 & 0.04 & 0.24 & 0.04 & 0.07 & 0.02 & 0.93 \\
&$\ttH$ & 0.08 & \multicolumn{1}{c}{$<$0.01} & 0.02 & 0.03 & 0.02 & \multicolumn{1}{c}{$<$0.01} & 0.35 & 0.50 \\
\multicolumn{2}{l}{Signal} & 40.77 & 9.69 & 4.24 & 2.08 & 0.38 & 0.11 & 0.51 & 57.79 \\
\multicolumn{2}{r}{uncertainties} &\multicolumn{1}{c}{$^{+3.69}_{-3.62}$} & \multicolumn{1}{c}{$^{+1.13}_{-1.17}$} & \multicolumn{1}{c}{$^{+ 0.55}_{-0.55}$} & \multicolumn{1}{c}{$^{+ 0.23}_{-0.23}$} & \multicolumn{1}{c}{$^{+ 0.03}_{-0.03}$} & \multicolumn{1}{c}{$^{+ 0.01}_{- 0.02}$} & \multicolumn{1}{c}{$^{+ 0.06}_{-0.06}$} & \multicolumn{1}{c}{$^{+4.89}_{-4.80}$} \\
\hline
\multicolumn{2}{l}{Total expected} & \multicolumn{1}{c}{72.41} & \multicolumn{1}{c}{12.88} & \multicolumn{1}{c}{5.32} & \multicolumn{1}{c}{2.71} &\multicolumn{1}{c}{0.86} & \multicolumn{1}{c}{0.43} & \multicolumn{1}{c}{0.79} & \multicolumn{1}{c}{95.41} \\
\multicolumn{2}{r}{uncertainties} &\multicolumn{1}{c}{ $^{+7.35}_{-6.27}$} & \multicolumn{1}{c}{$^{+1.25}_{-1.21}$} & \multicolumn{1}{c}{$^{+0.78}_{-0.65}$} & \multicolumn{1}{c}{$^{+0.34}_{-0.28}$} & \multicolumn{1}{c}{$^{+ 0.10}_{- 0.09}$} & \multicolumn{1}{c}{$^{+0.15}_{-0.12}$} & \multicolumn{1}{c}{$^{+ 0.14}_{-0.12}$} & \multicolumn{1}{c}{$^{+9.86}_{-8.32}$} \\
\hline
\multicolumn{2}{l}{Observed}  &\multicolumn{1}{c}{ 73 } &\multicolumn{1}{c}{ 13 } &\multicolumn{1}{c}{ 4  } &\multicolumn{1}{c}{ 2  } &\multicolumn{1}{c}{ 1  } &\multicolumn{1}{c}{ 1  } &\multicolumn{1}{c}{ 0  } &\multicolumn{1}{c}{ 94 }\\
\hline
\end{tabular}
}
\end{table}

\begin{figure}[!htb]
\vspace*{0.3cm}
\centering
\includegraphics[width=0.32\textwidth]{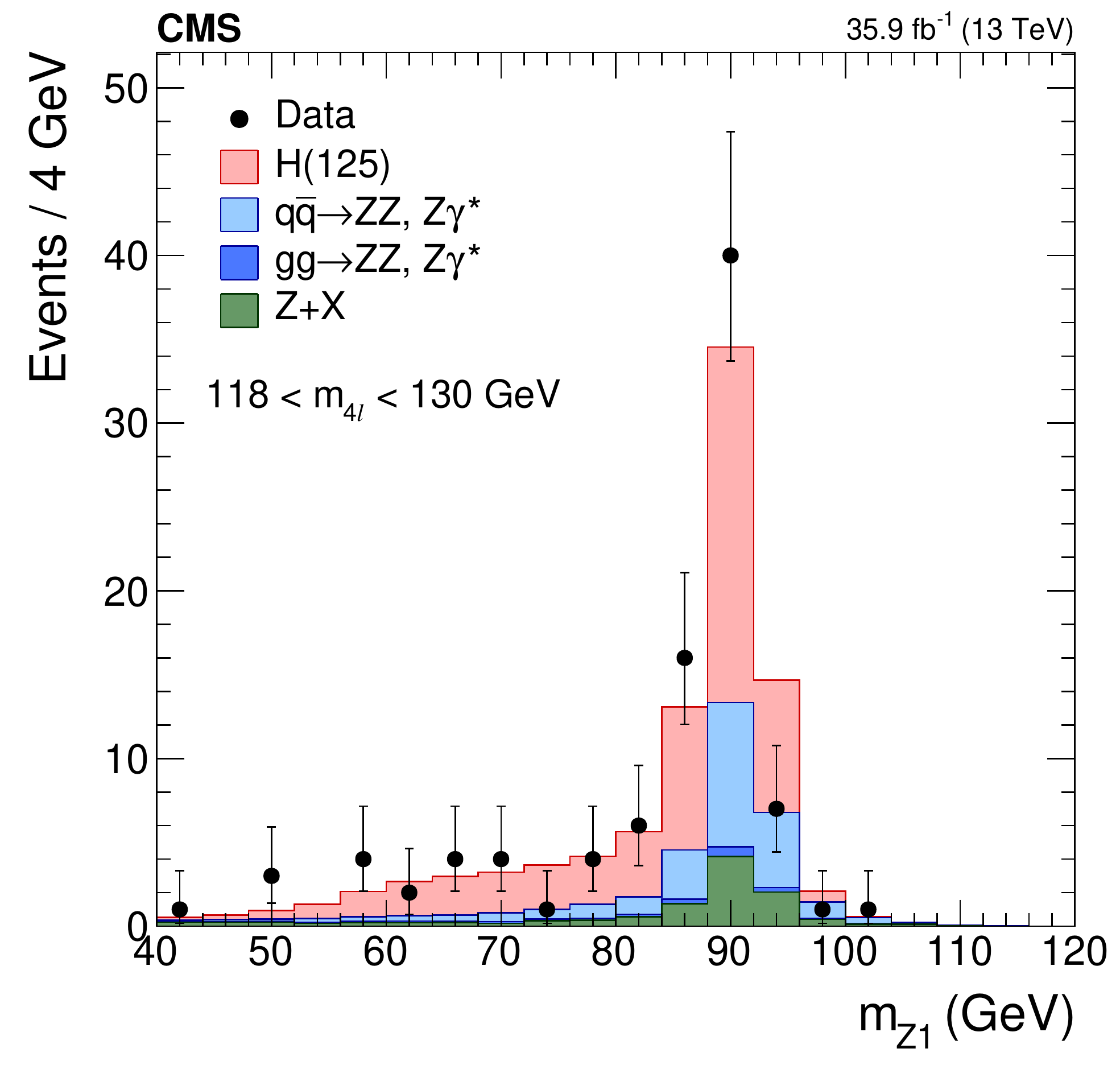}
\includegraphics[width=0.32\textwidth]{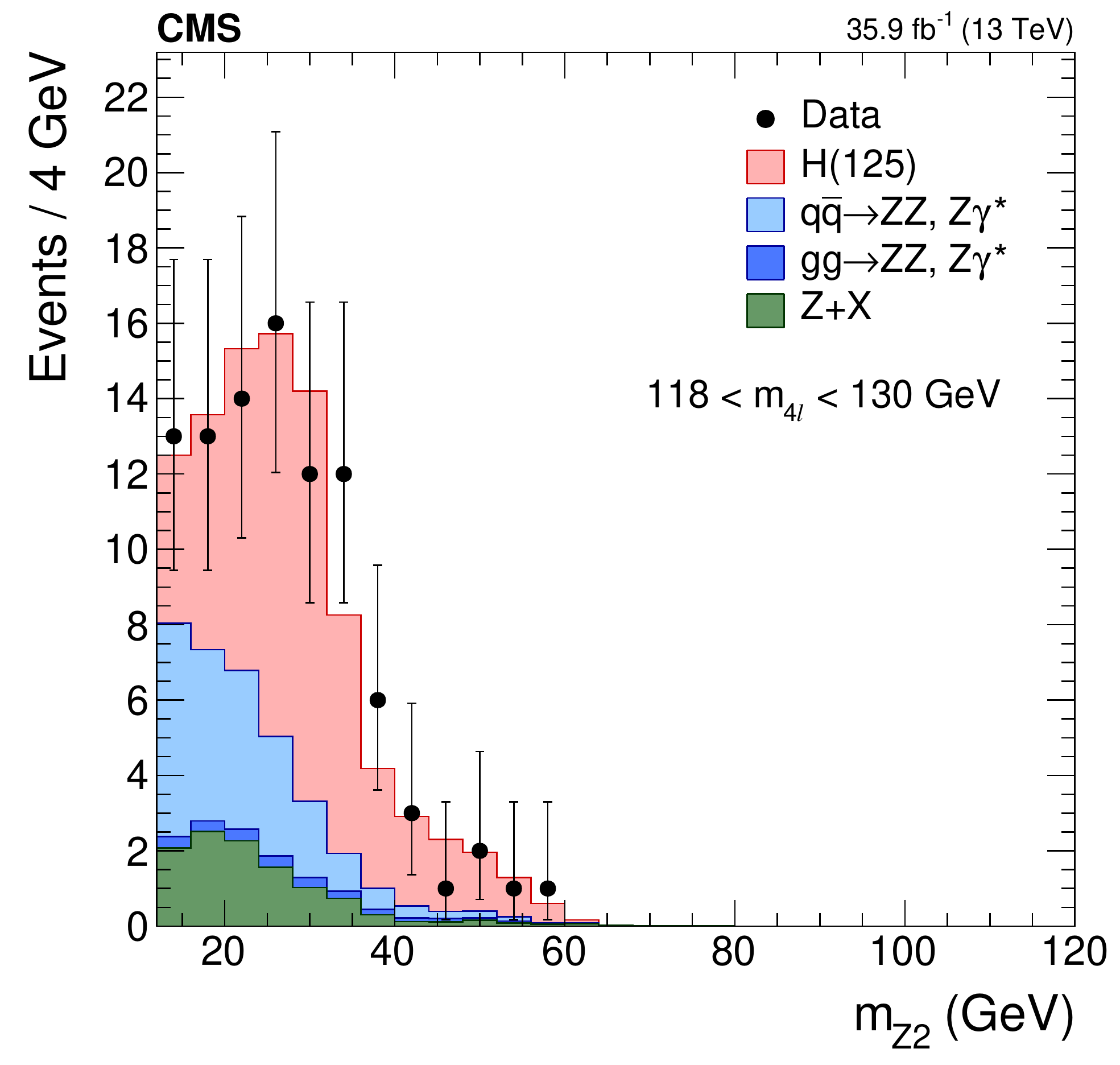}
\includegraphics[width=0.32\textwidth]{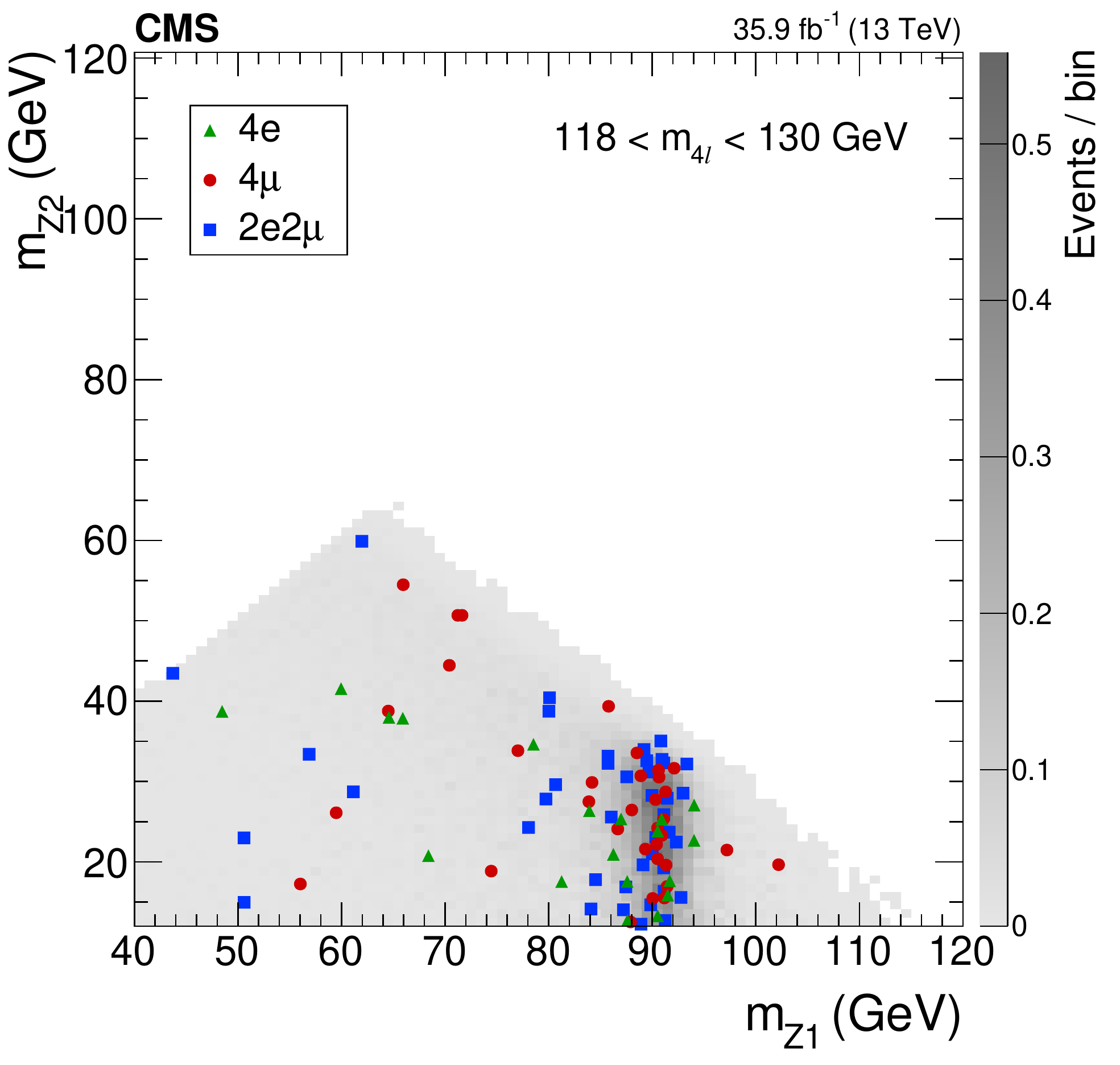}
\caption{Distribution of the $\cPZ_1$ (left) and $\cPZ_2$ (middle) reconstructed invariant masses and two-dimensional distribution of these two variables (right) in the mass region $118<\mllll<130\GeV$. The stacked histograms and the gray scale represent the expected signal and background distributions, and points represent the data. The SM Higgs boson signal with $\mH=125\GeV$, denoted as $\PH(125)$, and the $\cPZ\cPZ$ backgrounds are normalized to the SM expectation, whilst the $\cPZ$+X background is normalized to the estimation from data.
The order in perturbation theory used for the normalization of the irreducible backgrounds is described in Section~\ref{sec:irrbkgd}.
\label{fig:MZ1MZ2}}

\end{figure}

\begin{figure}[!htb]
\vspace*{0.3cm}
\centering
\includegraphics[width=0.6\textwidth]{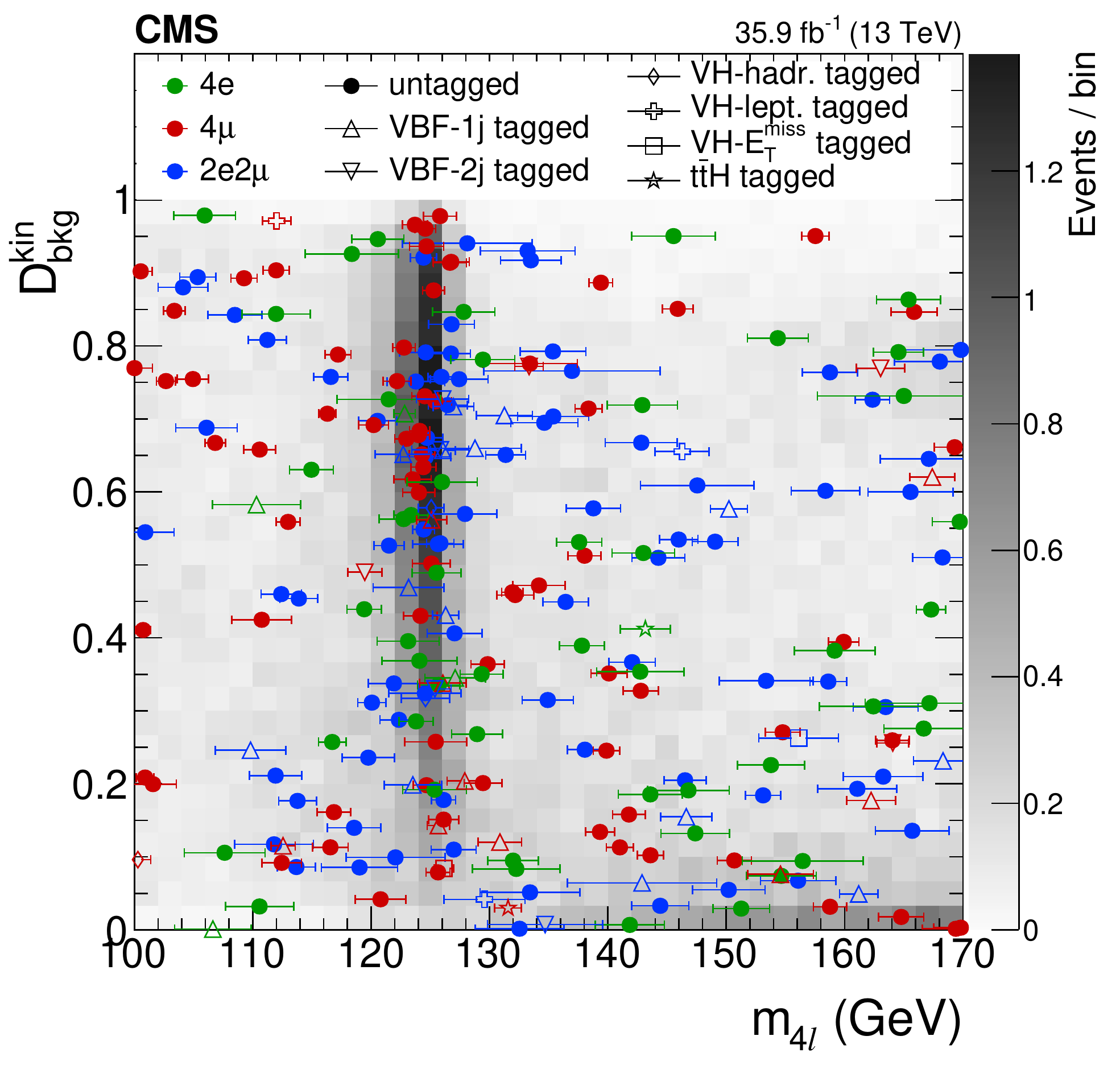}
\caption{Distribution of $\KD$ versus $\mllll$ in the mass region $100<\mllll<170\GeV$. The gray scale represents the expected total number of $\cPZ\cPZ$ background and SM Higgs boson signal events for $\mH=125\GeV$. The points show the data and the horizontal bars represent $\MassD$. Different marker colors and styles are used to denote final state and the categorization of the events, respectively.
\label{fig:KD}}

\end{figure}

\begin{figure}[!htb]
\vspace*{0.3cm}
\centering
\includegraphics[width=0.32\textwidth]{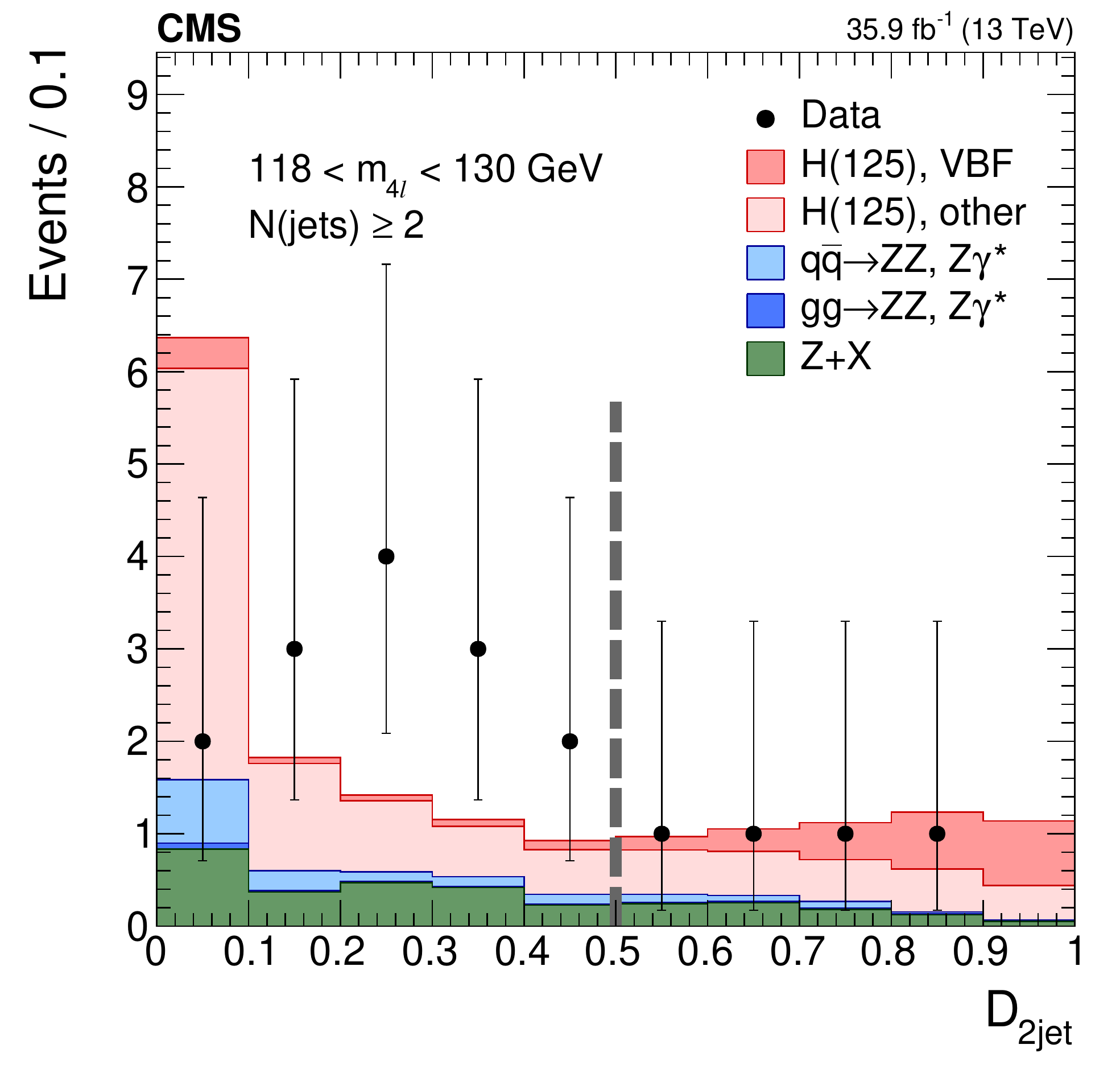}
\includegraphics[width=0.32\textwidth]{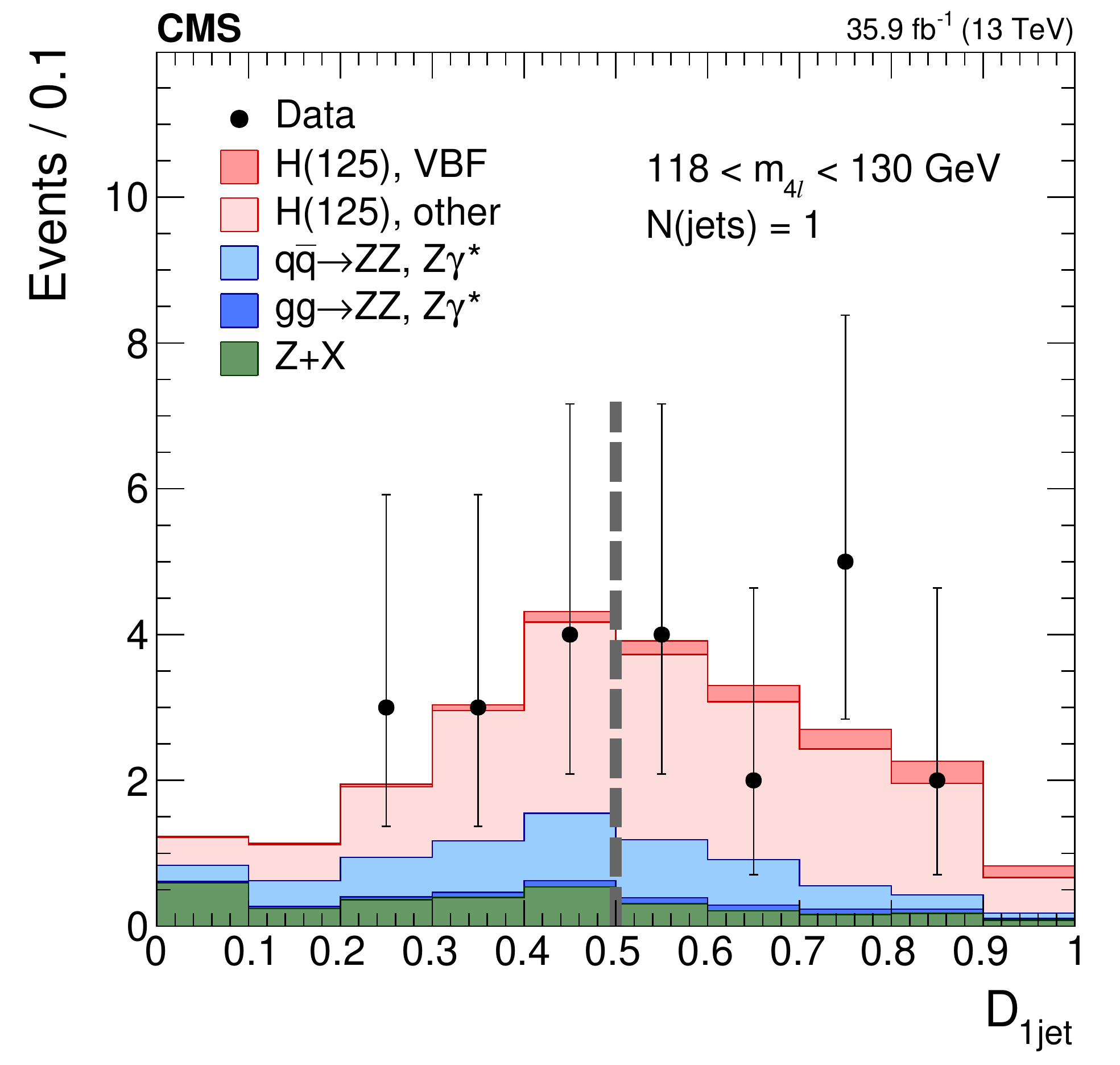}
\includegraphics[width=0.32\textwidth]{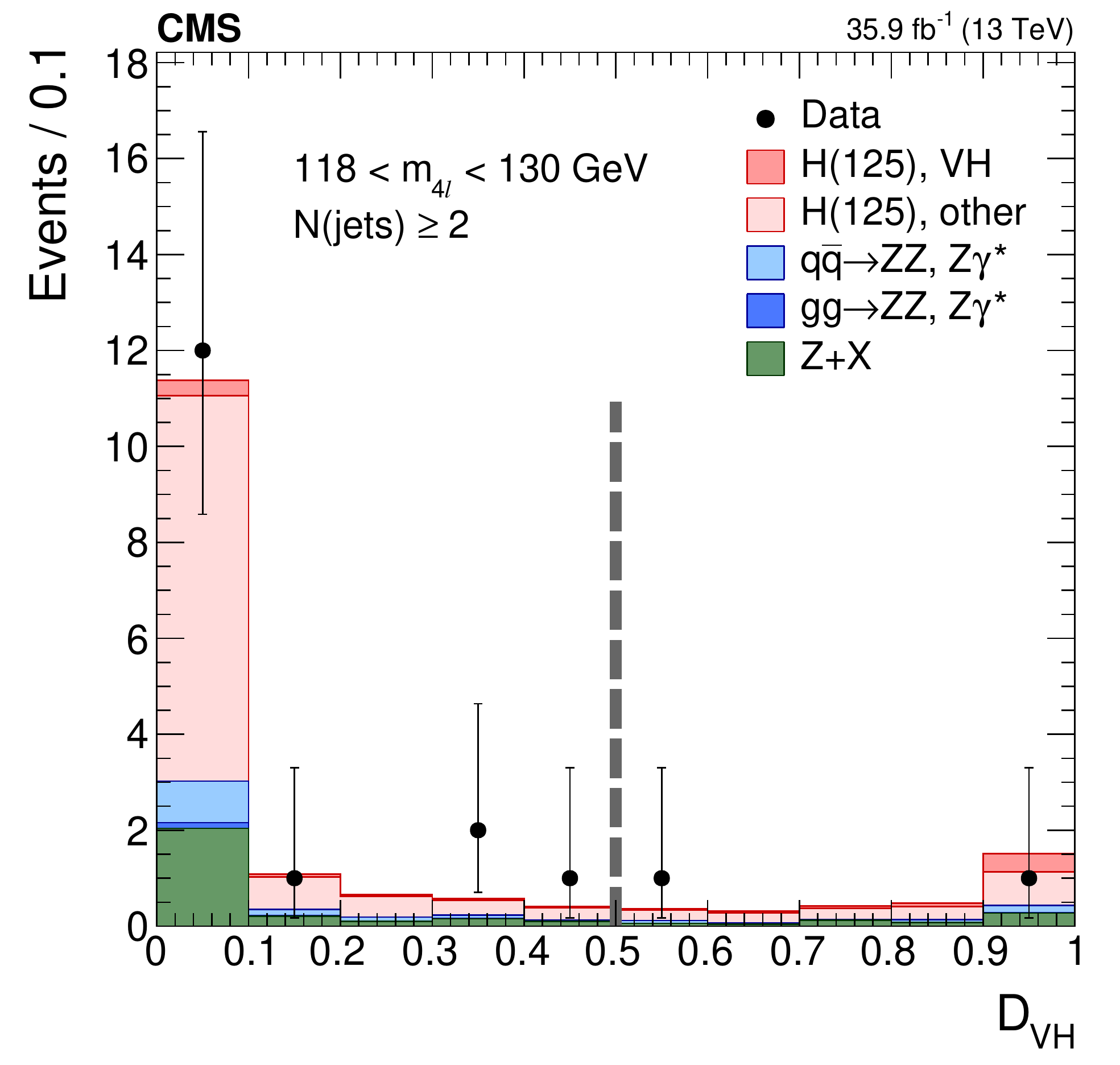}
\caption{Distribution of categorization discriminants in the mass region $118<\mllll<130\GeV$.
(Left) \DMeVbfjj.
(Middle) \DMeVbfj.
(Right) \DMeVh = max(\DMeWh,\DMeZh).
Points with error bars represent the data and stacked histograms represent expected signal and background distributions. The SM Higgs boson signal with $\mH=125\GeV$, denoted as $\PH(125)$, and the $\cPZ\cPZ$ backgrounds are normalized to the SM expectation,  whilst the $\cPZ$+X background is normalized to the estimation from data. The vertical gray dashed lines denote the working points used in the event categorization. The SM Higgs boson signal is separated into two components: the production mode that
is targeted by the specific discriminant, and other production modes, where the gluon
fusion dominates.
The order in perturbation theory used for the normalization of the irreducible backgrounds is described in Section~\ref{sec:irrbkgd}.
\label{fig:Djet}}

\end{figure}

The reconstructed dilepton invariant masses for the selected $\cPZ_1$ and $\cPZ_2$ candidates are shown in Fig.~\ref{fig:MZ1MZ2} for $118<\mllll<130\GeV$, along with their correlation.
Figure~\ref{fig:KD} shows the correlation between the kinematic discriminant $\KD$ with the four-lepton invariant mass, the two variables used in the likelihood fit to extract the results (see Section~\ref{res:sigstre}).
The gray scale represents the expected combined relative density of the ZZ background and the Higgs boson signal.
The points show the data and the measured four-lepton mass uncertainties $\MassD$ as horizontal bars.
Different marker colors and styles are used to denote the final state and the categorization of the events, respectively.
This distribution shows that the two observed events around 125\GeV in the VH-$\ETmiss$-tagged and ttH-tagged categories (empty star and square markers) have low values of $\KD$,
implying that these events are more compatible with the background than the signal hypothesis.
The distribution of the discriminants used for event categorization and the corresponding working point values are shown in Fig.~\ref{fig:Djet}.

\clearpage
\subsection{Signal strength modifiers}
\label{res:sigstre}

To extract the signal strength modifier we perform a multi-dimensional fit that relies on two variables: the four-lepton invariant mass $\mllll$ and the $\KD$ discriminant.
We define the two-dimensional likelihood function as:
\begin{equation}
\mathcal{L}_{2D}(\mllll,\KD) = \mathcal{L}(\mllll) \mathcal{L}(\KD|\mllll) .
\end{equation}
The mass dimension is unbinned and uses the model described in Section~\ref{sec:signal}.
The conditional term is implemented by creating a two-dimensional template of $\mllll$ vs. $\KD$ normalized to 1 for each bin of $\mllll$.
Based on the seven event categories and the three final states ($4\Pe$, $4\mu$, $2\Pe2\mu$), the $(\mllll, \KD)$ unbinned distributions are split into 21 categories.

\begin{figure}[!htb]
\centering
\includegraphics[width=0.46\linewidth]{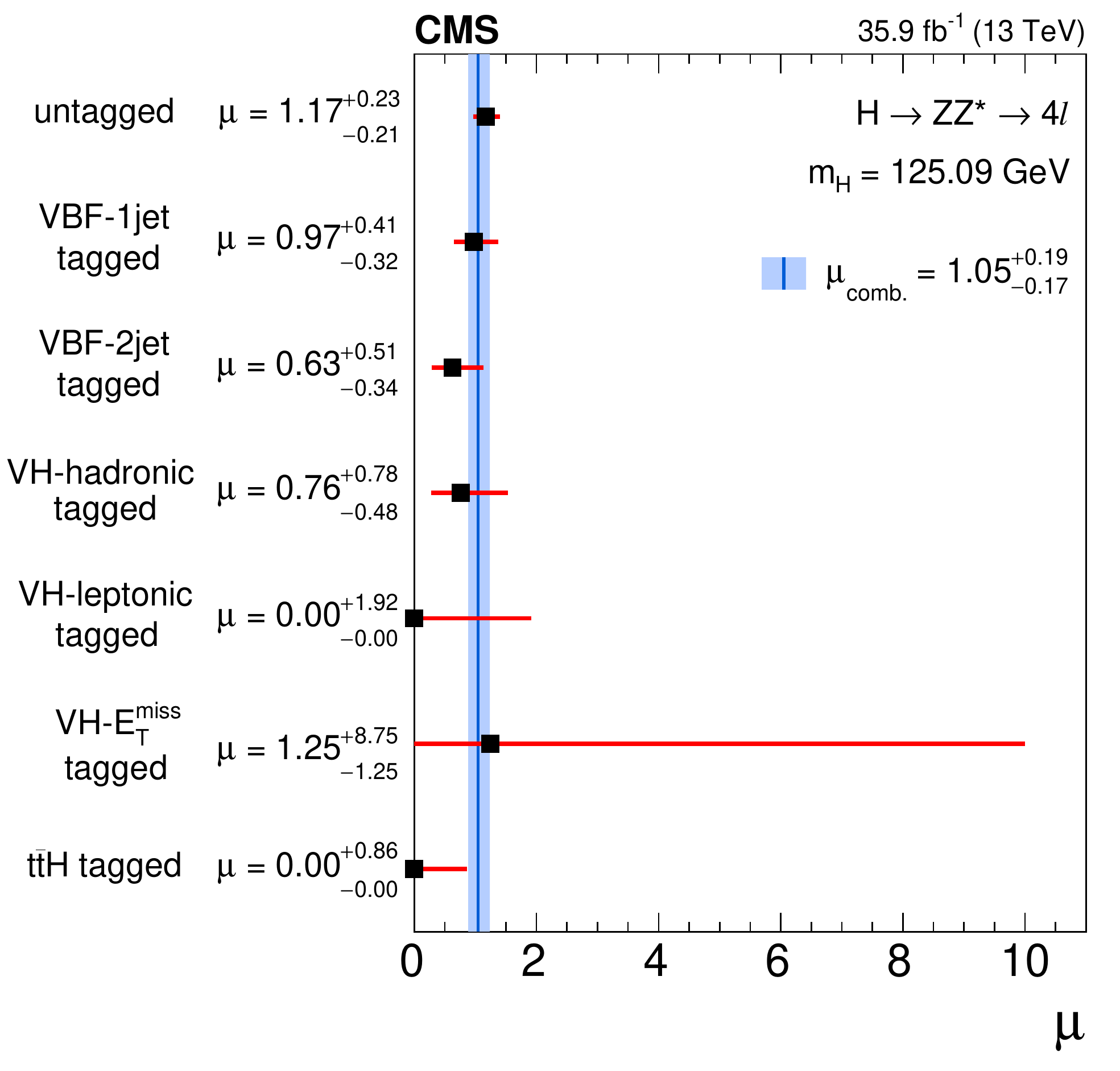}
\includegraphics[width=0.46\linewidth]{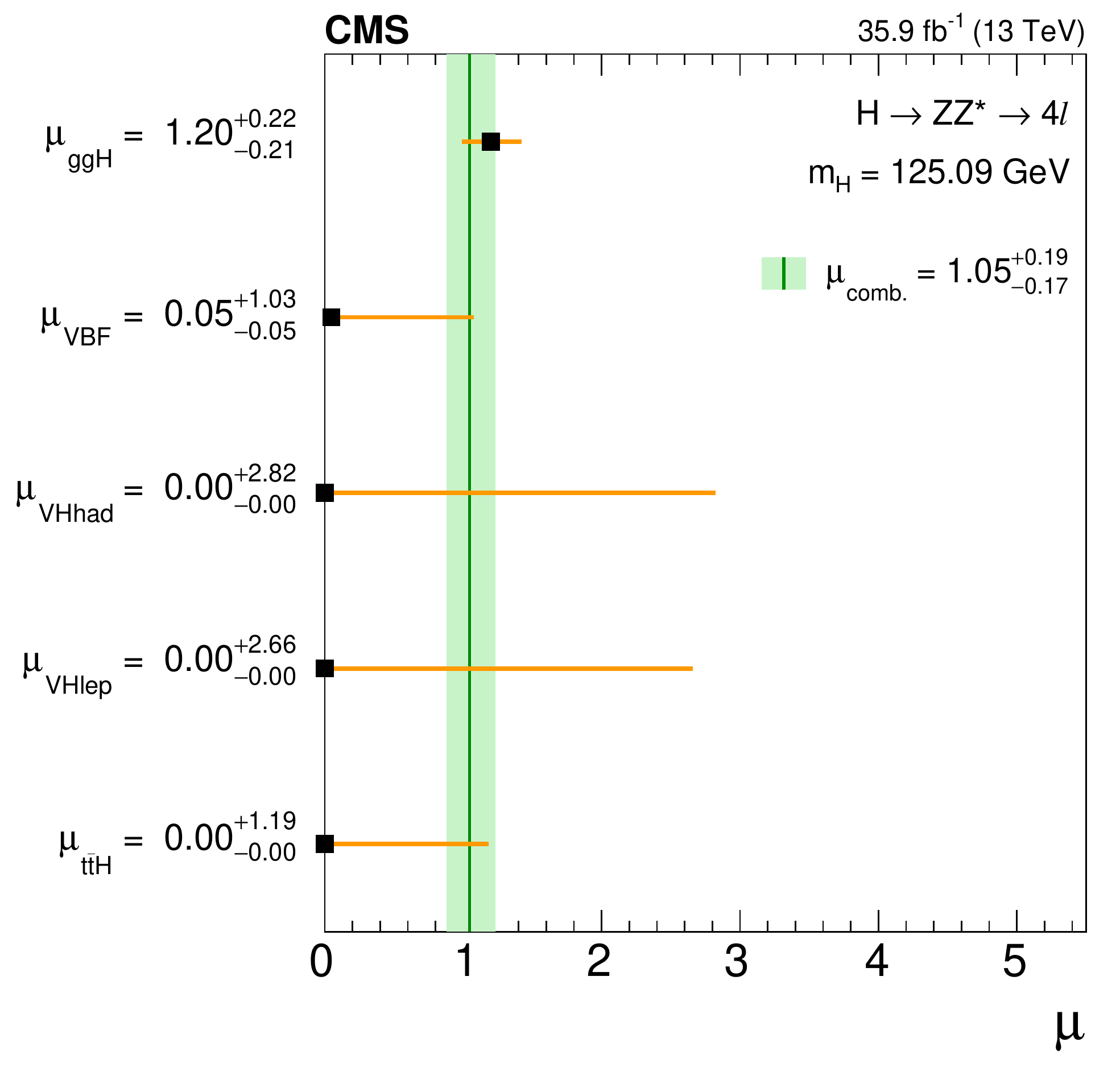} \\
\includegraphics[width=0.46\linewidth]{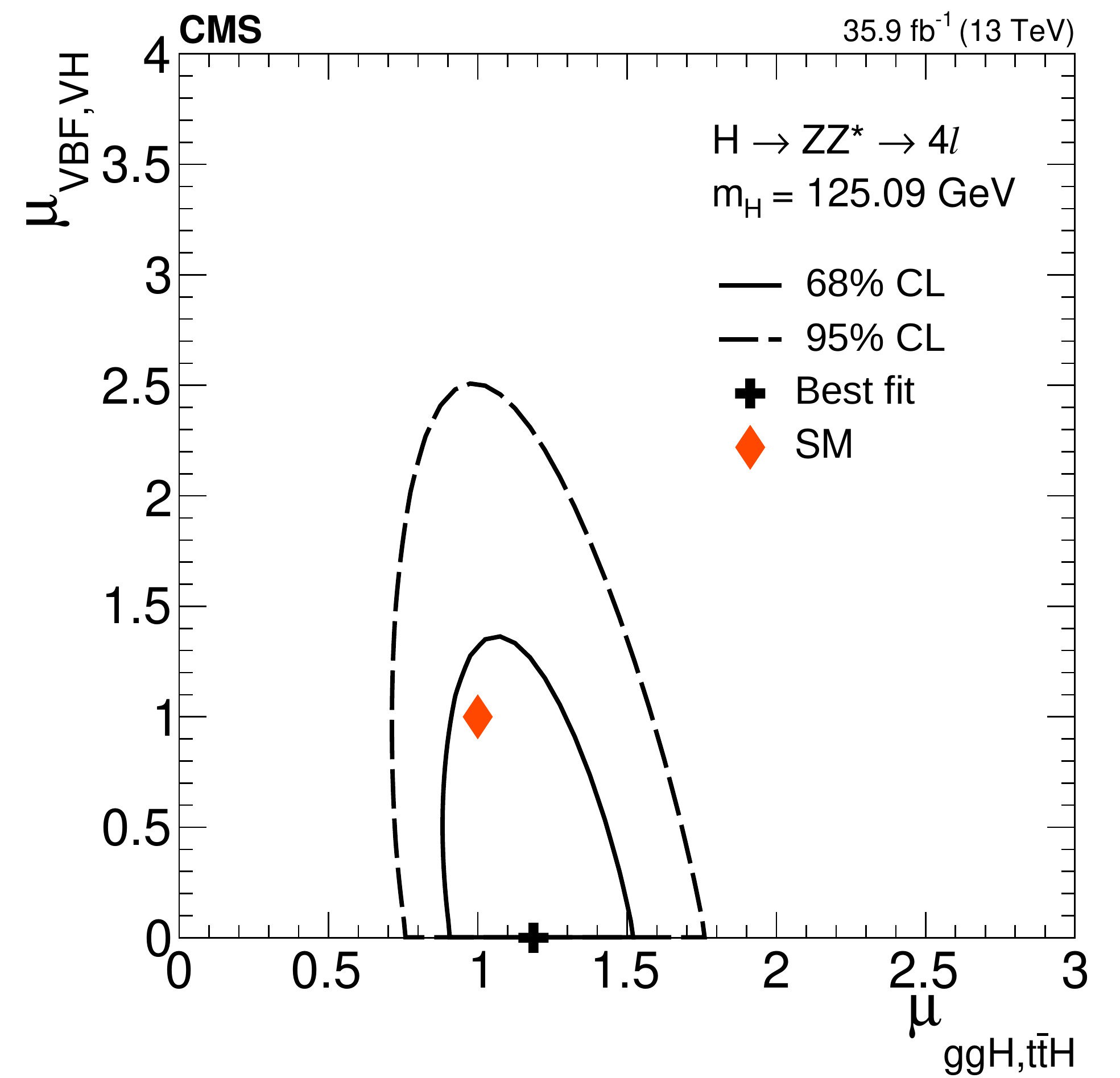}
\caption{
 (Top left) Observed values of the signal strength modifier $\mu=\sigma/\sigma_{SM}$ for the seven event categories, compared to the combined $\mu$ shown as a vertical line. The horizontal bars and the filled band indicate the $\pm$ 1$\sigma$ uncertainties.
(Top right) Results of likelihood scans for the signal strength modifiers corresponding to the main SM Higgs boson production modes, compared to the combined $\mu$ shown as a vertical line.
The horizontal bars and the filled band indicate the $\pm$ 1$\sigma$ uncertainties.
The uncertainties include both statistical and systematic sources.
(Bottom) Result of the 2D likelihood scan for the $\muF$ and $\muV$ signal strength modifiers.
The solid and dashed contours show the 68\% and 95\% CL regions, respectively.
The cross indicates the best fit values, and the diamond represents the expected values for the SM Higgs boson.
\label{fig:mucat}}

\end{figure}

A simultaneous fit to all categories is performed to extract the signal strength modifier.
The relative fraction of $4\Pe$, $4\mu$, and $2\Pe2\mu$ signal events is fixed to the SM prediction.
Systematic uncertainties are included in the form of nuisance parameters and the results are obtained using an
asymptotic approach with a test statistic based on the profile likelihood ratio~\cite{LHC-HCG,Cowan:2010js}.
The individual contributions of statistical and systematic uncertainties are separated by performing a
likelihood scan removing the systematic uncertainties to determine the statistical uncertainty.
The systematic uncertainty is then taken as the
difference in quadrature between the total uncertainty and the statistical uncertainty.
At the ATLAS and CMS Run 1 combined mass value of $\mH=125.09\GeV$, the signal strength modifier is $\mu = 1.05~^{+0.15}_{-0.14}\stat~^{+0.11}_{-0.09}\syst =\valMuAtRunIMass$.
It is compared to the measurement for each of the seven event categories in Fig.~\ref{fig:mucat} (top left).
The observed values are consistent with the SM prediction of $\mu = 1$ within the uncertainties.
The dominant sources of experimental systematic uncertainty are the uncertainties in the lepton identification efficiencies and integrated luminosity measurement, while the dominant theoretical sources are the uncertainty in the total gluon fusion cross section as well as the uncertainty in the category migration for the gluon fusion process.
The contributions to the total uncertainty from experimental and theoretical sources are found to be similar in magnitude.

A fit is performed for five signal strength modifiers ($\mu_{\Pg\Pg\PH}$, $\mu_{\mathrm{VBF}}$, $\mu_{\mathrm{V}\PH\text{had}}$, $\mu_{\mathrm{V}\PH\text{lep}}$, and $\mu_{\ttH}$, all constrained to positive values) that control the contributions of the main SM Higgs boson production modes.
The WH and ZH processes are merged, and then split based on the decay of the associated vector boson into either hadronic decays (VHhad) or leptonic decays (VHlep).
The results are reported in Fig.~\ref{fig:mucat} (top right) and compared to the expected signal strength modifiers in Table~\ref{tab:sigstr}.
The expected uncertainties are evaluated by generating an Asimov data set~\cite{Cowan:2010js}, which is a representative event sample that
provides both the median expectation for an experimental result and its expected statistical variation, in the asymptotic approximation.
The coverage of the quoted intervals has been verified for a subset of results using the
Feldman-Cousins method~\cite{FC}.
The low observed signal strengths for the VBF, VH, and $\ttH$ processes can be explained by the mild excess in the untagged category, which leads to a higher than expected signal strength for the $\ggH$ process that contributes significantly to the total signal yield in categories that are based on the hadronic activity in the event.
In the categories that are not based on hadronic event activity, events with $\mllll$ near 125\GeV have low $\KD$ values, and are therefore more compatible with the background than the signal hypothesis.

\begin{table}[!hb]
\centering
\topcaption{Expected and observed signal strength modifiers.\label{tab:sigstr}}
\renewcommand{\arraystretch}{1.5}
\begin{tabular}{lcccccc}
\hline
&  Inclusive & $\mu_{\Pg\Pg\PH}$ & $\mu_{\mathrm{VBF}}$ &  $\mu_{\mathrm{V}\PH\text{had}}$ & $\mu_{\mathrm{V}\PH\text{lep}}$ & $\mu_{\ttH}$ \\
\hline
Expected  & $1.00^{+0.15}_{-0.14}\stat^{+0.10}_{-0.08}\syst$ & $1.00^{+0.23}_{-0.21}$ & $1.00^{+1.25}_{-0.97}$ & $1.00^{+3.96}_{-1.00}$  & $1.00^{+3.76}_{-1.00}$ & $1.00^{+3.23}_{-1.00}$ \\
Observed  & $1.05^{+0.15}_{-0.14}\stat^{+0.11}_{-0.09}\syst$ & $1.20^{+0.22}_{-0.21}$ & $0.05^{+1.03}_{-0.05}$  & $0.00^{+2.83}_{-0.00}$ & $0.00^{+2.66}_{-0.00}$ &  $0.00^{+1.19}_{-0.00}$\\
\hline
\end{tabular}
\end{table}

Two signal strength modifiers $\muF$ and $\muV$ are introduced as scale factors for the fermion- and vector-boson induced contribution to the expected SM cross section.
A two-parameter fit is performed simultaneously to all categories assuming a mass of $\mH = 125.09\GeV$, leading to the measurements of  $\muF=\valMuFAtRunIMass$ and $\muV=\valMuVAtRunIMass$.
The 68\% and 95\% CL contours in the ($\muF,\muV$) plane are shown in Fig.~\ref{fig:mucat} (bottom).
The SM predictions of $\muF = 1$ and $\muV = 1$ lie within the 68\% CL regions of this measurement.

\clearpage
\subsection{Cross section measurements}
\label{res:crosssec}

\begin{figure}[!htb]
\centering
\includegraphics[width=0.51\linewidth]{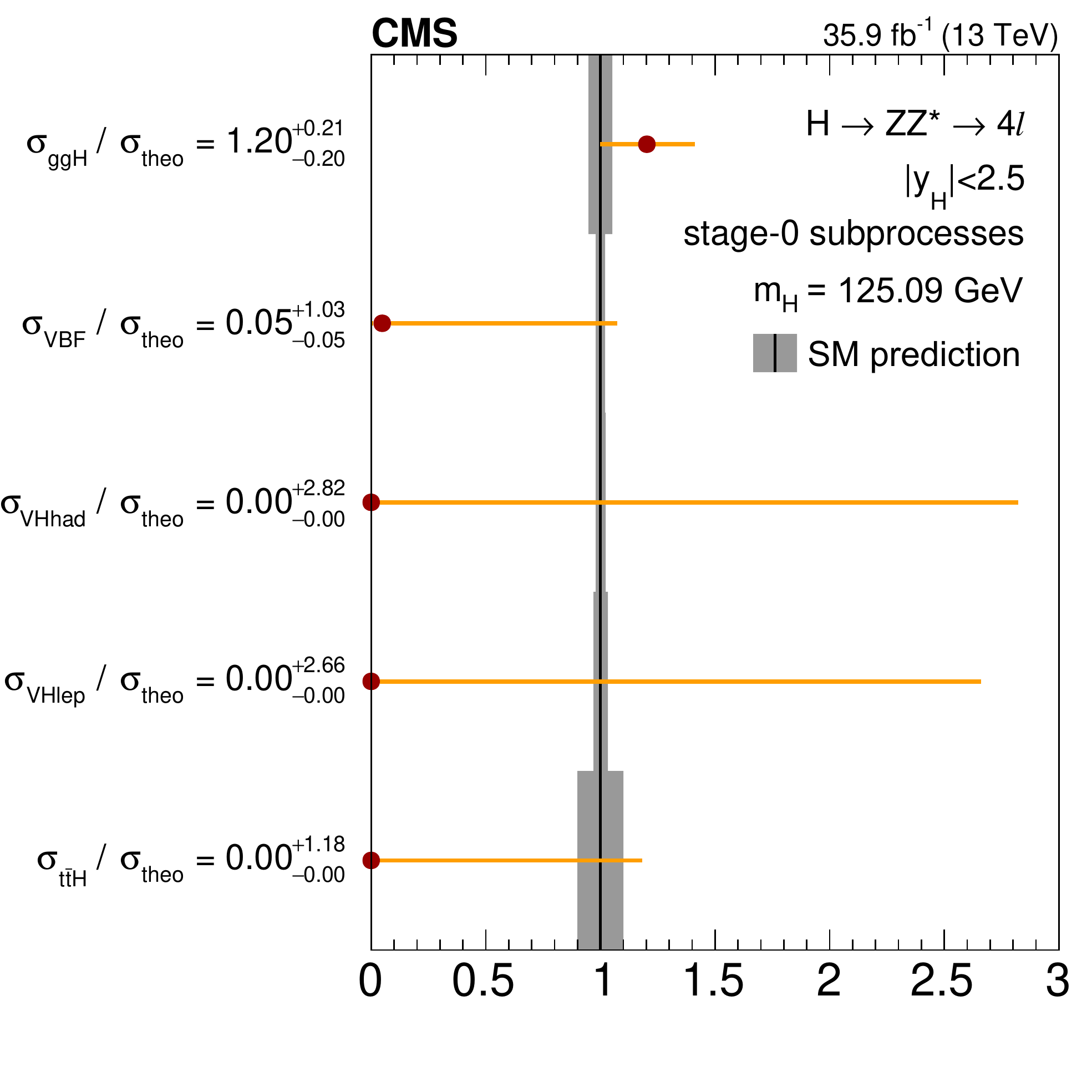}
\caption{
Results of the fit for simplified template cross sections for the `stage-0 subprocesses', normalized to the SM predictions. The grey bands indicate the theoretical uncertainties in the SM predictions.
The orange error bars show the full uncertainty, including experimental uncertainties and theoretical uncertainties causing migration of events between the various categories.
See Ref.~\cite{YR4} for further details of this approach.
\label{fig:stxs}}

\end{figure}

In this section we present various measurements of the cross section for Higgs boson production.
First we show cross section measurements for five SM Higgs boson production processes
($\sigma_{\Pg\Pg\PH}$, $\sigma_{\mathrm{VBF}}$, $\sigma_{\mathrm{V}\PH\text{had}}$, $\sigma_{\mathrm{V}\PH\text{lep}}$, and $\sigma_{\ttH}$) in a simplified fiducial
volume defined using a selection on the Higgs boson rapidity $\abs{y_\PH}<2.5$. Outside of this volume the analysis has a negligible acceptance.
The separation of the production processes is achieved through the categorization of events described in Section~\ref{sec:categorization}.
This measurement corresponds to the `stage-0' simplified template cross sections from Ref.~\cite{YR4}.
This approach allows one to reduce the dependence of the measurements
on the theoretical uncertainties in the SM predictions, avoiding extrapolation
of the measurements to the full phase space which carries nontrivial or sizeable theoretical uncertainties.
The measured cross sections, normalized to the SM prediction~\cite{YR4},
which is denoted as $\sigma_\text{theo}$, are shown in Fig.~\ref{fig:stxs}.
The dominant sources of experimental systematic uncertainty are the same as in the measurement of the signal strength modifier, while the
dominant theoretical source is the uncertainty in the category migration for the gluon fusion process.

The cross section for the production and decay $\Pp\Pp\to\Hllll$ in a tight fiducial phase space is also presented.
This measurement has minimal dependence on the assumptions of the relative fraction or kinematic
distributions of the separate production modes.
The definition of the generator-level fiducial volume, chosen to match closely the reconstruction-level selection, is very similar to the definition used in Ref.~\cite{CMSH4lFiducial8TeV}.
The differences with respect to Ref.~\cite{CMSH4lFiducial8TeV} are that
leptons are defined as ``dressed'' leptons, as opposed to Born-level leptons, and the lepton isolation criterion is
updated to match the reconstruction-level selection.
Leptons are ``dressed'' by adding the four-momenta of photons within $\Delta R<0.3$ to the bare leptons,
and leptons are considered isolated if the scalar sum of transverse momenta of all stable particles, excluding electrons, muons, and neutrinos, within $\Delta R < 0.3$ from the lepton is less than $0.35 \pt\,(\GeVns{})$.
For the measurement of differential cross sections related to jet observables, only well measured central jets
with $\pt>30\GeV$ and $\abs{\eta}<2.5$ are considered in both the fiducial and reconstruction-level selections.
To simplify the definition of the fiducial volume, the $\KD$ discriminant is not used
to select the ZZ candidate at the generator level. Instead  the $\cPZ_{1}$ candidate is chosen to be the one with $m(\cPZ_1)$ closest to the
nominal $\cPZ$ boson mass, and in cases where multiple $\cPZ_{2}$ candidates satisfy all criteria, the pair of leptons with the highest sum of the transverse momenta is chosen.
The same candidate selection is also used at the reconstruction level for the fiducial cross section measurements
to align the reconstruction- and fiducial-level selections as closely as possible.
The full definition of the fiducial volume is detailed in Table~\ref{tab:FidDef} and the acceptance $\mathcal{A}_\text{fid}$
for various SM production modes is given in Table~\ref{tab:summarySM}.

\begin{table}[!htb]
\centering
\topcaption{
Summary of requirements and selections used in the definition of the fiducial phase space for the $\Pp\Pp\to\Hllll$ cross section measurements.
\label{tab:FidDef}
}
\begin{tabular}{lc}
\hline
\multicolumn{2}{c}{Lepton kinematics and isolation} \\
\hline
Leading lepton $\pt$ & $\pt > 20$\GeV \\
Subleading lepton $\pt$ & $\pt > 10$\GeV \\
Additional electrons (muons) $\pt$ & $\pt > 7\,(5)$\GeV \\
Pseudorapidity of electrons (muons) & $\abs{\eta} < 2.5\,(2.4)$ \\
Sum $\pt$ of all stable particles within $\Delta R < 0.3$ from lepton & ${<}0.35\pt$ \\[1.2ex]
\hline
\multicolumn{2}{c}{Event topology} \\
\hline
\multicolumn{2}{l}{Existence of at least two same-flavor OS lepton pairs, where leptons satisfy criteria above} \\
Invariant mass of the Z$_1$ candidate & $40 < m_{\cPZ_{1}} < 120 \GeV$ \\
Invariant mass of the Z$_2$ candidate & $12 < m_{\cPZ_{2}} < 120 \GeV$ \\
Distance between selected four leptons & $\Delta R(\ell_{i},\ell_{j})>0.02$ for any $i\neq j$  \\
Invariant mass of any opposite-sign lepton pair & $m_{\ell^{+}\ell'^{-}}>4 \GeV$ \\ \\
Invariant mass of the selected four leptons & $105 < \mllll < 140 \GeV$  \\
\hline
\end{tabular}
\end{table}

A maximum likelihood fit of the signal and  background parameterizations to the observed $4\ell$
mass distribution, $N_{\mathrm{obs}}(\mllll)$, is performed to extract
the integrated fiducial cross section $\sigma_{\text{fid}}$ for $\Pp\Pp\to\PH\to4\ell$.
The fit is done without any event categorization targeting different production modes and does not use the $\KD$ observable to minimize the model dependence.
The fit is performed simultaneously in all final states assuming a Higgs boson mass of $\mH = 125.09\GeV$, and
the branching fraction of the Higgs boson decays to different final states ($4\Pe,4\mu,2\Pe2\mu$) is allowed to float.

The number of expected events in each final state $\mathrm{f}$ and in each bin $i$ of an observable considered is expressed as a
function of $\mllll$ as:
\begin{equation}
\label{eqn:m4l}
\begin{aligned}
N_{\text{exp}}^{\mathrm{f},i}(m_{4\ell}) &= N_{\text{fid}}^{\mathrm{f},i}(m_{4\ell})+N_{\text{nonfid}}^{\mathrm{f},i}(m_{4\ell})+N_{\text{nonres}}^{\mathrm{f},i}(m_{4\ell})+N_{\text{bkg}}^{\mathrm{f},i}(m_{4\ell}) \\
&=\sum_j\epsilon_{i,j}^{\mathrm{f}} \,  \left(1+f_\text{nonfid}^{\mathrm{f},i} \right)\, \sigma_{\text{fid}}^{\mathrm{f},j} \,  \mathcal{L}\, \mathcal{P}_{\mathrm{res}}(m_{4\ell}) \\
&~~~~~+ N_{\text{nonres}}^{\mathrm{f},i}\, \mathcal{P}_{\text{nonres}}(m_{4\ell})+N_{\text{bkg}}^{\mathrm{f},i}\, \mathcal{P}_{\text{bkg}}(m_{4\ell}).
\end{aligned}
\end{equation}

The shape of the resonant signal contribution, $\mathcal{P}_{\mathrm{res}}(m_{4\ell})$, is modelled by a double-sided Crystal Ball function, as described in Section~\ref{sec:signal}, and the normalization is proportional to the fiducial cross section. The non-resonant contribution from $\WH$, $\ZH$, and $\ttH$ production, $N_{\text{nonres}}$, is modeled by a Landau distribution, $\mathcal{P}_{\text{nonres}}(m_{4\ell})$, whose shape parameters are constrained in the fit to be within a range determined from signal samples with full detector simulation and is treated as a background in this measurement.

The $\epsilon_{i,j}^{\mathrm{f}}$ factor represents the detector response matrix that maps the number of expected events in a given observable
bin $j$ at the fiducial level to the number of expected events in the bin $i$ at the reconstruction level. This response matrix
is measured using signal samples with full detector simulation and corrected for residual differences between data and simulation.
This procedure accounts for the unfolding of detector effects from the observed distributions and is the same as in
Refs.~\cite{CMSHggFiducial8TeV} and \cite{CMSH4lFiducial8TeV}.
In the case of the integrated fiducial cross section measurement the efficiencies reduce to single values, which for different
SM production modes are listed in Table~\ref{tab:summarySM}.

An  additional resonant contribution arises from events which are reconstructed, but do not originate from the fiducial phase space, $N_{\text{nonfid}}$. These events are due to detector effects that cause differences between the quantities used for the fiducial phase space definition and the analogous quantities at the reconstruction level. This contribution is treated as background and is referred to as the ``nonfiducial signal'' contribution. The shape of these events is verified using signal samples with full detector simulation to be identical to the shape of the fiducial signal, and its normalization is fixed to be a fraction of the fiducial signal component. The value of this fraction, which we denote as $f_{\text{nonfid}}$, has been determined from signal samples with full detector simulation for each of the signal production modes studied. The value of $f_{\text{nonfid}}$ for different signal production modes is shown in Table~\ref{tab:summarySM}.

\begin{table}[!htb]
\centering
\topcaption{
Summary of the fraction of signal events for different SM signal production modes within the fiducial phase space (acceptance $\mathcal{A}_\text{fid}$), reconstruction efficiency ($\epsilon$) for signal events from within the fiducial phase space, and ratio of reconstructed events which are from outside the fiducial phase space to reconstructed events
which are from within the fiducial phase space ($f_\text{nonfid}$).
For all production modes the values given are for $m_\PH=125\GeV$.
Also shown in the last column is the factor $(1+f_{\text{nonfid}})\epsilon$
which regulates the signal yield for a given fiducial cross section, as shown in Eq.~\ref{eqn:m4l}.
The uncertainties listed are statistical only. The theoretical uncertainty in $\mathcal{A}_\text{fid}$ for the SM is less than $1\%$.
\label{tab:summarySM}
}
\begin{tabular}{lcccc} \hline
Signal process & $\mathcal{A}_\text{fid}$ & $\epsilon$ & $f_\text{nonfid}$  & $(1+f_\text{nonfid})\epsilon$ \\
\hline
gg$\to$H (\POWHEG)& 0.398 $\pm$ 0.001 & 0.592 $\pm$ 0.001 & 0.049 $\pm$ 0.001 & 0.621 $\pm$ 0.001 \\
VBF (\POWHEG)& 0.445 $\pm$ 0.001 & 0.601 $\pm$ 0.002 & 0.038 $\pm$ 0.001 & 0.624 $\pm$ 0.002 \\
WH (\POWHEG \textsc{minlo}) & 0.314 $\pm$ 0.001 & 0.577 $\pm$ 0.002 & 0.068 $\pm$ 0.001 & 0.616 $\pm$ 0.002 \\
ZH (\POWHEG \textsc{minlo}) & 0.342 $\pm$ 0.002 & 0.592 $\pm$ 0.003 & 0.071 $\pm$ 0.002 & 0.634 $\pm$ 0.003 \\
ttH (\POWHEG)& 0.311 $\pm$ 0.002 & 0.572 $\pm$ 0.003 & 0.136 $\pm$ 0.003 & 0.650 $\pm$ 0.004 \\
\hline
\end{tabular}
\end{table}

The results are compared to the predictions obtained from \POWHEG and \textsc{nnlops}~\cite{NNLOPS} which have NLO and
NNLO accuracy in pQCD for inclusive distributions, respectively. In both cases the total gluon fusion
cross section is taken from Ref.~\cite{Anastasiou2016}.

\begin{figure}[!htb]
\centering
\includegraphics[width=0.45\linewidth]{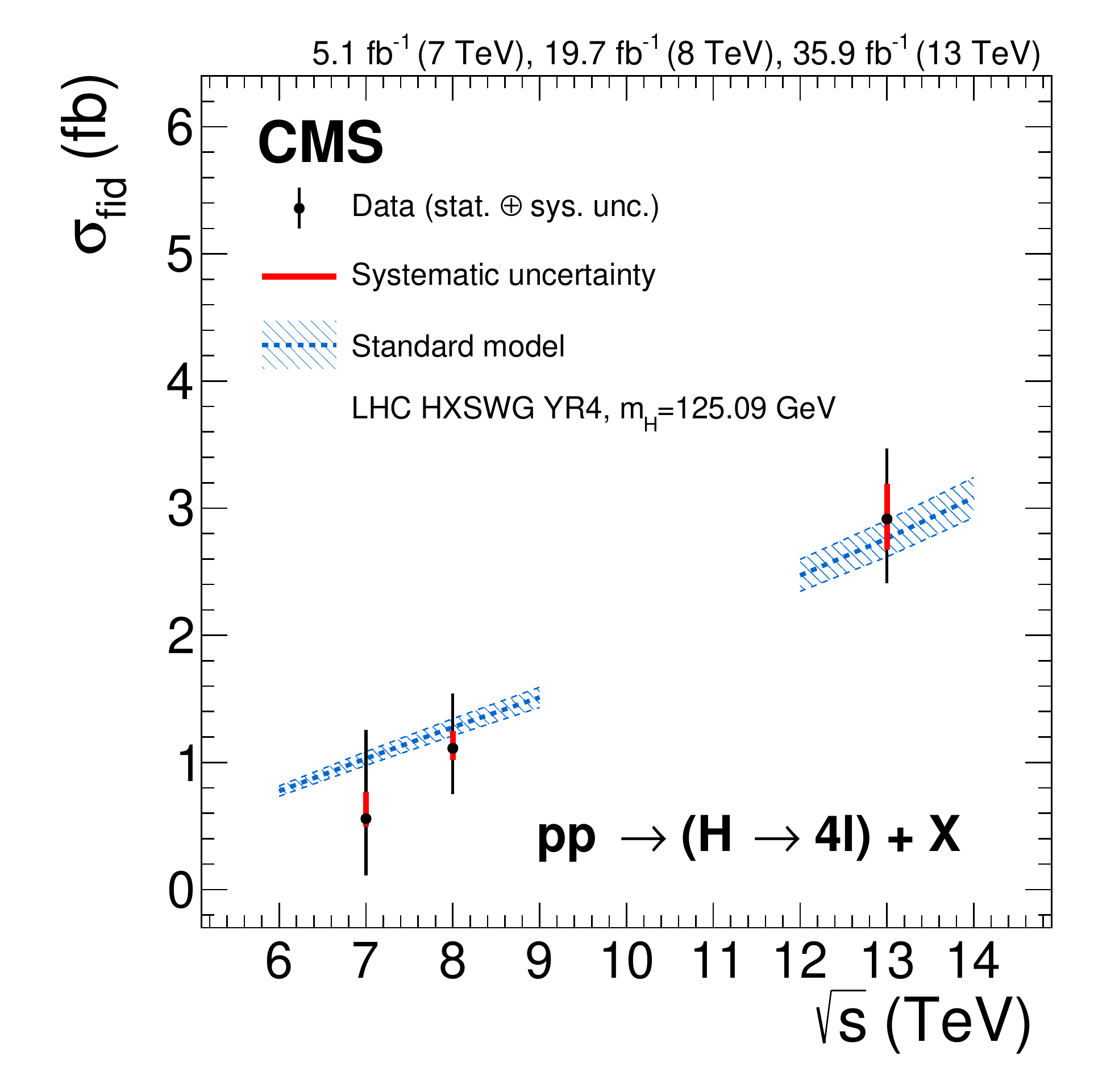}
\includegraphics[width=0.45\linewidth]{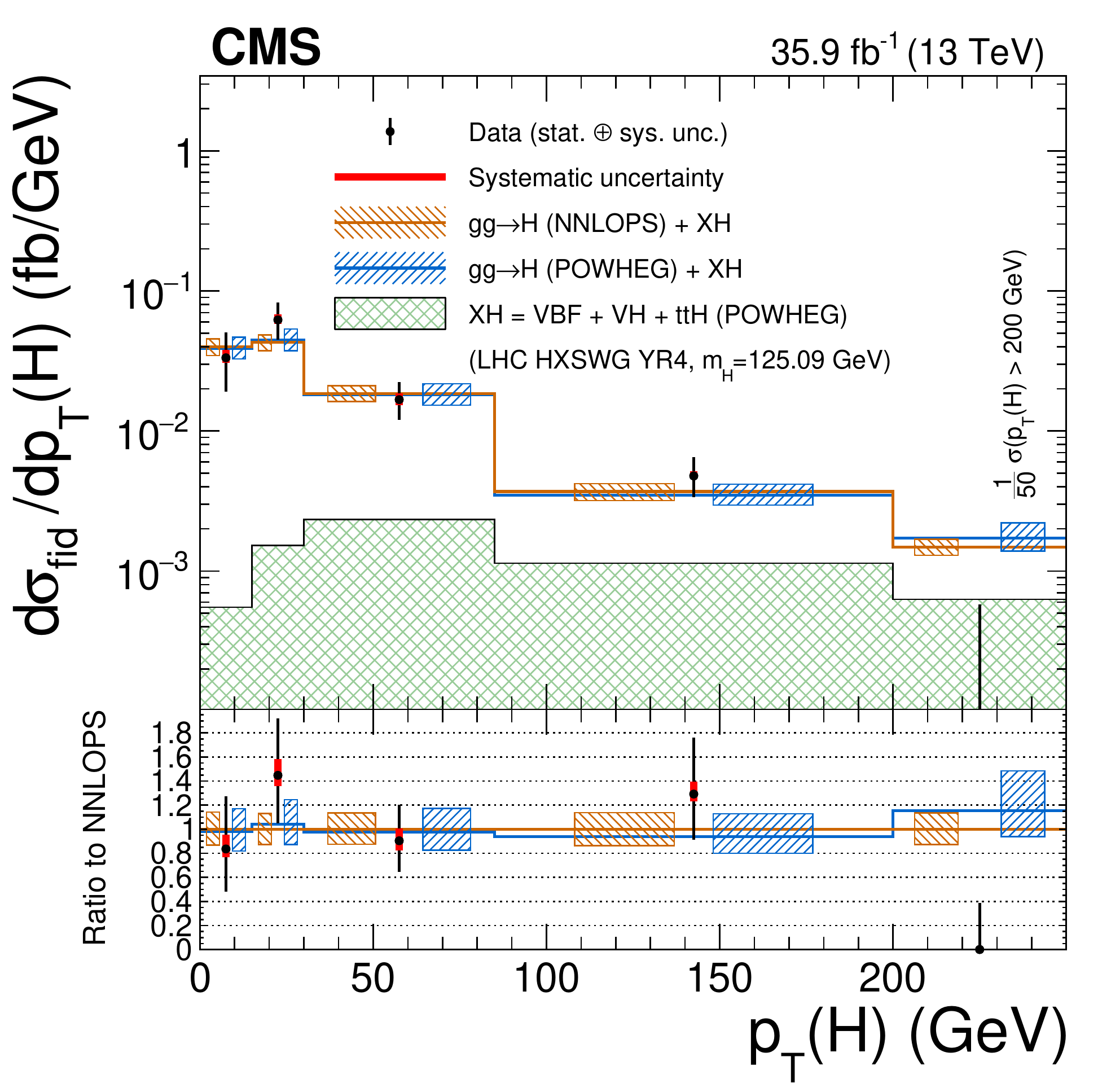} \\
\includegraphics[width=0.45\linewidth]{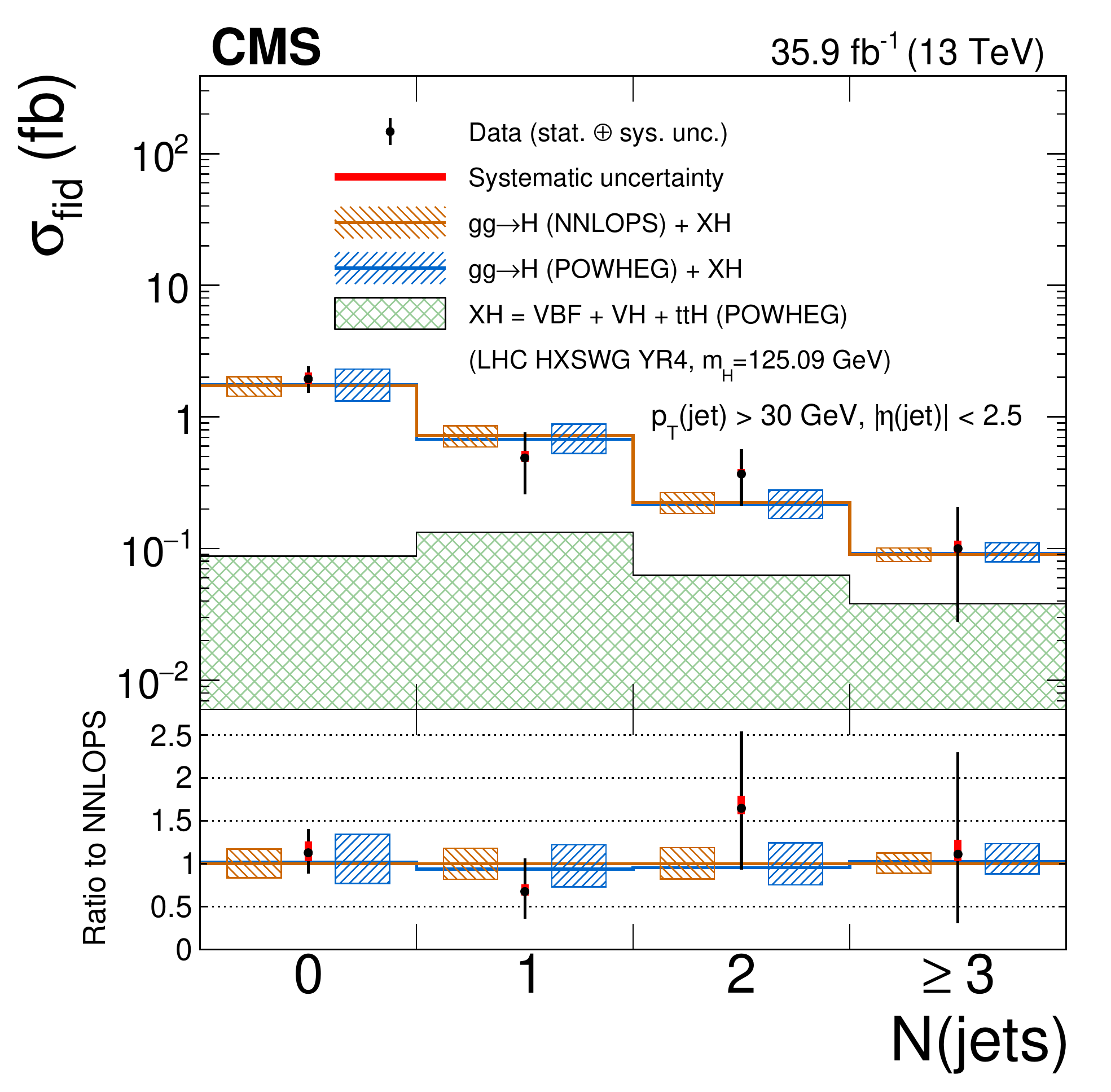}
\includegraphics[width=0.45\linewidth]{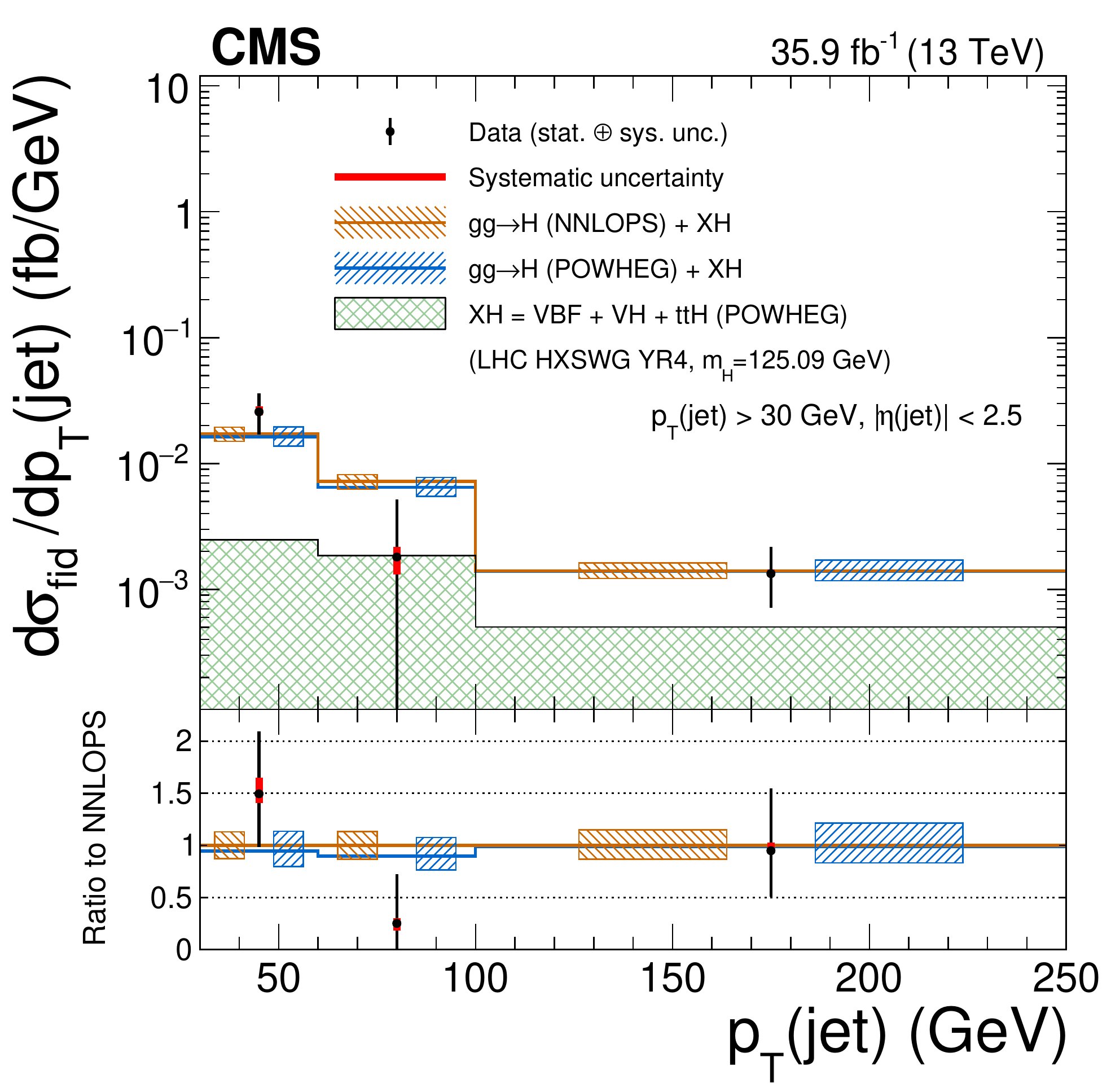}
\caption{The measured fiducial cross section as a function of $\sqrt{s}$ (top left).
The acceptance is calculated using \textsc{nnlops} at  $\sqrt{s} = 13$\TeV and {\sc hres}~\cite{Grazzini:2013mca,deFlorian:2012mx}
at  $\sqrt{s} = 7$ and 8\TeV and the total cross sections and uncertainties are taken from Ref.~\cite{YR4}. The fiducial volume for $\sqrt{s} = 7$ and 8\TeV uses the lepton isolation definition from
Ref.~\cite{CMSH4lFiducial8TeV}, while for $\sqrt{s} = 13$\TeV the definition described in the text is used.
The results of the differential cross section measurements are shown for $\pt(\PH)$ (top right), N(jets) (bottom left) and $\pt(\text{jet})$ of the leading associated jet (bottom right).
The acceptance and theoretical uncertainties in the differential bins are calculated using \POWHEG and {\sc nnlops}.
The subdominant component of the signal ($\mathrm{VBF} + \mathrm{V}\PH +\ttH$) is denoted as XH. In the differential cross section measurement for $\pt(\PH)$,
the last bin represents the integrated cross section for  $\pt(\PH) >200$\GeV and is scaled by 1/50 for presentation purposes.
No events are observed with $\pt(\PH) >200$\GeV.
\label{fig:fiducialresult}}
\end{figure}

The integrated fiducial cross section is measured to be
$\sigma_{\text{fid}}=2.92~^{+0.48}_{-0.44}\stat~^{+0.28}_{-0.24}\syst\unit{fb}$.
This can be compared to the SM expectation obtained from \textsc{nnlops} of $\sigma_{\text{fid}}^\mathrm{SM}=2.76\pm0.14\unit{fb}$. The integrated fiducial
cross section as a function of $\sqrt{s}$ is also shown in Fig.~\ref{fig:fiducialresult}.
The compatibility of the integrated fiducial cross sections measured in the $4\Pe$, $4\mu$, and $2\Pe2\mu$ final states with the SM prediction is estimated
 using a likelihood ratio with the three cross sections at their best fit values in the numerator and the three cross sections fixed to the SM predictions in
the denominator. The compatibility, defined as the asymptotic $p$-value of the fit, is found to be 88\%.
The measured differential cross
section results for $\pt(\PH)$, N(jets), and $\pt(\text{jet})$ of the leading associated jet can also be seen in Fig.~\ref{fig:fiducialresult}.
The dominant sources of systematic uncertainty are the experimental uncertainties in the lepton identification efficiencies and
integrated luminosity measurement, and the theoretical sources of uncertainty are found to be subdominant.
To estimate the model dependence of the measurement, the unfolding procedure is repeated
using different response matrices created by varying the relative fraction of each SM production mode
within its experimental constraints. The uncertainty is determined
to be negligible with respect to the experimental systematic uncertainties.

\clearpage
\subsection{Higgs boson mass measurement}

In this section we show the results of the measurement of the mass of
the resonance, using additional information in the likelihood fit with respect to the
signal strength and cross section measurements.

To improve the four-lepton invariant mass resolution, a kinematic fit is performed
using a mass constraint on the intermediate Z resonance. Previous studies~\cite{CMSH4lLegacy} of the Higgs boson mass
show that the selected $\cPZ_1$ has a significant on-shell component, while the invariant mass
distribution for the selected $\cPZ_2$ is wider than the detector resolution.
Therefore only the $\cPZ_1$ candidate is considered when performing the kinematic constraint.

The likelihood to be maximized is constructed as follows:
\begin{equation}
\mathcal{L}(\hat{p}_\mathrm{T}^1,\hat{p}_\mathrm{T}^2|\pt^{1}, \sigma_{\pt^1}, \pt^{2}, \sigma_{\pt^2}) =\mathrm{Gauss}(\pt^{1}| \hat{p}_\mathrm{T}^1, \sigma_{\pt^1}) \, \mathrm{Gauss}(\pt^{2}| \hat{p}_\mathrm{T}^2, \sigma_{\pt^2}) \,  \mathcal{L}( m_{12} | m_{\Z}, m_\PH),
\end{equation}
where $\pt^{1}$ and $a^{2}$ are the reconstructed transverse momenta of the two leptons forming
the $\cPZ_1$ candidate, $\sigma_{\pt^1}$ and $\sigma_{\pt^2}$ are the corresponding per-lepton resolutions,
$\hat{p}_\mathrm{T}^1$ and $\hat{p}_\mathrm{T}^2$ are the refitted transverse momenta, and $m_{12}$ is the invariant mass calculated
from the refitted four-momenta. The term $\mathcal{L}( m_{12} | m_{\Z}, m_\PH)$ is the mass constraint term.
For a Higgs boson mass near 125 \GeV, the selected $\cPZ_1$ is not always on-shell, so a Breit--Wigner shape does not
perfectly describe the $\cPZ_1$ shape at the generator level. We therefore choose $\mathcal{L}( m_{12} | m_{\Z}, m_\PH)$ to be
the $m(\cPZ_1)$ shape at the generator level from the SM Higgs boson sample with $m_\PH=125\GeV$, where the same
algorithm for selecting the $\cPZ_1$ and $\cPZ_2$ candidates, as described in Section~\ref{sec:objects}, is used.
For each event, the likelihood is maximized and the refitted transverse momenta are used to recalculate the
four-lepton mass and mass uncertainty, which are denoted as $m_{4\ell}'$ and  $\MassDprime$, respectively.
These distributions are then used to build the likelihood used to extract the Higgs boson mass.

\begin{figure}[!htb]
  \centering
    \includegraphics[width=0.46\linewidth]{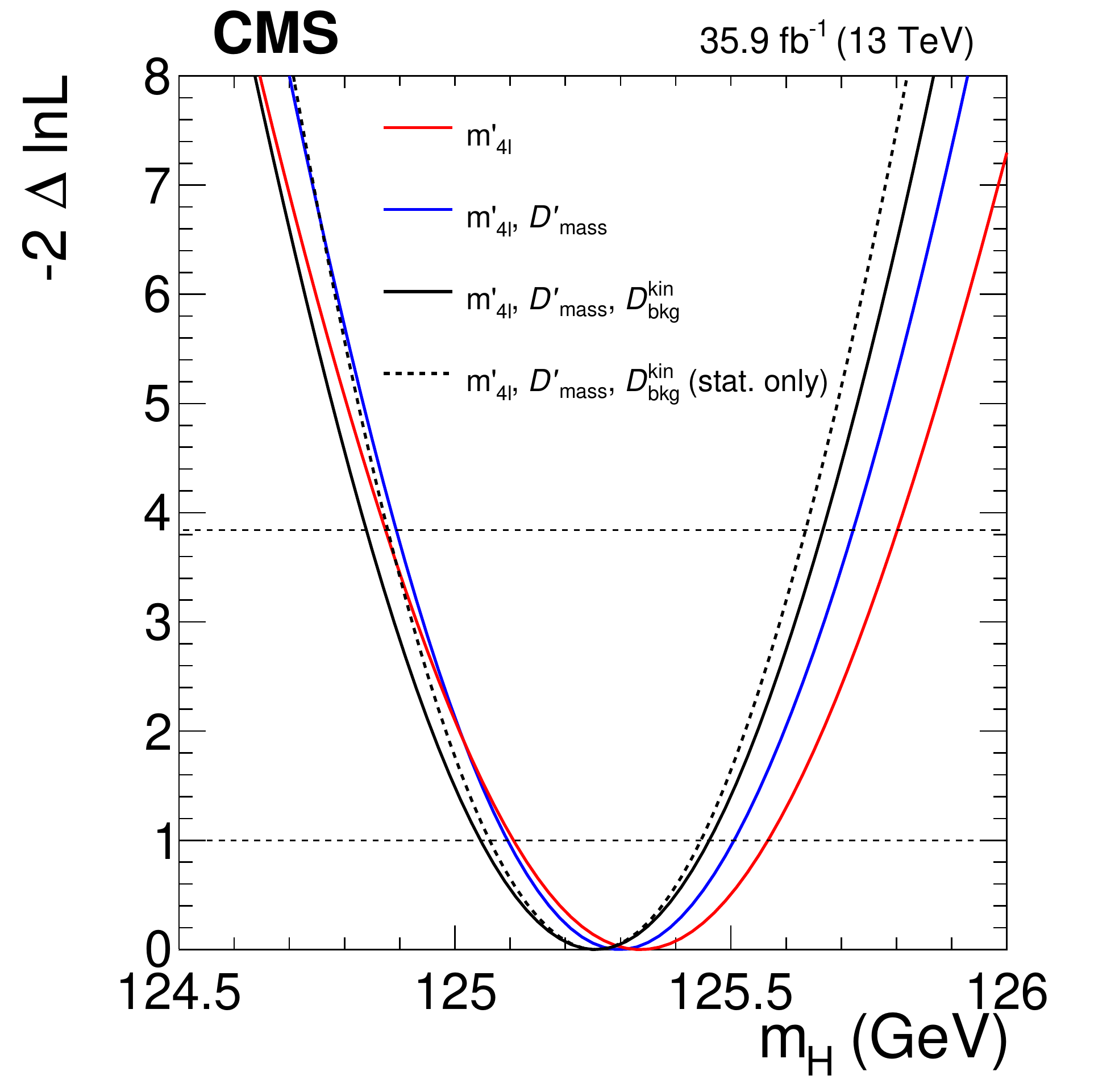}
    \includegraphics[width=0.46\linewidth]{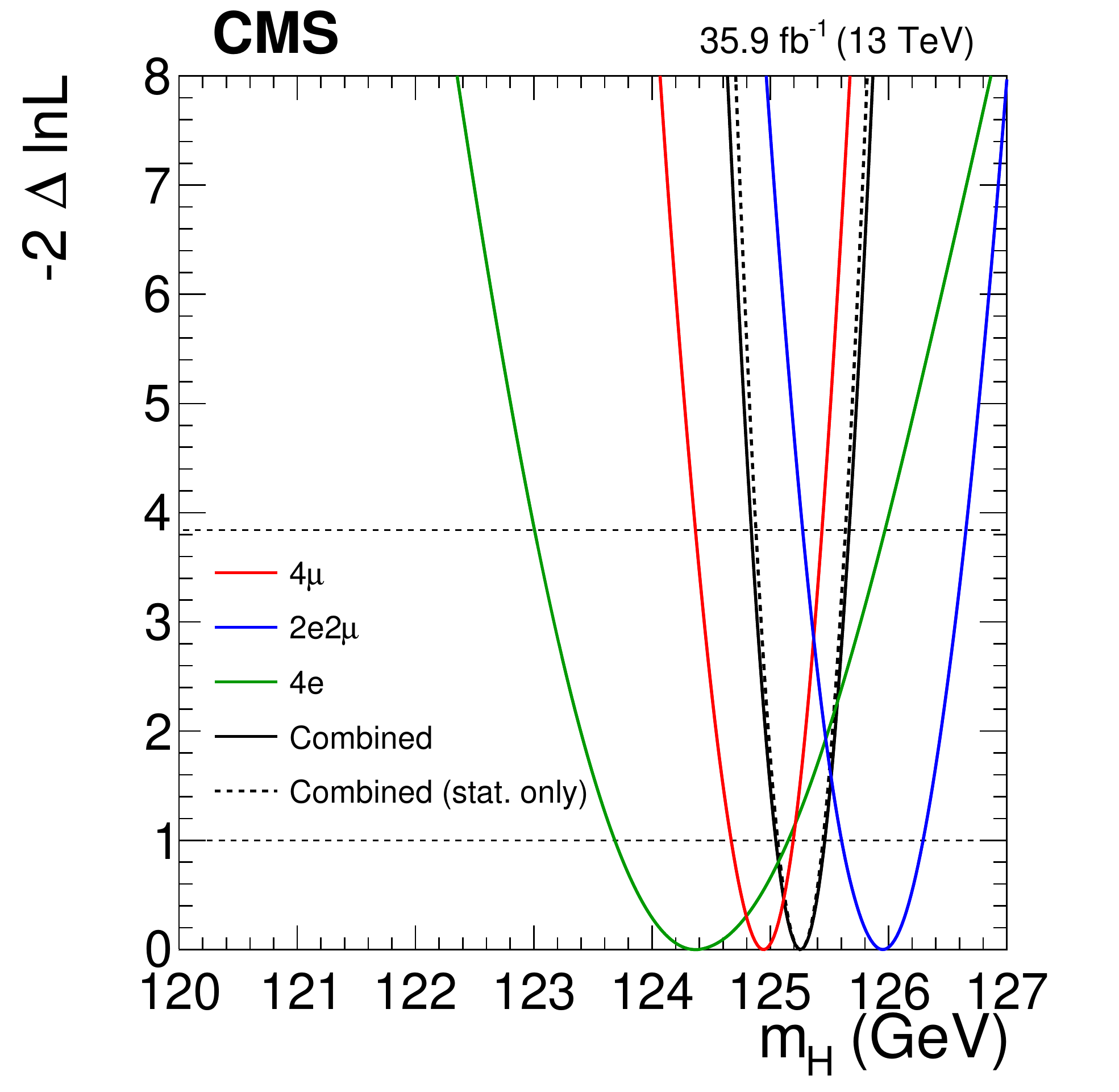}
    \caption{
      Left: 1D likelihood scans as a function of the Higgs boson mass for the 1D, 2D, and 3D measurement.
      Right: 1D likelihood scans as a function of mass for the
      different final states and the combination of all final states for the 3D mass measurement.
      The likelihood scans are shown for the mass measurement using the refitted mass distribution with the $m(\cPZ_1)$ constraint.
      Solid lines represent scans with all uncertainties included, dashed lines those with only statistical uncertainties.
      \label{fig:LL_mass_bydim_comb}}

\end{figure}

The 1D likelihood scans vs. $\mH$, while profiling the signal strength modifier
$\mu$ along with all other nuisance parameters for the 1D $\mathcal{L}(m_{4\ell}')$, 2D $\mathcal{L}(m_{4\ell}',\MassDprime)$,
and 3D $\mathcal{L}(m_{4\ell}',\MassDprime,\KD)$ fits, including the $m(\cPZ_1)$ constraint,
are shown in Fig.~\ref{fig:LL_mass_bydim_comb}. All systematic uncertainties described in Section~\ref{sec:systematics} are
included. When estimating separately the systematic and statistical uncertainties, the signal strength is profiled in the
likelihood scan with the systematic uncertainties removed, so that its uncertainty is included in the statistical uncertainty.
As in the measurement of the signal strengths, the relative fraction of $4\Pe$, $4\mu$, and $2\Pe2\mu$ signal events is fixed to the SM prediction.
If the relative fractions are allowed to float, the change in the fitted mass value is much smaller than the uncertainty.

The best fit masses and the expected increase in the uncertainty relative to
the 3D fit with the $m(\cPZ_1)$ constraint for each of the six fits are shown in Table \ref{tab:mass}.
The nominal result for the mass measurement is obtained from the 3D fit with the $m(\cPZ_1)$ constraint,
for which the fitted value of $m_\PH$ in the three subchannels
is $m_{\PH}^{4\mu}=124.94\pm0.25\stat\pm0.08\syst\GeV$,
$m_{\PH}^{4\Pe}=124.37\pm0.62\stat\pm0.38\syst\GeV$,
and $m_{\PH}^{2\Pe2\mu}=125.95\pm0.32\stat\pm0.14\syst\GeV$
leading to a combined value $m_{\PH} = \valMassThreeDRefit\GeV$.
The systematic uncertainty in the mass measurement
is completely dominated by the uncertainty in the lepton momentum scale. The expected uncertainty in the mass measurement using the
3D fit with the $m(\cPZ_1)$ constraint is evaluated with two Asimov data sets.
The ``prefit'' expected uncertainty is $\pm0.24\stat\pm0.09\syst\GeV$.  Here $m_\PH=125\GeV$, $\mu=1$, and all nuisance parameters
are fixed to their nominal values.
The ``postfit'' expected uncertainty with $m_\PH$, $\mu$, and all nuisance parameters
fixed to their best-fit estimates from the data is $\pm0.23\stat\pm0.08\syst\GeV$.
The probability of the ``prefit'' uncertainty being less than or equal to the observed value is determined from an ensemble of
pseudo-experiments to be about 18\%.
The mutual compatibility of the $m_\PH$ results from the three individual channels is tested using a likelihood ratio with three masses
in the numerator and a common mass in the denominator, and thus two degrees of freedom. The signal strength is profiled in both the
numerator and denominator. The resulting compatibility, defined as the asymptotic $p$-value of the fit, is 2.5\%.
The tension between the three individual channels is driven by the difference between the $4\mu$ and $2\Pe2\mu$ channels, where
the compatibility of the 1D mass measurements without the $m(\cPZ_1)$ constraint is 8\%.
In the 1D mass measurement the main potential source of systematic bias is the lepton momentum scale; this possibility is disfavored
by the fact that the measured mass in the $2\Pe2\mu$ channel is not in between the measurements
in the $4\Pe$ and $4\mu$ channels. This bias has also been checked by performing the 1D mass measurements without
the $m(\cPZ_1)$ constraint using $\cPZ\to4\ell$ events, and the resulting mass
is measured to be $m_{\cPZ}^{4\mu}=90.85\pm0.27\stat\pm0.04\syst\GeV$,
$m_{\cPZ}^{4\Pe}=90.85\pm0.74\stat\pm0.28\syst\GeV$,
and $m_{\cPZ}^{2\Pe2\mu}=90.61\pm0.48\stat\pm0.10\syst\GeV$
leading to a combined value $m_{\cPZ}=90.84\pm0.23\stat\pm0.07\syst\GeV$. The compatibility with the nominal Z-boson mass from Ref.~\cite{Zmass}
is 14\% and the mutual compatibility between the three individual channels is 90\%.
The modelling of the event-by-event mass uncertainties is a possible source of systematic bias in the 2D and 3D measurements.
It is checked by performing a Kolmogorov-Smirnov compatibility test of
the expected and observed distributions in
an expanded $\mllll$ range yielding $p$-values of 10\% for the $2\Pe2\mu$ channel, 55\% for the $4\Pe$ channel, and $94\%$ for the $4\mu$ channel.

{\renewcommand{\arraystretch}{1.2}
\begin{table}[!hbt]
\centering
\topcaption{ Best fit values for the mass of the Higgs boson
  measured in the $4\ell$ final
  states, with the 1D, 2D, and 3D fit, respectively, as described in the
  text. All mass values are given in \GeV.  The uncertainties include both the statistical
  and systematic components. The expected $\mH$ uncertainty change shows the change in
  the expected precision on the measurement for the different fit scenarios, relative to 3D $\mathcal{L}(m'_{4\ell},\MassDprime,\KD)$.
\label{tab:mass}}
\begin{tabular}{lccc}
\hline
No $m(\cPZ_{1})$ constraint     &   3D: $\mathcal{L}(m_{4\ell},\MassD,\KD)$ & 2D: $\mathcal{L}(m_{4\ell},\MassD)$    &   1D: $\mathcal{L}(m_{4\ell})$   \\
\hline
Expected $\mH$ uncertainty change  &   $+8.1\%$               & $+11\%$         &  $+21\%$  \\
Observed $\mH$ (\GeVns{})      &   125.28$\pm0.22$   & 125.36$\pm0.24$  &  125.39$\pm0.25$  \\
\hline
With $m(\cPZ_{1})$ constraint   &    3D: $\mathcal{L}(m_{4\ell}',\MassDprime,\KD)$  & 2D: $\mathcal{L}(m_{4\ell}',\MassDprime)$ & 1D: $\mathcal{L}(m_{4\ell}')$   \\
\hline
Expected $\mH$ uncertainty change   &   ---         & $+3.2\%$         &  $+$11\%   \\
Observed $\mH$ (\GeVns{})        &   125.26$\pm0.21$  & 125.30$\pm0.21$  &  125.34$\pm0.23$   \\
\hline
\end{tabular}
\end{table}
}

\clearpage

\subsection{Measurement of the Higgs boson width using on-shell production}
\label{sec:widthresult}

In this section, we describe a model-independent measurement of the width performed using the $m_{4\ell}$ distribution
in the range $105 < m_{4\ell} < 140$\GeV. This measurement is limited by the four-lepton invariant mass
resolution and is therefore sensitive to a width of about 1\GeV.
Therefore, we take into account the interference between the signal and background production of the $4\ell$ final state in this analysis.

An unbinned maximum likelihood fit to the $m_{4\ell}$ distribution is performed.
The strengths of fermion and vector boson induced couplings are independent and are left unconstrained
in the fit.
By splitting events into two categories, namely those with a VBF-like two-jet topology and the rest,
it is possible to constrain the two sets of couplings.
The general parameterization of the probability density function is described in Section~\ref{sec:signal}.

\begin{figure}[!htb]
\centering
\includegraphics[width=0.52\linewidth]{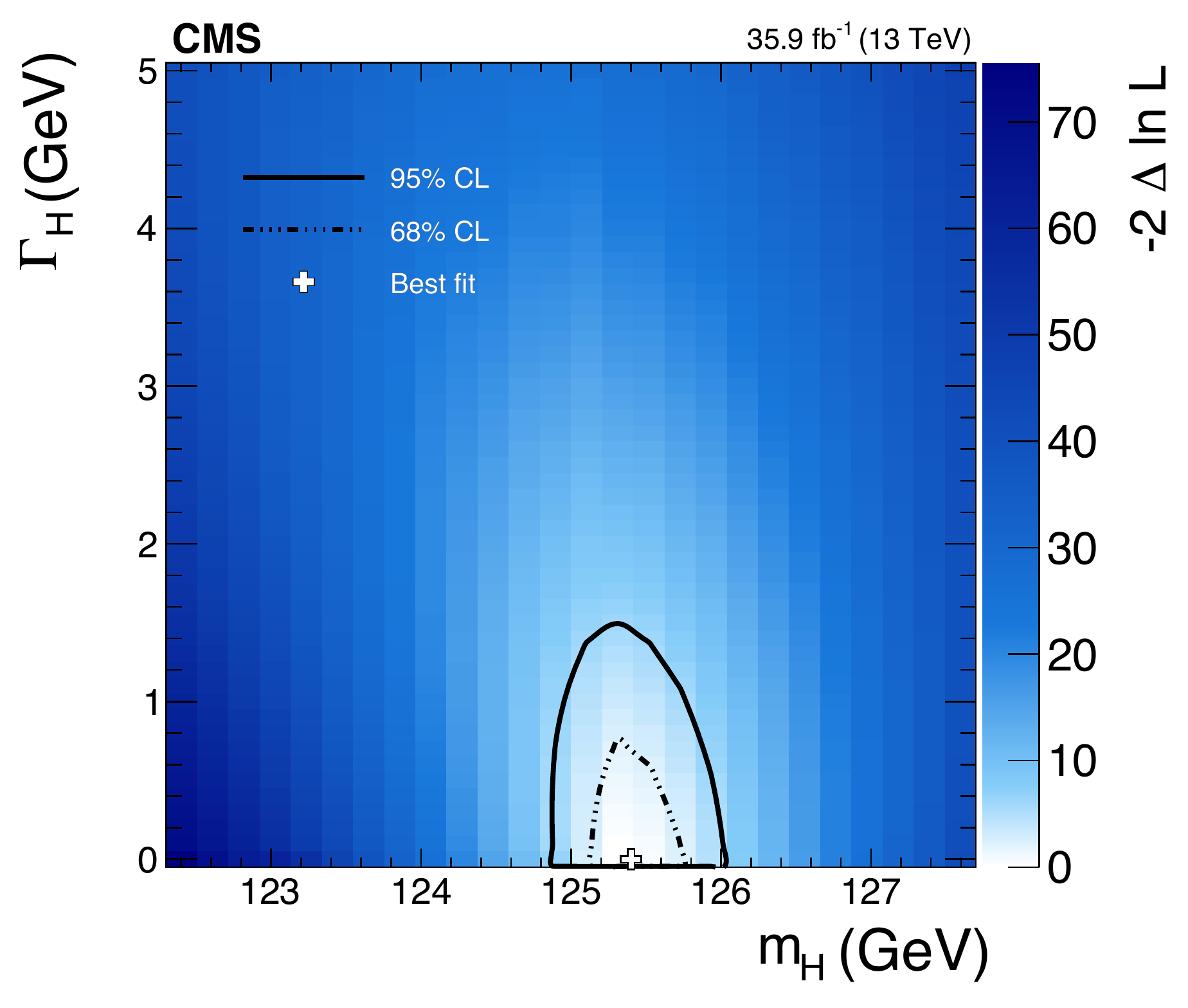}
\includegraphics[width=0.43\linewidth]{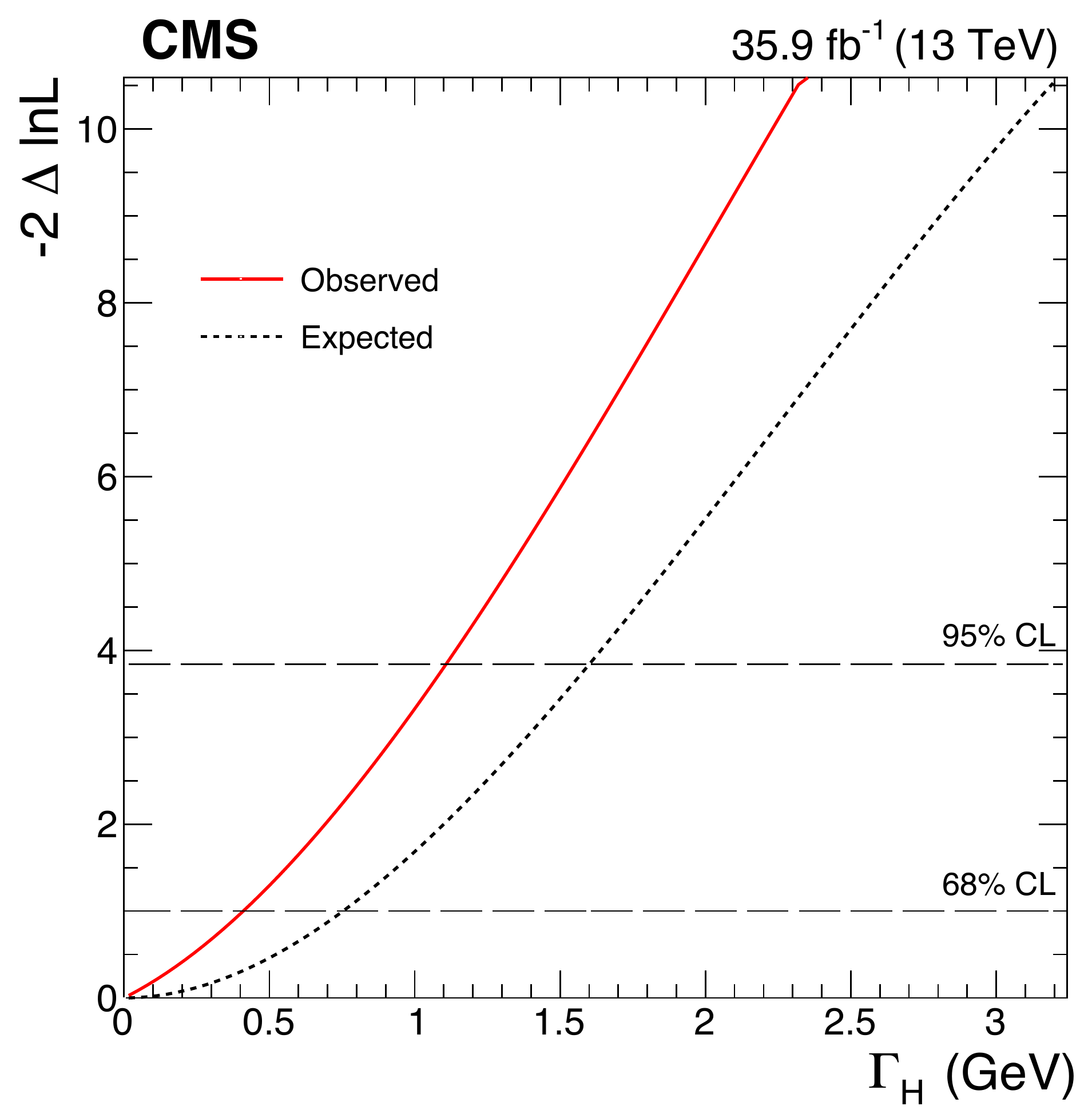}
\caption{
	(Left) Observed likelihood scan of $m_\PH$ and $\Gamma_\PH$ using the signal range $105 < m_{4\ell} < 140$\GeV.
	(Right) Observed and expected likelihood scan of $\Gamma_\PH$ using the signal range $105 < m_{4\ell} < 140$\GeV, with $m_\PH$ profiled.
    \label{fig:WidthScans}
}
\end{figure}

The joint constraint on the width $\Gamma_\PH$ and mass $m_\PH$ of the Higgs boson is shown
in Fig.~\ref{fig:WidthScans} (left).
Figure~\ref{fig:WidthScans} (right) shows the likelihood as a function of $\Gamma_\PH$ with the $m_\PH$ parameter unconstrained.
The width is constrained to be $\Gamma_{\PH}<1.10\GeV$ at 95\% CL.
The observed and expected results are summarized in Table~\ref{tab:summary_width} and are consistent with the expected detector resolution.
The dominant sources of uncertainty are the uncertainty in the lepton momentum scale when determining the mass and the uncertainty in the four-lepton mass resolution when determining the width.

\begin{table*}[!htb]
\centering
\topcaption{
Summary of allowed 68\%~CL (central values with uncertainties) and 95\%~CL (ranges in square brackets)
intervals on the width $\Gamma_\PH$  of the Higgs boson.
The expected results are quoted for the SM signal production cross section ($\mu_{\mathrm{VBF},\mathrm{V}\PH}=\mu_{\Pg\Pg\PH,\ttbar\PH}=1$)
and the values of $m_\PH=125$\GeV.
In the observed results $\mu_{\mathrm{VBF},\mathrm{V}\PH}$ and $\mu_{\Pg\Pg\PH,\ttbar\PH}$ are left unconstrained in the fit.
}
\renewcommand{\arraystretch}{1.25}
\begin{tabular}{cccccc}
\hline
Parameter &     $m_{4\ell}$ range                 &  \multicolumn{2}{c}{Expected} &  \multicolumn{2}{c}{Observed}   \\
\hline
$\Gamma_\PH$ (\GeVns{})  & $[105, 140]$      & \multicolumn{2}{c}{$0.00^{+0.75}_{-0.00}$ $ [0.00,1.60]$} & \multicolumn{2}{c}{$0.00^{+0.41}_{-0.00}$ $ [0.00,1.10]$}   \\
\hline
\end{tabular}
\label{tab:summary_width}
\end{table*}

\clearpage
\section{Summary}
\label{sec:summary}

The first results on Higgs boson production in the four-lepton final state at $\sqrt{s}=13\TeV$ have been presented,
using $\usedLumi$ of pp collisions collected by the CMS experiment at the LHC.
The signal strength modifier $\mu$, defined as the ratio of the observed Higgs boson rate in the $\PH\to\Z\Z\to4\ell$
decay channel to the standard model expectation, is measured to be
$\mu = 1.05~^{+0.15}_{-0.14}\stat~^{+0.11}_{-0.09}\syst = \valMuAtRunIMass$
at $m_{\PH}=125.09\GeV$, the combined ATLAS and CMS measurement of the Higgs boson mass.
Two signal strength modifiers associated with the fermion- and vector-boson induced contributions to the expected standard model cross section
are measured to be $\muF=\valMuFAtRunIMass$ and $\muV=\valMuVAtRunIMass$, respectively.
The cross section at $\sqrt{s}=13\TeV$ in a fiducial phase space defined to match the
experimental acceptance in terms of the lepton kinematics and event topology, predicted in the standard model to be $2.76\pm0.14\unit{fb}$,
is measured to be $2.92~^{+0.48}_{-0.44}\stat~^{+0.28}_{-0.24}\syst\unit{fb}$. Differential cross sections are reported as a
function of the transverse momentum of the Higgs boson, the number of associated jets, and the transverse momentum of the leading associated jet.
The mass is measured to be $m_{\PH}=\valMassThreeDRefit~$GeV and the width is constrained to
be $\Gamma_{\PH}<1.10\GeV$ at 95\% confidence level.
The production and decay properties of the Higgs boson are consistent, within their uncertainties, with the expectations for the standard model Higgs boson.

\begin{acknowledgments}
\hyphenation{Bundes-ministerium Forschungs-gemeinschaft Forschungs-zentren Rachada-pisek} We congratulate our colleagues in the CERN accelerator departments for the excellent performance of the LHC and thank the technical and administrative staffs at CERN and at other CMS institutes for their contributions to the success of the CMS effort. In addition, we gratefully acknowledge the computing centers and personnel of the Worldwide LHC Computing Grid for delivering so effectively the computing infrastructure essential to our analyses. Finally, we acknowledge the enduring support for the construction and operation of the LHC and the CMS detector provided by the following funding agencies: the Austrian Federal Ministry of Science, Research and Economy and the Austrian Science Fund; the Belgian Fonds de la Recherche Scientifique, and Fonds voor Wetenschappelijk Onderzoek; the Brazilian Funding Agencies (CNPq, CAPES, FAPERJ, and FAPESP); the Bulgarian Ministry of Education and Science; CERN; the Chinese Academy of Sciences, Ministry of Science and Technology, and National Natural Science Foundation of China; the Colombian Funding Agency (COLCIENCIAS); the Croatian Ministry of Science, Education and Sport, and the Croatian Science Foundation; the Research Promotion Foundation, Cyprus; the Secretariat for Higher Education, Science, Technology and Innovation, Ecuador; the Ministry of Education and Research, Estonian Research Council via IUT23-4 and IUT23-6 and European Regional Development Fund, Estonia; the Academy of Finland, Finnish Ministry of Education and Culture, and Helsinki Institute of Physics; the Institut National de Physique Nucl\'eaire et de Physique des Particules~/~CNRS, and Commissariat \`a l'\'Energie Atomique et aux \'Energies Alternatives~/~CEA, France; the Bundesministerium f\"ur Bildung und Forschung, Deutsche Forschungsgemeinschaft, and Helmholtz-Gemeinschaft Deutscher Forschungszentren, Germany; the General Secretariat for Research and Technology, Greece; the National Scientific Research Foundation, and National Innovation Office, Hungary; the Department of Atomic Energy and the Department of Science and Technology, India; the Institute for Studies in Theoretical Physics and Mathematics, Iran; the Science Foundation, Ireland; the Istituto Nazionale di Fisica Nucleare, Italy; the Ministry of Science, ICT and Future Planning, and National Research Foundation (NRF), Republic of Korea; the Lithuanian Academy of Sciences; the Ministry of Education, and University of Malaya (Malaysia); the Mexican Funding Agencies (BUAP, CINVESTAV, CONACYT, LNS, SEP, and UASLP-FAI); the Ministry of Business, Innovation and Employment, New Zealand; the Pakistan Atomic Energy Commission; the Ministry of Science and Higher Education and the National Science Centre, Poland; the Funda\c{c}\~ao para a Ci\^encia e a Tecnologia, Portugal; JINR, Dubna; the Ministry of Education and Science of the Russian Federation, the Federal Agency of Atomic Energy of the Russian Federation, Russian Academy of Sciences, the Russian Foundation for Basic Research and the Russian Competitiveness Program of NRNU ``MEPhI"; the Ministry of Education, Science and Technological Development of Serbia; the Secretar\'{\i}a de Estado de Investigaci\'on, Desarrollo e Innovaci\'on, Programa Consolider-Ingenio 2010, Plan de Ciencia, Tecnolog\'{i}a e Innovaci\'on 2013-2017 del Principado de Asturias and Fondo Europeo de Desarrollo Regional, Spain; the Swiss Funding Agencies (ETH Board, ETH Zurich, PSI, SNF, UniZH, Canton Zurich, and SER); the Ministry of Science and Technology, Taipei; the Thailand Center of Excellence in Physics, the Institute for the Promotion of Teaching Science and Technology of Thailand, Special Task Force for Activating Research and the National Science and Technology Development Agency of Thailand; the Scientific and Technical Research Council of Turkey, and Turkish Atomic Energy Authority; the National Academy of Sciences of Ukraine, and State Fund for Fundamental Researches, Ukraine; the Science and Technology Facilities Council, UK; the US Department of Energy, and the US National Science Foundation.

{\tolerance=800
Individuals have received support from the Marie-Curie program and the European Research Council and Horizon 2020 Grant, contract No. 675440 (European Union); the Leventis Foundation; the A. P. Sloan Foundation; the Alexander von Humboldt Foundation; the Belgian Federal Science Policy Office; the Fonds pour la Formation \`a la Recherche dans l'Industrie et dans l'Agriculture (FRIA-Belgium); the Agentschap voor Innovatie door Wetenschap en Technologie (IWT-Belgium); the Ministry of Education, Youth and Sports (MEYS) of the Czech Republic; the Council of Scientific and Industrial Research, India; the HOMING PLUS program of the Foundation for Polish Science, cofinanced from European Union, Regional Development Fund, the Mobility Plus program of the Ministry of Science and Higher Education, the National Science Center (Poland), contracts Harmonia 2014/14/M/ST2/00428, Opus 2014/13/B/ST2/02543, 2014/15/B/ST2/03998, and 2015/19/B/ST2/02861, Sonata-bis 2012/07/E/ST2/01406; the National Priorities Research Program by Qatar National Research Fund; the Programa Clar\'in-COFUND del Principado de Asturias; the Thalis and Aristeia programs cofinanced by EU-ESF and the Greek NSRF; the Rachadapisek Sompot Fund for Postdoctoral Fellowship, Chulalongkorn University and the Chulalongkorn Academic into Its 2nd Century Project Advancement Project (Thailand); and the Welch Foundation, contract C-1845.
\par}
\end{acknowledgments}

\bibliography{auto_generated}

\providecommand{\href}[2]{#2}\begingroup\raggedright\begin{thebibliography}{10}%
\makeatletter
\providecommand{\hrefCMSnoop }[0]{\@secondoftwo}%
\makeatother
\providecommand{\doi}{\texttt{doi:}\begingroup \urlstyle{tt}\Url}

\bibitem{Aad:2012tfa}
\hrefCMSnoop {}{{ATLAS Collaboration}, ``{Observation of a new particle in the
  search for the Standard Model Higgs boson with the ATLAS detector at the
  LHC}'',} \textit{ Phys. Lett. B} \textbf{ 716} (2012) 1,
  \href{http://dx.doi.org/10.1016/j.physletb.2012.08.020}{\doi{10.1016/j.physletb.2012.08.020}},
\href{http://www.arXiv.org/abs/1207.7214}{\texttt{arXiv:1207.7214}}.

\bibitem{Chatrchyan:2012ufa}
\hrefCMSnoop {}{{CMS Collaboration}, ``{Observation of a new boson at a mass of
  125 GeV with the CMS experiment at the LHC}'',} \textit{ Phys. Lett. B}
  \textbf{ 716} (2012) 30,
  \href{http://dx.doi.org/10.1016/j.physletb.2012.08.021}{\doi{10.1016/j.physletb.2012.08.021}},
\href{http://www.arXiv.org/abs/1207.7235}{\texttt{arXiv:1207.7235}}.

\bibitem{Chatrchyan:2013lba}
\hrefCMSnoop {}{{CMS Collaboration}, ``{Observation of a new boson with mass
  near 125 GeV in pp collisions at $\sqrt{s}$ = 7 and 8 TeV}'',} \textit{ JHEP}
  \textbf{ 06} (2013) 081,
  \href{http://dx.doi.org/10.1007/JHEP06(2013)081}{\doi{10.1007/JHEP06(2013)081}},
\href{http://www.arXiv.org/abs/1303.4571}{\texttt{arXiv:1303.4571}}.

\bibitem{CMS:2014ega}
\hrefCMSnoop {}{{CMS Collaboration}, ``{Precise determination of the mass of
  the Higgs boson and tests of compatibility of its couplings with the standard
  model predictions using proton collisions at 7 and 8 $\,\text {TeV}$}'',}
  \textit{ Eur. Phys. J. C} \textbf{ 75} (2015) 212,
  \href{http://dx.doi.org/10.1140/epjc/s10052-015-3351-7}{\doi{10.1140/epjc/s10052-015-3351-7}},
\href{http://www.arXiv.org/abs/1412.8662}{\texttt{arXiv:1412.8662}}.

\bibitem{AtlasProperties}
\hrefCMSnoop {}{{ATLAS Collaboration}, ``{Measurements of the Higgs boson
  production and decay rates and coupling strengths using $pp$ collision data
  at $\sqrt{s}=7$ and 8 TeV in the ATLAS experiment}'',} \textit{ Eur. Phys. J.
  C} \textbf{ 76} (2016) 6,
  \href{http://dx.doi.org/10.1140/epjc/s10052-015-3769-y}{\doi{10.1140/epjc/s10052-015-3769-y}},
\href{http://www.arXiv.org/abs/1507.04548}{\texttt{arXiv:1507.04548}}.

\bibitem{CMS:2015kwa}
\hrefCMSnoop {}{{ATLAS and CMS Collaborations}, ``Measurements of the {H}iggs
  boson production and decay rates and constraints on its couplings from a
  combined {ATLAS and CMS} analysis of the {LHC} pp collision data at
  $\sqrt{s}=7$ and 8 {TeV}'',} \textit{ JHEP} \textbf{ 08} (2016) 45,
  \href{http://dx.doi.org/10.1007/JHEP08(2016)045}{\doi{10.1007/JHEP08(2016)045}},
  \href{http://www.arXiv.org/abs/1606.02266}{\texttt{arXiv:1606.02266}}.

\bibitem{Englert:1964et}
\hrefCMSnoop {}{F.~Englert and R.~Brout, ``Broken symmetry and the mass of
  gauge vector mesons'',} \textit{ Phys. Rev. Lett.} \textbf{ 13} (1964) 321,
  \href{http://dx.doi.org/10.1103/PhysRevLett.13.321}{\doi{10.1103/PhysRevLett.13.321}}.

\bibitem{Higgs:1964ia}
\hrefCMSnoop {}{P.~W. Higgs, ``{Broken symmetries, massless particles and gauge
  fields}'',} \textit{ Phys. Lett.} \textbf{ 12} (1964) 132,
  \href{http://dx.doi.org/10.1016/0031-9163(64)91136-9}{\doi{10.1016/0031-9163(64)91136-9}}.

\bibitem{Higgs:1964pj}
\hrefCMSnoop {}{P.~W. Higgs, ``Broken symmetries and the masses of gauge
  bosons'',} \textit{ Phys. Rev. Lett.} \textbf{ 13} (1964) 508,
  \href{http://dx.doi.org/10.1103/PhysRevLett.13.508}{\doi{10.1103/PhysRevLett.13.508}}.

\bibitem{Guralnik:1964eu}
\hrefCMSnoop {}{G.~S. Guralnik, C.~R. Hagen, and T.~W.~B. Kibble, ``Global
  conservation laws and massless particles'',} \textit{ Phys. Rev. Lett.}
  \textbf{ 13} (1964) 585,
  \href{http://dx.doi.org/10.1103/PhysRevLett.13.585}{\doi{10.1103/PhysRevLett.13.585}}.

\bibitem{Higgs:1966ev}
\hrefCMSnoop {}{P.~W. Higgs, ``Spontaneous symmetry breakdown without massless
  bosons'',} \textit{ Phys. Rev.} \textbf{ 145} (1966) 1156,
  \href{http://dx.doi.org/10.1103/PhysRev.145.1156}{\doi{10.1103/PhysRev.145.1156}}.

\bibitem{Kibble:1967sv}
\hrefCMSnoop {}{T.~W.~B. Kibble, ``Symmetry breaking in non-abelian gauge
  theories'',} \textit{ Phys. Rev.} \textbf{ 155} (1967) 1554,
  \href{http://dx.doi.org/10.1103/PhysRev.155.1554}{\doi{10.1103/PhysRev.155.1554}}.

\bibitem{Aad:2015zhl}
\hrefCMSnoop {}{{ATLAS and CMS Collaborations}, ``Combined measurement of the
  {H}iggs boson mass in $pp$ collisions at $\sqrt{s}=7$ and 8 {TeV} with the
  {ATLAS} and {CMS} experiments'',} \textit{ Phys. Rev. Lett.} \textbf{ 114}
  (2015) 191803,
  \href{http://dx.doi.org/10.1103/PhysRevLett.114.191803}{\doi{10.1103/PhysRevLett.114.191803}},
\href{http://www.arXiv.org/abs/1503.07589}{\texttt{arXiv:1503.07589}}.

\bibitem{CMSH4lLegacy}
\hrefCMSnoop {}{{CMS Collaboration}, ``{Measurement of the properties of a
  Higgs boson in the four-lepton final state}'',} \textit{ Phys. Rev. D}
  \textbf{ 89} (2014) 092007,
  \href{http://dx.doi.org/10.1103/PhysRevD.89.092007}{\doi{10.1103/PhysRevD.89.092007}},
\href{http://www.arXiv.org/abs/1312.5353}{\texttt{arXiv:1312.5353}}.

\bibitem{CMSH4lSpinParity}
\hrefCMSnoop {}{{CMS Collaboration}, ``Study of the mass and spin-parity of the
  {H}iggs boson candidate via its decays to {$Z$} boson pairs'',} \textit{
  Phys. Rev. Lett.} \textbf{ 110} (2013) 081803,
  \href{http://dx.doi.org/10.1103/PhysRevLett.110.081803}{\doi{10.1103/PhysRevLett.110.081803}},
\href{http://www.arXiv.org/abs/1212.6639}{\texttt{arXiv:1212.6639}}.

\bibitem{CMSH4lAnomalousCouplings}
\hrefCMSnoop {}{{CMS Collaboration}, ``{Constraints on the spin-parity and
  anomalous HVV couplings of the Higgs boson in proton collisions at 7 and
  8\TeV}'',} \textit{ Phys. Rev. D} \textbf{ 92} (2015) 012004,
  \href{http://dx.doi.org/10.1103/PhysRevD.92.012004}{\doi{10.1103/PhysRevD.92.012004}},
  \href{http://www.arXiv.org/abs/1411.3441}{\texttt{arXiv:1411.3441}}.

\bibitem{Aad:2014eva}
\hrefCMSnoop {}{{ATLAS Collaboration}, ``{Measurements of Higgs boson
  production and couplings in the four-lepton channel in pp collisions at
  center-of-mass energies of 7 and 8 TeV with the ATLAS detector}'',} \textit{
  Phys. Rev. D} \textbf{ 91} (2015) 012006,
  \href{http://dx.doi.org/10.1103/PhysRevD.91.012006}{\doi{10.1103/PhysRevD.91.012006}},
\href{http://www.arXiv.org/abs/1408.5191}{\texttt{arXiv:1408.5191}}.

\bibitem{Aad:2015mxa}
\hrefCMSnoop {}{{ATLAS Collaboration}, ``Study of the spin and parity of the
  {H}iggs boson in diboson decays with the {ATLAS} detector'',} \textit{ Eur.
  Phys. J. C} \textbf{ 75} (2015) 476,
  \href{http://dx.doi.org/10.1140/epjc/s10052-015-3685-1}{\doi{10.1140/epjc/s10052-015-3685-1}},
  \href{http://www.arXiv.org/abs/1506.05669}{\texttt{arXiv:1506.05669}}.
[Erratum: \DOI{10.1140/epjc/s10052-016-3934-y}].

\bibitem{CMSH4lWidth}
\hrefCMSnoop {}{{CMS Collaboration}, ``{Constraints on the Higgs boson width
  from off-shell production and decay to {$\cPZ$}-boson pairs}'',} \textit{
  Phys. Lett. B} \textbf{ 736} (2014) 64,
  \href{http://dx.doi.org/10.1016/j.physletb.2014.06.077}{\doi{10.1016/j.physletb.2014.06.077}},
\href{http://www.arXiv.org/abs/1405.3455}{\texttt{arXiv:1405.3455}}.

\bibitem{CMSH4lLifetime}
\hrefCMSnoop {}{{CMS Collaboration}, ``{Limits on the Higgs boson lifetime and
  width from its decay to four charged leptons}'',} \textit{ Phys. Rev. D}
  \textbf{ 92} (2015) 072010,
  \href{http://dx.doi.org/10.1103/PhysRevD.92.072010}{\doi{10.1103/PhysRevD.92.072010}},
\href{http://www.arXiv.org/abs/1507.06656}{\texttt{arXiv:1507.06656}}.

\bibitem{Aad:2015xua}
\hrefCMSnoop {}{{ATLAS Collaboration}, ``{Constraints on the off-shell Higgs
  boson signal strength in the high-mass $ZZ$ and $WW$ final states with the
  ATLAS detector}'',} \textit{ Eur. Phys. J. C} \textbf{ 75} (2015) 335,
  \href{http://dx.doi.org/10.1140/epjc/s10052-015-3542-2}{\doi{10.1140/epjc/s10052-015-3542-2}},
\href{http://www.arXiv.org/abs/1503.01060}{\texttt{arXiv:1503.01060}}.

\bibitem{CMSH4lFiducial8TeV}
\hrefCMSnoop {}{{CMS Collaboration}, ``Measurement of differential and
  integrated fiducial cross sections for {H}iggs boson production in the
  four-lepton decay channel in pp collisions at $\sqrt{s} = 7$ and {8\TeV}'',}
  \textit{ JHEP} \textbf{ 04} (2016) 005,
  \href{http://dx.doi.org/10.1007/JHEP04(2016)005}{\doi{10.1007/JHEP04(2016)005}},
  \href{http://www.arXiv.org/abs/1512.08377}{\texttt{arXiv:1512.08377}}.

\bibitem{Aad:2015lha}
\hrefCMSnoop {}{{ATLAS Collaboration}, ``Measurements of the total and
  differential {H}iggs boson production cross sections combining the {$H \to
  \gamma \gamma$} and {$H \to ZZ^{*}\to 4l$} decay channels at {$\sqrt{s}=8$
  TeV} with the {ATLAS} detector'',} \textit{ Phys. Rev. Lett.} \textbf{ 115}
  (2015) 091801,
  \href{http://dx.doi.org/10.1103/PhysRevLett.115.091801}{\doi{10.1103/PhysRevLett.115.091801}},
\href{http://www.arXiv.org/abs/1504.05833}{\texttt{arXiv:1504.05833}}.

\bibitem{Chatrchyan:2008zzk}
\hrefCMSnoop {}{{CMS Collaboration}, ``The {CMS} experiment at the {CERN}
  {LHC}'',} \textit{ JINST} \textbf{ 3} (2008) S08004,
\href{http://dx.doi.org/10.1088/1748-0221/3/08/S08004}{\doi{10.1088/1748-0221/3/08/S08004}}.

\bibitem{TRK-11-001}
\hrefCMSnoop {}{{CMS Collaboration}, ``{Description and performance of track
  and primary-vertex reconstruction with the CMS tracker}'',} \textit{ JINST}
  \textbf{ 9} (2014) P10009,
  \href{http://dx.doi.org/10.1088/1748-0221/9/10/P10009}{\doi{10.1088/1748-0221/9/10/P10009}},
\href{http://www.arXiv.org/abs/1405.6569}{\texttt{arXiv:1405.6569}}.

\bibitem{Khachatryan:2015hwa}
\hrefCMSnoop {}{{CMS Collaboration}, ``{Performance of electron reconstruction
  and selection with the CMS detector in proton-proton collisions at $\sqrt{s}
  = 8$\TeV}'',} \textit{ JINST} \textbf{ 10} (2015) P06005,
  \href{http://dx.doi.org/10.1088/1748-0221/10/06/P06005}{\doi{10.1088/1748-0221/10/06/P06005}},
\href{http://www.arXiv.org/abs/1502.02701}{\texttt{arXiv:1502.02701}}.

\bibitem{Chatrchyan:2012xi}
\hrefCMSnoop {}{{CMS Collaboration}, ``{Performance of CMS muon reconstruction
  in pp collision events at $\sqrt{s} = 7$\TeV}'',} \textit{ JINST} \textbf{ 7}
  (2012) P10002,
  \href{http://dx.doi.org/10.1088/1748-0221/7/10/P10002}{\doi{10.1088/1748-0221/7/10/P10002}},
\href{http://www.arXiv.org/abs/1206.4071}{\texttt{arXiv:1206.4071}}.

\bibitem{CMS-TRG-12-001}
\hrefCMSnoop {}{{CMS Collaboration}, ``The {CMS} trigger system'',} \textit{
  JINST} \textbf{ 12} (2017) P01020,
  \href{http://dx.doi.org/10.1088/1748-0221/12/01/P01020}{\doi{10.1088/1748-0221/12/01/P01020}},
  \href{http://www.arXiv.org/abs/1609.02366}{\texttt{arXiv:1609.02366}}.

\bibitem{CMS:2011aa}
\hrefCMSnoop {}{{CMS Collaboration}, ``{Measurement of the inclusive W and Z
  production cross sections in pp collisions at $\sqrt{s}=7$ TeV}'',} \textit{
  JHEP} \textbf{ 10} (2011) 132,
  \href{http://dx.doi.org/10.1007/JHEP10(2011)132}{\doi{10.1007/JHEP10(2011)132}},
  \href{http://www.arXiv.org/abs/1107.4789}{\texttt{arXiv:1107.4789}}.

\bibitem{Alioli:2008gx}
\hrefCMSnoop {}{S.~Alioli, P.~Nason, C.~Oleari, and E.~Re, ``{NLO vector-boson
  production matched with shower in POWHEG}'',} \textit{ JHEP} \textbf{ 07}
  (2008) 060,
  \href{http://dx.doi.org/10.1088/1126-6708/2008/07/060}{\doi{10.1088/1126-6708/2008/07/060}},
\href{http://www.arXiv.org/abs/0805.4802}{\texttt{arXiv:0805.4802}}.

\bibitem{Nason:2004rx}
\hrefCMSnoop {}{P.~Nason, ``{A new method for combining NLO QCD with shower
  Monte Carlo algorithms}'',} \textit{ JHEP} \textbf{ 11} (2004) 040,
  \href{http://dx.doi.org/10.1088/1126-6708/2004/11/040}{\doi{10.1088/1126-6708/2004/11/040}},
\href{http://www.arXiv.org/abs/hep-ph/0409146}{\texttt{arXiv:hep-ph/0409146}}.

\bibitem{Frixione:2007vw}
\hrefCMSnoop {}{S.~Frixione, P.~Nason, and C.~Oleari, ``{Matching NLO QCD
  computations with parton shower simulations: the POWHEG method}'',} \textit{
  JHEP} \textbf{ 11} (2007) 070,
  \href{http://dx.doi.org/10.1088/1126-6708/2007/11/070}{\doi{10.1088/1126-6708/2007/11/070}},
\href{http://www.arXiv.org/abs/0709.2092}{\texttt{arXiv:0709.2092}}.

\bibitem{Luisoni2013}
\hrefCMSnoop {}{G.~Luisoni, P.~Nason, C.~Oleari, and F.~Tramontano,
  ``{HW$^{\pm}$/HZ + 0} and 1 jet at {NLO} with the {POWHEG BOX} interfaced to
  {GoSam} and their merging within {MiNLO}'',} \textit{ JHEP} \textbf{ 10}
  (2013) 1,
  \href{http://dx.doi.org/10.1007/JHEP10(2013)083}{\doi{10.1007/JHEP10(2013)083}},
  \href{http://www.arXiv.org/abs/1306.2542}{\texttt{arXiv:1306.2542}}.

\bibitem{YR4}
\hrefCMSnoop {}{D.~de~Florian {et~al.}, ``Handbook of {LHC} {H}iggs cross
  sections: 4. deciphering the nature of the {H}iggs sector'',} CERN Report
  CERN-2017-002-M, 2016.
\newblock
  \href{http://dx.doi.org/10.23731/CYRM-2017-002}{\doi{10.23731/CYRM-2017-002}},
  \href{http://www.arXiv.org/abs/1610.07922}{\texttt{arXiv:1610.07922}}.

\bibitem{Anastasiou2016}
C.~Anastasiou\hrefCMSnoop {}{ {et~al.}, ``High precision determination of the
  gluon fusion {Higgs} boson cross-section at the {LHC}'',} \textit{ JHEP}
  \textbf{ 05} (2016) 058,
  \href{http://dx.doi.org/10.1007/JHEP05(2016)058}{\doi{10.1007/JHEP05(2016)058}},
  \href{http://www.arXiv.org/abs/1602.00695}{\texttt{arXiv:1602.00695}}.

\bibitem{Ball2012153}
R.~D. Ball\hrefCMSnoop {}{ {et~al.}, ``Unbiased global determination of parton
  distributions and their uncertainties at {NNLO} and at {LO}'',} \textit{
  Nucl. Phys. B} \textbf{ 855} (2012) 153,
  \href{http://dx.doi.org/10.1016/j.nuclphysb.2011.09.024}{\doi{10.1016/j.nuclphysb.2011.09.024}},
  \href{http://www.arXiv.org/abs/1107.2652}{\texttt{arXiv:1107.2652}}.

\bibitem{Gao:2010qx}
Y.~Gao\hrefCMSnoop {}{ {et~al.}, ``{Spin determination of single-produced
  resonances at hadron colliders}'',} \textit{ Phys. Rev. D} \textbf{ 81}
  (2010) 075022,
  \href{http://dx.doi.org/10.1103/PhysRevD.81.075022}{\doi{10.1103/PhysRevD.81.075022}},
  \href{http://www.arXiv.org/abs/1001.3396}{\texttt{arXiv:1001.3396}}.
[Erratum: \DOI{10.1103/PhysRevD.81.079905}].

\bibitem{Bolognesi:2012mm}
S.~Bolognesi\hrefCMSnoop {}{ {et~al.}, ``{On the spin and parity of a
  single-produced resonance at the LHC}'',} \textit{ Phys. Rev. D} \textbf{ 86}
  (2012) 095031,
  \href{http://dx.doi.org/10.1103/PhysRevD.86.095031}{\doi{10.1103/PhysRevD.86.095031}},
\href{http://www.arXiv.org/abs/1208.4018}{\texttt{arXiv:1208.4018}}.

\bibitem{deFlorian:2012mx}
\hrefCMSnoop {}{D.~de~Florian, G.~Ferrera, M.~Grazzini, and D.~Tommasini,
  ``{Higgs boson production at the LHC: transverse momentum resummation effects
  in the $H \to \gamma \gamma$, $H \to WW \to \ell\nu\ell\nu$ and $H \to ZZ \to
  4\ell$ decay modes}'',} \textit{ JHEP} \textbf{ 06} (2012) 132,
  \href{http://dx.doi.org/10.1007/JHEP06(2012)132}{\doi{10.1007/JHEP06(2012)132}},
\href{http://www.arXiv.org/abs/1203.6321}{\texttt{arXiv:1203.6321}}.

\bibitem{Grazzini:2013mca}
\hrefCMSnoop {}{M.~Grazzini and H.~Sargsyan, ``{Heavy-quark mass effects in
  Higgs boson production at the LHC}'',} \textit{ JHEP} \textbf{ 09} (2013)
  129,
  \href{http://dx.doi.org/10.1007/JHEP09(2013)129}{\doi{10.1007/JHEP09(2013)129}},
\href{http://www.arXiv.org/abs/1306.4581}{\texttt{arXiv:1306.4581}}.

\bibitem{MCFM}
\hrefCMSnoop {}{J.~M. Campbell and R.~K. Ellis, ``{MCFM for the Tevatron and
  the LHC}'',} \textit{ Nucl. Phys. Proc. Suppl.} \textbf{ 205} (2010) 10,
  \href{http://dx.doi.org/10.1016/j.nuclphysbps.2010.08.011}{\doi{10.1016/j.nuclphysbps.2010.08.011}},
\href{http://www.arXiv.org/abs/1007.3492}{\texttt{arXiv:1007.3492}}.

\bibitem{Sjostrand2015159}
T.~Sj{\"o}strand\hrefCMSnoop {}{ {et~al.}, ``An introduction to {PYTHIA}
  8.2'',} \textit{ Comput. Phys. Commun.} \textbf{ 191} (2015) 159,
  \href{http://dx.doi.org/10.1016/j.cpc.2015.01.024}{\doi{10.1016/j.cpc.2015.01.024}},
\href{http://www.arXiv.org/abs/1410.3012}{\texttt{arXiv:1410.3012}}.

\bibitem{Khachatryan:2015pea}
\hrefCMSnoop {}{{CMS Collaboration}, ``{Event generator tunes obtained from
  underlying event and multiparton scattering measurements}'',} \textit{ Eur.
  Phys. J. C} \textbf{ 76} (2016) 155,
  \href{http://dx.doi.org/10.1140/epjc/s10052-016-3988-x}{\doi{10.1140/epjc/s10052-016-3988-x}},
\href{http://www.arXiv.org/abs/1512.00815}{\texttt{arXiv:1512.00815}}.

\bibitem{Agostinelli:2002hh}
\hrefCMSnoop {}{{GEANT4} Collaboration, ``{GEANT4}---a simulation toolkit'',}
  \textit{ Nucl. Instrum. Meth. A} \textbf{ 506} (2003) 250,
\href{http://dx.doi.org/10.1016/S0168-9002(03)01368-8}{\doi{10.1016/S0168-9002(03)01368-8}}.

\bibitem{GEANT}
\hrefCMSnoop {}{J.~Allison {et~al.}, ``{Geant4 developments and
  applications}'',} \textit{ IEEE Trans. Nucl. Sci.} \textbf{ 53} (2006) 270,
\href{http://dx.doi.org/10.1109/TNS.2006.869826}{\doi{10.1109/TNS.2006.869826}}.

\bibitem{Sirunyan:2017ulk}
\hrefCMSnoop {}{{CMS Collaboration}, ``Particle-flow reconstruction and global
  event description with the {CMS} detector'',} (2017).
  \href{http://www.arXiv.org/abs/1706.04965}{\texttt{arXiv:1706.04965}}.
Submitted to {JINST}.

\bibitem{Cacciari:2008gp}
\hrefCMSnoop {}{M.~Cacciari, G.~P. Salam, and G.~Soyez, ``{The anti-$k_t$ jet
  clustering algorithm}'',} \textit{ JHEP} \textbf{ 04} (2008) 063,
  \href{http://dx.doi.org/10.1088/1126-6708/2008/04/063}{\doi{10.1088/1126-6708/2008/04/063}},
  \href{http://www.arXiv.org/abs/0802.1189}{\texttt{arXiv:0802.1189}}.

\bibitem{Cacciari:2011ma}
\hrefCMSnoop {}{M.~Cacciari, G.~P. Salam, and G.~Soyez, ``{FastJet user
  manual}'',} \textit{ Eur. Phys. J. C} \textbf{ 72} (2012) 1896,
  \href{http://dx.doi.org/10.1140/epjc/s10052-012-1896-2}{\doi{10.1140/epjc/s10052-012-1896-2}},
\href{http://www.arXiv.org/abs/1111.6097}{\texttt{arXiv:1111.6097}}.

\bibitem{Cacciari:2007fd}
\hrefCMSnoop {}{M.~Cacciari and G.~P. Salam, ``{Pileup subtraction using jet
  areas}'',} \textit{ Phys. Lett. B} \textbf{ 659} (2008) 119,
  \href{http://dx.doi.org/10.1016/j.physletb.2007.09.077}{\doi{10.1016/j.physletb.2007.09.077}},
\href{http://www.arXiv.org/abs/0707.1378}{\texttt{arXiv:0707.1378}}.

\bibitem{Cacciari:2008gn}
\hrefCMSnoop {}{M.~Cacciari, G.~P. Salam, and G.~Soyez, ``The catchment area of
  jets'',} \textit{ JHEP} \textbf{ 04} (2008) 005,
  \href{http://dx.doi.org/10.1088/1126-6708/2008/04/005}{\doi{10.1088/1126-6708/2008/04/005}},
\href{http://www.arXiv.org/abs/0802.1188}{\texttt{arXiv:0802.1188}}.

\bibitem{CMS:EGM-14-001}
\hrefCMSnoop {}{{CMS Collaboration}, ``{Performance of photon reconstruction
  and identification with the CMS detector in proton-proton collisions at
  $\sqrt{s} = 8$\TeV}'',} \textit{ JINST} \textbf{ 10} (2015) P08010,
  \href{http://dx.doi.org/10.1088/1748-0221/10/08/P08010}{\doi{10.1088/1748-0221/10/08/P08010}},
\href{http://www.arXiv.org/abs/1502.02702}{\texttt{arXiv:1502.02702}}.

\bibitem{Fruhwirth:1987fm}
\hrefCMSnoop {}{R.~Fruhwirth, ``{Application of Kalman filtering to track and
  vertex fitting}'',} \textit{ Nucl. Instrum. Meth. A} \textbf{ 262} (1987)
  444,
\href{http://dx.doi.org/10.1016/0168-9002(87)90887-4}{\doi{10.1016/0168-9002(87)90887-4}}.

\bibitem{Chatrchyan:2011ds}
\hrefCMSnoop {}{{CMS Collaboration}, ``{Determination of jet energy calibration
  and transverse momentum resolution in CMS}'',} \textit{ JINST} \textbf{ 6}
  (2011) P11002,
  \href{http://dx.doi.org/10.1088/1748-0221/6/11/P11002}{\doi{10.1088/1748-0221/6/11/P11002}},
\href{http://www.arXiv.org/abs/1107.4277}{\texttt{arXiv:1107.4277}}.

\bibitem{Khachatryan:2016kdb}
\hrefCMSnoop {}{{CMS Collaboration}, ``{Jet energy scale and resolution in the
  CMS experiment in pp collisions at 8 TeV}'',} \textit{ JINST} \textbf{ 12}
  (2017) P02014,
  \href{http://dx.doi.org/10.1088/1748-0221/12/02/P02014}{\doi{10.1088/1748-0221/12/02/P02014}},
\href{http://www.arXiv.org/abs/1607.03663}{\texttt{arXiv:1607.03663}}.

\bibitem{Chatrchyan:2012jua}
\hrefCMSnoop {}{{CMS Collaboration}, ``{Identification of b-quark jets with the
  CMS experiment}'',} \textit{ JINST} \textbf{ 8} (2013) P04013,
  \href{http://dx.doi.org/10.1088/1748-0221/8/04/P04013}{\doi{10.1088/1748-0221/8/04/P04013}},
\href{http://www.arXiv.org/abs/1211.4462}{\texttt{arXiv:1211.4462}}.

\bibitem{CMS-PAS-BTV-15-001}
\href {https://cds.cern.ch/record/2138504}{{CMS Collaboration},
  ``{Identification of b quark jets at the CMS Experiment in the LHC Run 2}'',}
  CMS Physics Analysis Summary CMS-PAS-BTV-15-001, 2016.

\bibitem{Zmass}
\hrefCMSnoop {}{{ALEPH, DELPHI, L3, OPAL, SLD Collaborations, LEP Electroweak
  Working Group, and SLD Electroweak and Heavy Flavour Groups}, ``Precision
  electroweak measurements on the {Z} resonance'',} \textit{ Phys. Rep.}
  \textbf{ 427} (2006) 257,
  \href{http://dx.doi.org/10.1016/j.physrep.2005.12.006}{\doi{10.1016/j.physrep.2005.12.006}}.

\bibitem{Anderson:2013afp}
I.~Anderson\hrefCMSnoop {}{ {et~al.}, ``{Constraining anomalous $HVV$
  interactions at proton and lepton colliders}'',} \textit{ Phys. Rev. D}
  \textbf{ 89} (2014) 035007,
  \href{http://dx.doi.org/10.1103/PhysRevD.89.035007}{\doi{10.1103/PhysRevD.89.035007}},
\href{http://www.arXiv.org/abs/1309.4819}{\texttt{arXiv:1309.4819}}.

\bibitem{CrystalBall}
\href
  {http://www.slac.stanford.edu/pubs/slacreports/slac-r-236.html}{M.~Oreglia,
  ``{A study of the reactions $\psi^\prime \to \gamma \gamma \psi$}''}.
\newblock PhD thesis, Stanford University, 1980.
\newblock {{SLAC} Report {SLAC-R-236}}.

\bibitem{Grazzini2015407}
\hrefCMSnoop {}{M.~Grazzini, S.~Kallweit, and D.~Rathlev, ``{ZZ} production at
  the {LHC}: Fiducial cross sections and distributions in {NNLO QCD}'',}
  \textit{ Phys. Lett. B} \textbf{ 750} (2015) 407,
  \href{http://dx.doi.org/10.1016/j.physletb.2015.09.055}{\doi{10.1016/j.physletb.2015.09.055}},
  \href{http://www.arXiv.org/abs/1507.06257}{\texttt{arXiv:1507.06257}}.

\bibitem{Bierweiler:1312}
\hrefCMSnoop {}{A.~Bierweiler, T.~Kasprzik, and J.~H. K{\"u}hn, ``{Vector-boson
  pair production at the LHC to ${\cal O}(\alpha^3)$ accuracy}'',} \textit{
  JHEP} \textbf{ 12} (2013) 071,
  \href{http://dx.doi.org/10.1007/JHEP12(2013)071}{\doi{10.1007/JHEP12(2013)071}},
\href{http://www.arXiv.org/abs/1305.5402}{\texttt{arXiv:1305.5402}}.

\bibitem{Bonvini:1304.3053}
M.~Bonvini\hrefCMSnoop {}{ {et~al.}, ``{Signal-background interference effects
  in $gg \to H \to WW$ beyond leading order}'',} \textit{ Phys. Rev. D}
  \textbf{ 88} (2013) 034032,
  \href{http://dx.doi.org/10.1103/PhysRevD.88.034032}{\doi{10.1103/PhysRevD.88.034032}},
\href{http://www.arXiv.org/abs/1304.3053}{\texttt{arXiv:1304.3053}}.

\bibitem{Melnikov:2015laa}
\hrefCMSnoop {}{K.~Melnikov and M.~Dowling, ``{Production of two Z-bosons in
  gluon fusion in the heavy top quark approximation}'',} \textit{ Phys. Lett.
  B} \textbf{ 744} (2015) 43,
  \href{http://dx.doi.org/10.1016/j.physletb.2015.03.030}{\doi{10.1016/j.physletb.2015.03.030}},
\href{http://www.arXiv.org/abs/1503.01274}{\texttt{arXiv:1503.01274}}.

\bibitem{Li:2015jva}
\hrefCMSnoop {}{C.~S. Li, H.~T. Li, D.~Y. Shao, and J.~Wang, ``Soft gluon
  resummation in the signal-background interference process of {$gg(\to h^*)
  \to ZZ$~}'',} \textit{ JHEP} \textbf{ 08} (2015) 065,
  \href{http://dx.doi.org/10.1007/JHEP08(2015)065}{\doi{10.1007/JHEP08(2015)065}},
\href{http://www.arXiv.org/abs/1504.02388}{\texttt{arXiv:1504.02388}}.

\bibitem{Passarino:1312.2397v1}
\hrefCMSnoop {}{G.~Passarino, ``{Higgs CAT}'',} \textit{ Eur. Phys. J. C}
  \textbf{ 74} (2014) 2866,
  \href{http://dx.doi.org/10.1140/epjc/s10052-014-2866-7}{\doi{10.1140/epjc/s10052-014-2866-7}},
\href{http://www.arXiv.org/abs/1312.2397}{\texttt{arXiv:1312.2397}}.

\bibitem{Catani:2007vq}
\hrefCMSnoop {}{S.~Catani and M.~Grazzini, ``{Next-to-Next-to-Leading-Order}
  subtraction formalism in hadron collisions and its application to
  {H}iggs-boson production at the {Large Hadron Collider}'',} \textit{ Phys.
  Rev. Lett.} \textbf{ 98} (2007) 222002,
  \href{http://dx.doi.org/10.1103/PhysRevLett.98.222002}{\doi{10.1103/PhysRevLett.98.222002}},
\href{http://www.arXiv.org/abs/hep-ph/0703012}{\texttt{arXiv:hep-ph/0703012}}.

\bibitem{Grazzini:2008tf}
\hrefCMSnoop {}{M.~Grazzini, ``{NNLO predictions for the Higgs boson signal in
  the H $\to$ WW $\to\ell\nu\ell\nu$ and H$\to$ ZZ $\to4\ell$ decay
  channels}'',} \textit{ JHEP} \textbf{ 02} (2008) 043,
  \href{http://dx.doi.org/10.1088/1126-6708/2008/02/043}{\doi{10.1088/1126-6708/2008/02/043}},
\href{http://www.arXiv.org/abs/0801.3232}{\texttt{arXiv:0801.3232}}.

\bibitem{Landau:1944if}
\hrefCMSnoop {}{L.~Landau, ``{On the energy loss of fast particles by
  ionization}'',} \textit{ J. Phys. (USSR)} \textbf{ 8} (1944)
201.

\bibitem{CMS-PAS-LUM-17-001}
\href {https://cds.cern.ch/record/2257069}{{CMS Collaboration}, ``{CMS}
  luminosity measurements for the 2016 data taking period'',} CMS Physics
  Analysis Summary CMS-PAS-LUM-17-001, 2017.

\bibitem{Butterworth:2015oua}
\hrefCMSnoop {}{J.~Butterworth {et~al.}, ``{PDF4LHC} recommendations for {LHC
  Run II}'',} \textit{ J. Phys. G} \textbf{ 43} (2016) 023001,
  \href{http://dx.doi.org/10.1088/0954-3899/43/2/023001}{\doi{10.1088/0954-3899/43/2/023001}},
\href{http://www.arXiv.org/abs/1510.03865}{\texttt{arXiv:1510.03865}}.

\bibitem{Garwood}
\hrefCMSnoop {}{F.~Garwood, ``Fiducial limits for the poisson distribution'',}
  \textit{ Biometrika} \textbf{ 28} (1936), no.~3-4, 437,
  \href{http://dx.doi.org/10.1093/biomet/28.3-4.437}{\doi{10.1093/biomet/28.3-4.437}}.

\bibitem{LHC-HCG}
\href {http://cdsweb.cern.ch/record/1379837}{{ATLAS and CMS Collaborations, LHC
  Higgs Combination Group}, ``Procedure for the {LHC} {Higgs} boson search
  combination in {Summer} 2011'',} ATL-PHYS-PUB/CMS NOTE 2011-11, 2011/005,
  CERN, 2011.

\bibitem{Cowan:2010js}
\hrefCMSnoop {}{G.~Cowan, K.~Cranmer, E.~Gross, and O.~Vitells, ``{Asymptotic
  formulae for likelihood-based tests of new physics}'',} \textit{ Eur. Phys.
  J. C} \textbf{ 71} (2011) 1554,
  \href{http://dx.doi.org/10.1140/epjc/s10052-011-1554-0}{\doi{10.1140/epjc/s10052-011-1554-0}},
\href{http://www.arXiv.org/abs/1007.1727}{\texttt{arXiv:1007.1727}}.

\bibitem{FC}
\hrefCMSnoop {}{G.~J. Feldman and R.~D. Cousins, ``Unified approach to the
  classical statistical analysis of small signals'',} \textit{ Phys. Rev. D}
  \textbf{ 57} (1998) 3873,
  \href{http://dx.doi.org/10.1103/PhysRevD.57.3873}{\doi{10.1103/PhysRevD.57.3873}},
\href{http://www.arXiv.org/abs/physics/9711021}{\texttt{arXiv:physics/9711021}}.

\bibitem{CMSHggFiducial8TeV}
\hrefCMSnoop {}{{CMS Collaboration}, ``Measurement of differential cross
  sections for {H}iggs boson production in the diphoton decay channel in pp
  collisions at {$\sqrt{s}=8\TeV$}'',} \textit{ Eur. Phys. J. C} \textbf{ 76}
  (2015) 13,
  \href{http://dx.doi.org/10.1140/epjc/s10052-015-3853-3}{\doi{10.1140/epjc/s10052-015-3853-3}}.

\bibitem{NNLOPS}
\hrefCMSnoop {}{K.~Hamilton, P.~Nason, E.~Re, and G.~Zanderighi, ``{NNLOPS}
  simulation of {Higgs} boson production'',} \textit{ JHEP} \textbf{ 10} (2013)
  222,
  \href{http://dx.doi.org/10.1007/JHEP10(2013)222}{\doi{10.1007/JHEP10(2013)222}},
  \href{http://www.arXiv.org/abs/1309.0017}{\texttt{arXiv:1309.0017}}.

\end{thebibliography}\endgroup

\cleardoublepage \appendix\section{The CMS Collaboration \label{app:collab}}\begin{sloppypar}\hyphenpenalty=5000\widowpenalty=500\clubpenalty=5000\textbf{Yerevan Physics Institute,  Yerevan,  Armenia}\\*[0pt]
A.M.~Sirunyan, A.~Tumasyan
\vskip\cmsinstskip
\textbf{Institut f\"{u}r Hochenergiephysik,  Wien,  Austria}\\*[0pt]
W.~Adam, F.~Ambrogi, E.~Asilar, T.~Bergauer, J.~Brandstetter, E.~Brondolin, M.~Dragicevic, J.~Er\"{o}, M.~Flechl, M.~Friedl, R.~Fr\"{u}hwirth\cmsAuthorMark{1}, V.M.~Ghete, J.~Grossmann, J.~Hrubec, M.~Jeitler\cmsAuthorMark{1}, A.~K\"{o}nig, N.~Krammer, I.~Kr\"{a}tschmer, D.~Liko, T.~Madlener, I.~Mikulec, E.~Pree, D.~Rabady, N.~Rad, H.~Rohringer, J.~Schieck\cmsAuthorMark{1}, R.~Sch\"{o}fbeck, M.~Spanring, D.~Spitzbart, W.~Waltenberger, J.~Wittmann, C.-E.~Wulz\cmsAuthorMark{1}, M.~Zarucki
% ...
\vskip\cmsinstskip
% ...
1:~~Also at Vienna University of Technology, Vienna, Austria\\
% ...

\end{sloppypar}
\end{document}